%% file: GroupProfiles.tex
\documentclass[fleqn,usenatbib]{mnras}

\usepackage{newtxtext,newtxmath}

\usepackage[T1]{fontenc}
\usepackage{ae,aecompl}

\DeclareRobustCommand{\VAN}[3]{#3}
\let\VANthebibliography\thebibliography
\def\thebibliography{\DeclareRobustCommand{\VAN}[3]{##3}\VANthebibliography}

\usepackage{graphicx}	
\usepackage{bm} 

\usepackage{makecell}
\usepackage{multirow}

\newcommand{\mass}{$\mathcal{M}$}
\newcommand{\hMpc}{h^{-1} \mathrm{Mpc}}
\newcommand{\hMsun}{h^{-1} \mathrm{M}_{\odot}}
\newcommand{\Msun}{\mathrm{M}_{\odot}}
\newcommand{\gtc}{\mbox{G$^3$C}}

\let\la=\lesssim 
\let\ga=\gtrsim

\title[Satellite radial distributions]{Exploring the effect of baryons on the radial distribution of satellite galaxies with GAMA and IllustrisTNG}

\include{authors}

\date{Accepted XXX. Received YYY; in original form ZZZ}

\pubyear{2022}

\begin{document}
\label{firstpage}
\pagerange{\pageref{firstpage}--\pageref{lastpage}}
\maketitle

\begin{abstract}
We explore the radial distribution of satellite galaxies in groups in the Galaxy and Mass Assembly (GAMA) survey and the IllustrisTNG simulations. Considering groups with masses $12.0 \leq \log_{10} (\mathcal{M}_h / \hMsun) < 14.8$ at $z<0.267$, we find a good agreement between GAMA and a sample of TNG300 groups and galaxies designed to match the GAMA selection. Both display a flat profile in the centre of groups, followed by a decline that becomes steeper towards the group edge, and normalised profiles show no dependence on group mass. Using matched satellites from TNG and dark matter-only TNG-Dark runs we investigate the effect of baryons on satellite radial location. At $z=0$, we find that the matched subhaloes from the TNG-Dark runs display a much flatter radial profile: namely, satellites selected above a minimum stellar mass exhibit both smaller halo-centric distances and longer survival times in the full-physics simulations compared to their dark-matter only analogues. We then divide the TNG satellites into those which possess TNG-Dark counterparts and those which do not, and develop models for the radial positions of each. We find the satellites with TNG-Dark counterparts are displaced towards the halo centre in the full-physics simulations, and this difference has a power-law behaviour with radius. For the `orphan' galaxies without TNG-Dark counterparts, we consider the shape of their radial distribution and provide a model for their motion over time, which can be used to improve the treatment of satellite galaxies in semi-analytic and semi-empirical models of galaxy formation.

\end{abstract}




\section{Introduction}
  \input{sections/introduction.tex}

\section{Data, simulations and methods}
\label{sec:GP2}
  \input{sections/data.tex}

  \input{sections/methods.tex}

\section{Radial profiles from GAMA and TNG300-1}
\label{sec:GP3}
  \input{sections/gama_profile.tex}
  
\section{Comparing full-physics and DMO distributions}
\label{sec:GP4}
  \input{sections/tngdark_profile.tex}

\section{Fitting differences between full-physics and DMO runs}
\label{sec:GP5}
  \input{sections/fitting_selection.tex}

  \input{sections/type1_model.tex}

\section{Fitting the locations of unmatched satellites}
\label{sec:GP6}
  \input{sections/type2_profile.tex}
  \input{sections/type2_model.tex}

\section{Discussion and caveats}
\label{sec:GP7}
  \input{sections/model_tests.tex}
  \input{sections/discussion.tex}

\section{Conclusions}
\label{sec:GP8}
  \input{sections/conclusions.tex}


\section*{Acknowledgements}

SDR is supported by a Science and Technology Facilities Council (STFC) studentship.
JL acknowledges support from the STFC (grant number ST/I000976/1).

We thank Andrew Hopkins for suggestions about the manuscript.
We also thank the anonymous referee who provided comments which improved the manuscript.

GAMA is a joint European-Australasian project based around a
spectroscopic campaign using the Anglo-Australian Telescope. The
GAMA input catalogue is based on data taken from the Sloan Digital
Sky Survey and the UKIRT Infrared Deep Sky Survey. Complementary
imaging of the GAMA regions is being obtained by
a number of independent survey programs including GALEX MIS,
VST KiDS, VISTA VIKING, WISE, Herschel-ATLAS, GMRT and
ASKAP providing UV to radio coverage. GAMA is funded by the
STFC (UK), the ARC (Australia), the AAO, and the participating
institutions. The GAMA website is http://www.gama-survey.org/.

\section*{Data availability}

The TNG simulations are publicly available from the IllustrisTNG repository: 
https://www.tng-project.org.
The processed data of this article will be shared on 
reasonable request to the corresponding author.



\bibliographystyle{mnras}
\bibliography{refs} 



\appendix
  \input{sections/appendix}


\bsp	
\label{lastpage}
\end{document}

%% file: authors.tex
\author[S.~D.~Riggs et al.]
{{\parbox{\textwidth}{\raggedright Stephen~D.~Riggs,$^{1}$\thanks{E-mail:~S.Riggs@sussex.ac.uk}
Jon~Loveday,$^{1}$
Peter~A.~Thomas,$^{1}$
Annalisa~Pillepich,$^{2}$
Dylan~Nelson,$^{3}$
Benne~W.~Holwerda$^{4}$
}}\\
\vspace{0.4cm}\\
{\parbox{\textwidth}{\raggedright 
$^{1}$Astronomy Centre, University of Sussex, Falmer, Brighton BN1 9QH, UK
\\
$^{2}$Max-Planck-Institut f\"{u}r Astronomie, K\"{o}nigstuhl 17, 69117 Heidelberg, Germany
\\
$^{3}$Universit\"{a}t Heidelberg, Zentrum f\"{u}r Astronomie, Institut für theoretische Astrophysik, Albert-Ueberle-Str. 2, 69120 Heidelberg, Germany
\\
$^{4}$Department of Physics and Astronomy, University of Louisville, Louisville, KY 40292, USA
\\
}}}

%% file: sections/introduction.tex
In the $\Lambda$CDM model of the Universe,
galaxies form in dark matter haloes.
The dark matter interacts only by gravity,
forming structures into which gas collapses to form
stars and thus galaxies.
However, this gravity-only model of structure is incomplete,
as the baryonic physics of the galaxies is known to affect 
the halo structures in which they reside.
One way in which this manifests is in the number and location
of substructures, which can host luminous satellite galaxies.
This can be explored through the clustering of galaxies
or by the radial profiles of satellite galaxy locations within groups.

Much of the importance of understanding the differences between 
a dark matter-only (DMO) view of the Universe and a full-physics view
comes from the use of galaxy formation models built upon DMO simulations.
Semi-analytic models of galaxy formation \citep[SAMs; e.g.][]{Henriques2015, Lacey2016, Lagos2018}
are one of these.
In many SAMs, satellite galaxies are split into two populations: Type 1s and Type 2s.
Type 1 satellites reside in resolved dark matter subhaloes, which have not been disrupted,
and it is assumed the locations of these are the same as in the underlying DMO simulation.
Type 2 satellites, or `orphan' galaxies, are those which have persisted beyond the 
lifetime of their host dark matter subhalo (see e.g. \citealt{Pujol2017}),
meaning the locations of these satellites 
are not available from the simulation itself, and require additional modelling.

These Type 2 satellites are necessary as 
it has been found that DMO simulations generically have too few subhaloes that would host galaxies
in the inner regions of haloes, compared to the number of galaxies seen in observations.
For example, this is seen by \cite{Angulo2009},
where it is also noted that more massive subhaloes are less centrally concentrated 
as they experience greater dynamical friction and merge quickly if they are near the centre,
and by \cite{Bose2020}, who are unable to reproduce the satellite population
of the Milky Way from DMO simulations without Type 2 satellites.
Further, \cite{Behroozi2019} argue that without orphans the stellar masses of 
the other satellite galaxies would need to be increased in a manner that is inconsistent with their known evolution.
However, Type 2s are often viewed as a resolution issue, 
and some studies \citep[e.g.][]{Manwadkar2021} have been able to avoid the need for them
by using only more massive subhaloes.

On the other hand, cosmological hydrodynamical galaxy simulations allow exploration of 
the effects of baryons on structures directly. 
The addition of baryons, hydrodynamics, and galaxy processes changes both the masses \citep[e.g.][]{Sawala2013, Despali2017, Lovell2018}
and the abundances \citep[e.g.][]{Schaller2015, Chua2017} of (sub)structures,
as well as the distributions \citep[e.g.][]{Marini2021}.
In the Illustris simulation, the distribution of satellite galaxies 
from the centre of their host halo has been considered by \cite{Vogelsberger2014a},
where they show 
that the number density of satellite galaxies is enhanced on small scales compared to subhaloes in a DMO simulation.
The distribution of galaxies around clusters is also shown to be different for DMO and full-physics simulations
by \cite{Haggar2021}, using \textsc{TheThreeHundred} project.
They show that DMO simulations both do not have a high enough subhalo density near the cluster centre
compared to the full-physics simulations,
and have a subhalo density that is too low within groups of satellites which reside at the cluster edge.
Further, \cite{Nagai2005} find that differences between simulations depend on the object selection
due to tidal stripping and that the addition of baryons slightly enhances satellite survival.
However, baryons can also reduce satellite survival due to disruption by a disc \citep[e.g.][]{GarrisonKimmel2017}.

The IllustrisTNG cosmological magnetohydrodynamical simulations 
\citep[TNG,][]{Marinacci2018, Naiman2018, Nelson2018, Nelson2019a, Pillepich2018, Springel2018} 
are a recent set of simulations consisting of 3 different box sizes, 
each run at 3 different resolutions. 
The existence of dark matter-only counterparts to each of these simulations
provides the opportunity to explore the effect of baryons on
satellite galaxies in more detail and
across a greater range of resolutions than has previously been possible.
This is particularly true for the highest resolution TNG50 simulation \citep{Nelson2019b, Pillepich2019},
which is designed to match the resolution of zoom simulations 
while providing a much greater volume, enabling a detailed look inside simulated galaxies and haloes.

Differences between the full-physics TNG and the DMO TNG-Dark runs have been found in a number of studies,
with \cite{Chua2021}, \cite{Emami2021} and \cite{Anbajagane2022} finding that the baryons 
change the properties of haloes, including the shapes.
Of most relevance to our study,
\cite{Bose2019} show that the distribution of satellite galaxies in the full-physics runs differs from
that of subhaloes in TNG-Dark, instead better matching the mass distribution of the host.
They also show that the distribution of full-physics satellites can be better reproduced
by only considering the few TNG-Dark subhaloes with the highest values of $V_{\mathrm{peak}}$,
the maximum circular velocity they had at any point in the past.

From an observational perspective, satellite galaxy radial distributions have been inferred in several studies.
With the Sloan Digital Sky Survey, \cite{Guo2012} explore the dependence of the profiles on luminosity limits,
and \cite{Wang2014} show there is a colour dependence,
while \cite{Tal2012} find the distributions can be best fit by including a baryonic contribution near the centre.
\cite{Budzynski2012} consider
the dependencies of cluster profiles on properties including halo mass and satellite luminosity, 
comparing the profiles binned by halo mass to some earlier SAMs.
This follows the work of \cite{Hansen2005} which additionally looked at the profiles as a function of group size.
More recently, cluster profiles were explored by \cite{Adhikari2020},
who show differences in the distributions of galaxies of different colours.

The Galaxy and Mass Assembly survey 
(GAMA; \citealt{Driver2009,Driver2011,Liske2015, Baldry2018, Driver2022a}) 
offers a suitable observational sample of groups to 
determine the radial distribution of satellites and to compare against simulations,
as it has a high completeness in high-density regions.
The stellar masses of galaxies in groups has been explored by \cite{VazquezMata2020},
and \cite{Kafle2016} find no evidence of variation in satellite galaxy masses with radial position.
Recently, \citet[][hereafter RBL21]{Riggs2021}, explored the group--galaxy clustering in GAMA,
finding evidence of a central core to the distribution of galaxies in groups,
and a good match between GAMA and TNG clustering results.

In this work we study the locations of satellite galaxies in the TNG simulations
and their DMO counterparts, 
comparing against observational results from the GAMA survey.
We do this by using the satellite profile of groups of galaxies,
i.e. the number density of satellites as a function of radial separation from the group centre.
We examine the differences between full-physics and TNG-Dark 
a) by selecting satellites above fixed stellar mass limits, 
b) by identifying their analogue dark-matter subhaloes in the DMO runs, 
and c) by  distinguishing between satellites with and without matched DMO subhaloes. 
We hence investigate the dependencies of these differences on host and subhalo properties. 
Finally, we develop models to account for these differences 
and to correct the satellite locations in DMO simulations.
In Section \ref{sec:GP2} of this paper we describe the GAMA and TNG data we use
and we explain the methods used to select galaxies and produce profiles;
showing the resultant profiles for GAMA, TNG and the TNG-Dark counterparts in Sections \ref{sec:GP3} and \ref{sec:GP4}.
We provide models for the differences in satellite locations in Sections \ref{sec:GP5} and \ref{sec:GP6}
and finally, in Sections \ref{sec:GP7} and \ref{sec:GP8}, we discuss our results and provide conclusions.

In this work group (halo) masses are expressed in 
$\log_{10} (\mathcal{M}_h / \hMsun)$,
taking $\mathcal{M}_h$ to be $M_{200 \rm m}$,
the mass enclosed by an overdensity 200 times the mean density of the Universe.
We denote the radius of a sphere associated with this overdensity as $R_{200 \rm m}$.
We generally express stellar masses from IllustrisTNG 
in $\log_{10} (\mathcal{M}_{\star} / \Msun)$
using the simulation value of $h=0.6774$,
for consistency with the mass limits given in \cite{Pillepich2018}.
The cosmology assumed for GAMA is a $\Lambda$CDM model with
$\Omega_{\Lambda} = 0.75$, $\Omega_{\textrm{m}} = 0.25$,
and $H_{0} = h 100 \ \textrm{km s}^{-1}\textrm{Mpc}^{-1}$.

%% file: sections/data.tex
\subsection{GAMA survey}

Our group sample from the GAMA survey is derived from the three 12 $\times$ 5 degrees equatorial fields,
G09, G12 and G15, of the GAMA-II survey \citep{Liske2015}.
GAMA-II has a Petrosian magnitude limit of $r < 19.8$ mag
and is well suited to group-finding as it is
96 per cent complete for all galaxies which have up to 5 neighbours within 40 arcsec.

The GAMA Galaxy Group Catalogue (\gtc v9) was produced from 
the GAMA-II spectroscopic survey using the same friends-of-friends (FoF) algorithm
used for GAMA-I by \citet[][hereafter RND11]{Robotham2011}.
Group masses are estimates from the total $r$-band luminosity of the group using the power-law scaling relation 
for $M_{200 \rm m}$ determined in \citet[equation 37]{Viola2015}.
This scaling relation is consistent with the one recently determined by \cite{Rana2022}.

We use the same selection of \gtc v9 groups as RBL21.
Groups with 5 or more members are selected, as RND11 find these richer groups to be most reliable.
We select these groups if they fulfill the requirements that they are 
at redshift $z<0.267$ and have a mass in the range $12.0 \leq \log_{10} (\mathcal{M}_h / \hMsun) < 14.8$.
Additionally, we impose the requirement that {\tt  GroupEdge} $> 0.9$,
selecting only those which are estimated to have at least 90\% of the group
within the GAMA-II survey boundaries.
This leaves us with a sample of 1,894 groups with 17,674 galaxies, detailed in Table \ref{tab:gama_groups}.

\begin{table*}
    \centering
    \caption{Numbers of groups and galaxies in each mass bin selected from GAMA, the mock catalogues and 
    the GAMA-matched TNG300-1 sample.
    Values given for the mock catalogues are the mean from the 9 realisations.}
\begin{tabular}{ccccccccccccccc}
\hline
& & \multicolumn{2}{c}{GAMA} & & \multicolumn{2}{c}{Halo Mocks} & &
\multicolumn{2}{c}{FoF Mocks} & & \multicolumn{2}{c}{TNG300-1} \\
\cline{3-4} \cline{6-7} \cline{9-10} \cline{12-13} \\[-2ex]
& $\log_{10} (\mathcal{M}_h / \hMsun)$ & $N_{\rm grps}$ & $N_{\rm gals}$ & &
$N_{\rm grps}$ & $N_{\rm gals}$ & &
$N_{\rm grps}$ & $N_{\rm gals}$ & &
$N_{\rm grps}$ & $N_{\rm gals}$ \\
\hline
\mass1 & [12.0, 13.1] & 380 & 2204 & & 352 & 2210 & & 346 & 2272 & & 368 & 2152 \\
\mass2 & [13.1, 13.4] & 547 & 3646 & & 383 & 2890 & & 401 & 2775 & & 404 & 2941 \\
\mass3 & [13.4, 13.7] & 566 & 4723 & & 366 & 3815 & & 523 & 4233 & & 413 & 3765 \\
\mass4 & [13.7, 14.8] & 401 & 7101 & & 306 & 8377 & & 430 & 8205 & & 467 & 8364 \\
\hline
Total & [12.0, 14.8] & 1894 & 17674 & & 1407 & 17291 & & 1699 & 17486 & & 1652 & 17222 \\
\hline
    \end{tabular}
    \label{tab:gama_groups}
\end{table*}

We select all galaxies within these groups,
and identify the centrals using the iterative central from RND11,
namely the galaxy which remains after iteratively removing the galaxy furthest from 
the centre of light of the remaining group members until only one is left.
All other galaxies within the groups are then satellites.
The iterative centre was shown to be most reliable in GAMA-I by RND11,
and RBL21 confirmed this is also the case in GAMA-II.
In most cases the iterative central is the brightest galaxy of the group.

\subsection{Mock group catalogue}

To determine systematics within GAMA we use the mock catalogues created for GAMA-I
(mocks for GAMA-II are in development).
The mock galaxy catalogues consist of 9 realisations of a lightcone created from the
GALFORM \citep{Bower2006} SAM run on the Millennium \cite{Springel2005} DMO simulation.
Further details about the creation of these mocks are given in RND11.

Two different catalogues of mock groups have been created from the GALFORM galaxy mocks, 
allowing us to explore any biases introduced by the group finding algorithm in GAMA:
\begin{itemize}
    \item The \textit{halo mocks} (\texttt{G3CMockHaloGroupv06}) contain the
intrinsic dark matter haloes of the Millennium simulation which the mock galaxies reside in.
    \item The \textit{FoF mocks} (\texttt{G3CMockFoFGroupv06}) contain
groups derived by applying the same FoF algorithm used for the GAMA groups to the mock galaxies.
\end{itemize}
Comparing the halo and FoF mocks allows us to explore 
how accurately the GAMA FoF algorithm detects the intrinsic haloes,
providing a way of qualifying the differences between the group finding methods in
observations (FoF mock) and simulations (halo mock).
This then informs us how directly comparable the GAMA observational sample
is to simulations such as TNG.

We select groups from the both the halo mocks and FoF mocks using the same criteria as GAMA,
requiring redshift $z<0.267$, halo mass in the range $12.0 \leq \log_{10} (\mathcal{M}_h / \hMsun) < 14.8$
and at least 5 members.

\subsection{TNG simulations}

We explore the effect of baryons with 
the IllustrisTNG cosmological magnetohydrodynamical simulations
\citep[TNG,][]{Marinacci2018, Naiman2018, Nelson2018, Nelson2019a, Nelson2019b, Pillepich2018, Pillepich2019, Springel2018},
and their matching TNG-Dark dark matter-only N-body simulations.
The TNG simulations were run using the \textsc{Arepo} code \citep{Springel2010}
and incorporate the key physical processes of galaxy formation,
including gas heating and cooling, star formation 
and feedback from supernovae and black holes.
For a full explanation of the processes included we refer
the reader to \cite{Pillepich2018a} and \cite{Weinberger2017}.

TNG consists of simulations
at three different box sizes, each run at a variety of resolutions.
We primarily use the runs with the best resolution;
TNG50-1 with box size $35 \hMpc$ and baryonic mass resolution $5.7 \times 10^4 \hMsun$,
TNG100-1 with box size $75 \hMpc$ and baryonic mass resolution $9.4 \times 10^5 \hMsun$,
and TNG300-1 with box size $205 \hMpc$ and baryonic mass resolution $7.6 \times 10^6 \hMsun$.
We additionally include the runs at worse resolution in some of our analysis.
The second tier of resolution, denoted with -2, has baryonic masses 8 times larger than the -1 runs,
and the third tier, denoted with -3, has baryonic masses 64 times larger than the -1 runs.

We select galaxies from these simulations where
\texttt{SubhaloFlag} equals 1, i.e. objects identified as cosmological in origin
(rather than a fragment or substructure formed within an existing galaxy),
and where
the stellar mass within twice the half mass radius
exceeds $10^7$, $10^8$ and $10^9 \Msun$ for TNG50-1, TNG100-1 and TNG300-1 respectively,
limits which correspond to $\approx 100$ stellar particles.
We take the stellar mass of galaxies to be that within twice the half mass radius,
and for the total subhalo mass we take 
the mass of all particles bound to the subhalo.

When comparing against GAMA we use TNG300-1,
as this gives the largest sample of high-mass groups.
We select galaxies from the simulation snapshot at $z=0.2$,
close to the GAMA mean redshift,
and bring the stellar masses into agreement with the TNG100-1 resolution (as well as with GAMA)
by multiplying by the resolution correction factor of 1.4 suggested by \cite{Pillepich2018}.

Elsewhere when looking at simulations of differing resolutions we use the snapshots at $z=0$ and
we do not apply resolution corrections as we are interested in the direct simulation outputs,
and we mainly instead use the better resolution of TNG50-1 and TNG100-1
to perform more detailed examinations of the satellite galaxies.

Each TNG run has a matching TNG-Dark run with the same box size and resolution.
These allow direct comparisons between the outcome of
modelling the Universe in DMO
and that of including hydrodynamics and galaxy physics to model the baryons.

%% file: sections/methods.tex
\subsection{Group radial profile calculation}

Profiles are derived for GAMA as a function of 
the projected radius $r_{\bot}$,
calculated in the standard way \citep[e.g.][]{Fisher1994}.
The vector separation of a satellite at position ${\bm r}_{\rm sat}$ from a group at ${\bm r}_{\rm grp}$,
is given by ${\bm s} = {\bm r}_{\rm sat} - {\bm r}_{\rm grp}$
and the vector to the midpoint of the pair from an observer at the origin by 
${\bm l} = ({\bm r}_{\rm sat} + {\bm r}_{\rm grp})/2$.
These are used to find the line-of-sight separation
$r_\| = |{\bm s}.\hat{{\bm l}}|$, with $\hat{{\bm l}}$ being
the unit vector in the direction of ${\bm l}$, 
and this leads to the projected separation
$r_{\bot} = \sqrt{{\bm s}.{\bm s} - r_\|^2}$.
We do not apply any limits on the line-of-sight distance,
instead simply including all galaxies allocated to the groups
(although we note that this choice implicitly introduces 
limits due to the line-of-sight linking condition in the RND11 FoF algorithm).

When measuring projected two-dimensional profiles for TNG
we take the projection to occur along the z-axis,
but have checked that our results are not sensitive to the choice of projection axis.
With TNG we can also measure three-dimensional profiles,
which we are unable to do for GAMA.
All satellite galaxies which are members of the FoF group are included,
and distances measured relative to the centre of the FoF group.

When calculating profiles, we additionally divide the data into bins of group masses, and include two different forms:

Firstly, the average group profiles,
which we define as the density of galaxies as a function of physical 
projected separation from the group centre.
This is calculated for each group mass bin by counting the number of satellites in radial bins
and dividing by the total number of groups in the mass bin.

Secondly, the normalised profiles,
which we define as the density of galaxies using separations as a fraction of the group $R_{200 \rm m}$.
The amplitudes of these are divided by the number of galaxies in the mass bin.
This can be used to look for differences in the shape of the satellite
distribution in different group mass bins,
as it normalises out the trend for more massive groups
to be more extended and include more galaxies.

We calculate the uncertainties on profiles using jackknife resampling for GAMA and TNG.
For GAMA we split the sample into 9 samples in RA,
and with TNG we divide the boxes into 8 sub-cubes,
showing uncertainties as the square root of the diagonal terms in each covariance matrix.
The mock catalogues contain 9 realisations of the GAMA survey,
and so we can estimate the uncertainties by using the scatter between the realisations.

%% file: sections/gama_profile.tex
In this section we examine the satellite distribution of GAMA groups,
and compare this against a sample of groups and galaxies from TNG300-1
designed to match the GAMA selection.

\subsection{Selecting groups from TNG300-1 to match GAMA}
\label{sec:group_sel}

When comparing against GAMA data, 
groups in TNG300-1 are chosen using a modified form
of the selection function in RBL21.
This modification is necessary as RBL21 
only identify if groups have at least 5 visible galaxies, 
whereas with the simulated data we can in principle identify all the visible galaxies in the chosen groups.

To select the group and galaxy sample we require galaxy luminosities,
for which we use the dust-corrected $r$-band luminosities of dust model C from \cite{Nelson2018}.
We then perform the following procedure for galaxy and group selection:
\begin{enumerate}
    \item Find the comoving distance at which each simulated galaxy has an observed magnitude of $m_r = 19.8$ mag.
    \item Determine the selection probability by finding the volume of the GAMA lightcone out to this comoving distance and dividing by the total GAMA volume for $z < 0.267$. We additionally multiply the selection probabilities by 0.95 to account for our GAMA sample (from RBL21) being 95\% complete.
    \item Assign each simulated group a random probability and select the galaxies whose selection probability is greater than or equal to the random probability assigned to their host group.
    \item Include groups (and their constituent visible galaxies) only if at least 5 galaxies have been identified as visible.
\end{enumerate}

We show the mass function of the groups 
we have selected from GAMA, the mocks and TNG300-1 in Appendix \ref{app:GMF},
demonstrating our group selection method for TNG300-1
reproduces the expected shape of the mass function,
although with differences in the detail due to different underlying galaxy populations.
Small differences between GAMA and TNG are partly caused by
nearby GAMA groups which contain some galaxies below the mass resolution limit of TNG300-1, 
although we have checked that the inclusion of these does not impact the derived profiles.

\subsection{Average group profiles}

\begin{figure*}
    \centering
    \includegraphics[width=\linewidth]{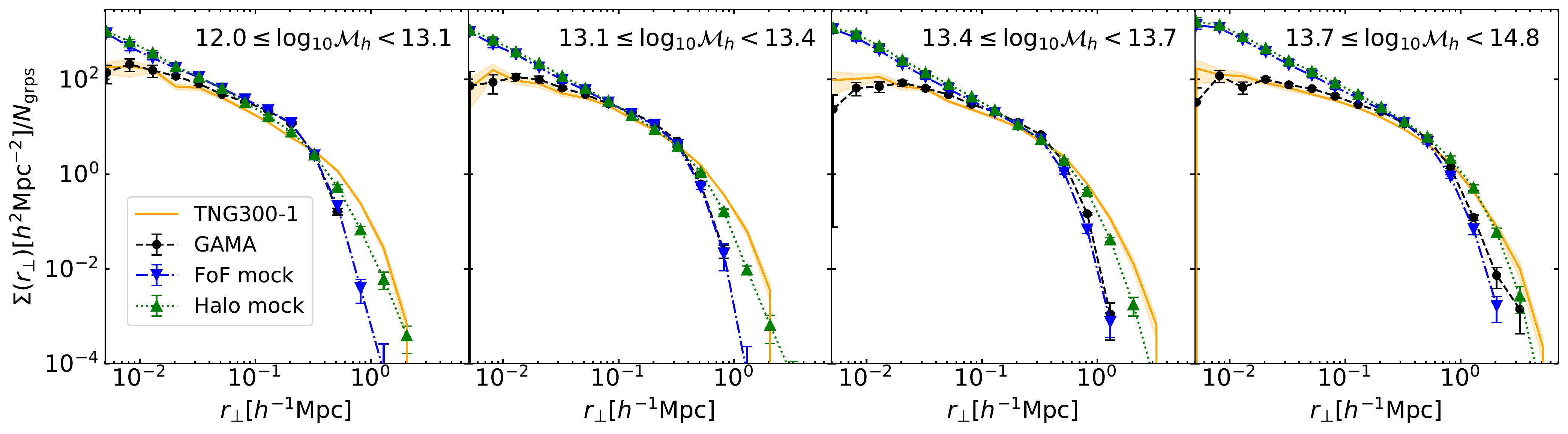}
    \caption{Satellite galaxy projected radial profiles
    in the four mass bins listed in Table \protect\ref{tab:gama_groups}
    for selected groups and galaxies from GAMA, the mock catalogues and GAMA-matched TNG300-1.
    In all panels black circles show the GAMA results, blue downwards triangles the FoF mocks,
    green upwards triangles the halo mocks and orange solid lines TNG300-1.}
    \label{fig:profiles}
\end{figure*}

In Fig.~\ref{fig:profiles} we show for the first time direct results
for radial distributions of satellite galaxies in GAMA groups,
calculating the average group profiles in the four mass bins considered.
In all the group samples used, increasing group mass
leads to a greater number of satellites and wider groups
due to halo radius increasing with mass.

The shape of the profiles is such that they are almost flat
on the smallest scales, $r_{\bot} < 0.02 \hMpc$.
With increasing scale there is then a gradual decrease in density
until $r_{\bot} \approx 0.5 \hMpc$, where a rapid drop is visible.

Comparing with the profiles obtained from the mock catalogues allows us to
investigate the effects introduced by the use of the FoF group finder for GAMA.
On small scales the profiles are similar for the halo and FoF mocks,
with the density increasing to the smallest scales considered.
The similarity of the mocks suggests that GAMA is reliable at small halo-centric distances, 
in agreement with the conclusions of \cite{Driver2022b}, 
as the FoF algorithm accurately reproduces the intrinsic haloes. 
However, it is noteworthy that the survey mocks and GAMA have a very different behaviour. 
This is most likely driven by inaccuracies in the locations of orphan satellites in GALFORM \cite[see e.g.][]{Pujol2017},
and this in turn provides further justification for our objective of correcting for issues in mocks based on DMO simulations.

At the turnover radius beyond which the density drops in the \mass1 bin
($r_{\bot} \approx 0.2 \hMpc$), the FoF mocks have slightly
more galaxies that the halo mocks,
probably due to chance alignments of galaxies on the sky.
Beyond this turnover radius, the FoF mocks drop off much faster than
the halo mocks in all bins, suggesting the outer edges of the groups
are missed by the FoF group finder.
At the outer edges, GAMA and the FoF mocks display very similar results,
and from this we suggest that the true profile of GAMA groups 
(that is comparable to simulations)
would lie about where that of the halo mocks is on these scales.

Overall the mock comparisons tell us that
GAMA profiles should be reliable on scales smaller than the turnover,
but likely underestimate the number density at the outer edge of the groups.

The projected satellite galaxy profile of the GAMA-matched TNG300-1 sample is consistent with GAMA
on small scales where GAMA is reliable, 
with the flattening of the profile at $r_{\bot} \approx 0.02 \hMpc$ 
being consistent between the two within uncertainties.

At face value the TNG300-1 profiles are always above the GAMA profiles 
on large scales ($r_{\bot} \ga 0.5 \hMpc$). 
However, the differences seen between the halo and FoF mock catalogues
show that there are significant differences between the group membership in
simulated and observed groups on these scales.
As we previously noted,
correcting for this difference in methods  
possibly leads to GAMA profiles which are
similar to the halo mocks on large scales.
This suggests that the distribution of galaxies around groups in TNG300-1 
is similar to the observations across all scales,
although with a slight excess of galaxies around the edges of low mass groups.

We note that the flattening of the profiles on the smallest scales in both GAMA and TNG300-1
could be affected by misidentification of the central galaxy in the groups,
although we do not see evidence of this.
In GAMA previous studies (RND11 and RBL21) find the iterative centrals we use are least impacted by mis-centring,
and the central usually corresponds to the brightest galaxy in the group.
In TNG, the central is defined as the galaxy at the minimum of potential 
of the halo and is usually the most massive galaxy in the group. 
Further, the fact that we see a flattening in both cases,
and the consistency between the mocks on these scales,
supports the idea that it is a physical effect.

\subsection{Normalised profiles}

\begin{figure}
    \centering
    \includegraphics[width=\linewidth]{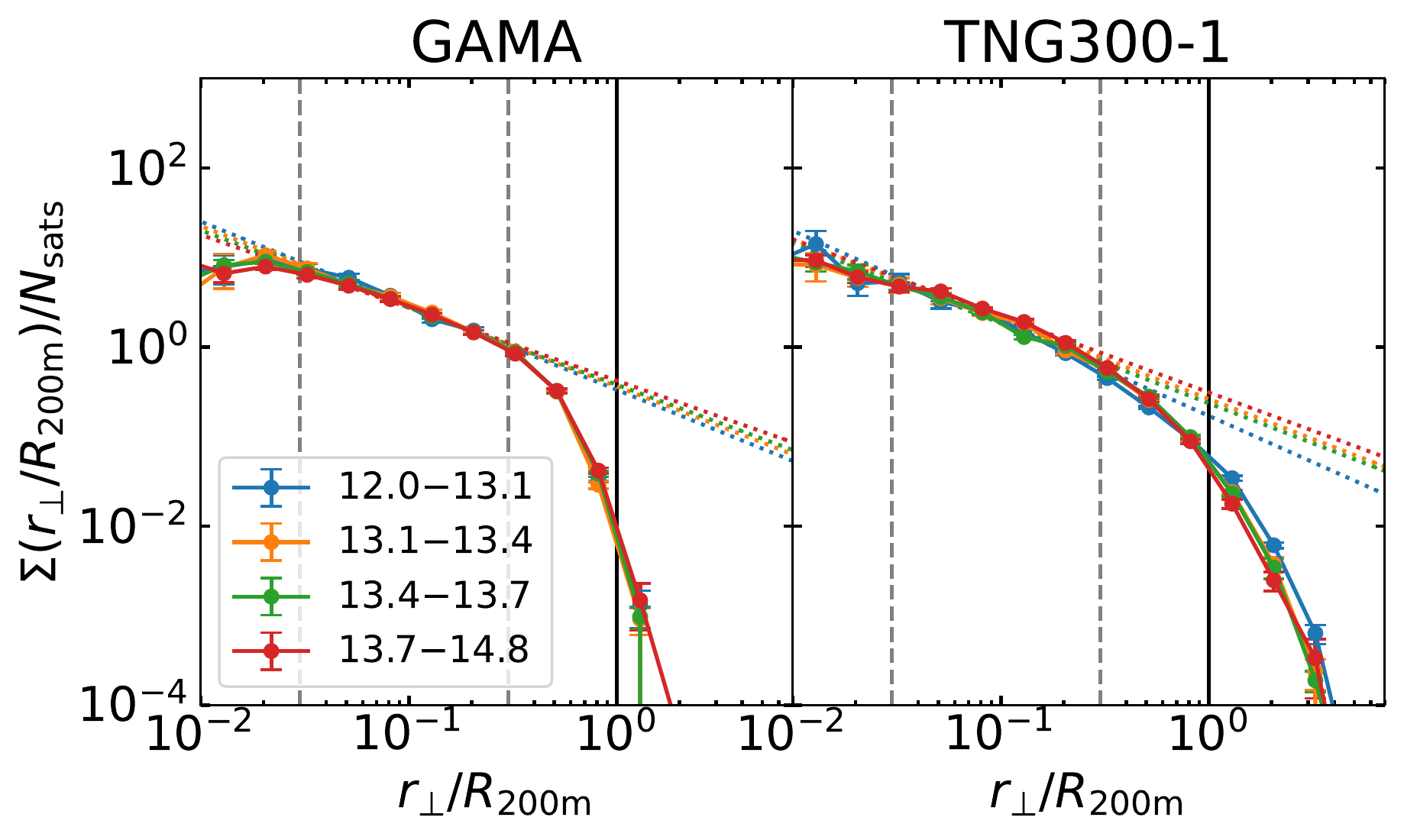}
    \caption{Satellite galaxy projected profiles 
        in the four group mass bins for our GAMA group sample and 
        GAMA-matched TNG300-1 sample,
        calculated as a function of normalised radius and then
        normalised by the total number of satellites.
        The vertical solid lines mark the radius $R_{200 \rm m}$.
        Power law fits are shown as dotted lines, 
        and the region these are fit over is marked with vertical dashed lines.}
    \label{fig:profiles_norm}
\end{figure}

We investigate changes in the profile with group mass by normalising the
satellite distances by group radius $R_{200 \rm m}$ and 
the profile amplitude by the total number of satellites.

The normalised profiles for GAMA and the GAMA-matched TNG300-1 sample are shown in Fig.~\ref{fig:profiles_norm}.
We have not included the mocks here as the conclusions from these remain the same as above,
that GAMA profiles should be reliable on small scales but drop too rapidly on large scales.

There is no mass dependence visible in the normalised profiles for GAMA,
with the profiles being consistent across all scales in the mass bins we use.
This suggests a universal shape to the satellite distribution in GAMA galaxy groups,
with the number and average radial separation of galaxies depending only on the group mass.

TNG300-1 shows exactly the same result of no group mass dependence to the profile shape.
We can also see more clearly here that this trend continues to the edge of the groups.
 
The normalised radial profiles approximately follow a power law over the range
$0.03 < r_\bot / R_{200\rm m} < 0.3$.
In Table \ref{tab:prof_slopes} we provide the power law slopes of GAMA and TNG300-1 profiles in this range.
Both GAMA and TNG300-1 have slopes of approximately -0.9,
and there is very little change in the values with mass given the uncertainties.

\begin{table}
    \centering
    \caption{The slopes of power law fits to the normalised radial profiles from Fig.~\protect\ref{fig:profiles_norm} in the range $0.03 < r_\bot / R_{200\rm m} < 0.3$.}
    \begin{tabular}{ccc}
        $\log_{10} (\mathcal{M}_h / \hMsun)$ & GAMA slope & TNG300-1 slope \\
        \hline
$[12.0, 13.1]$ &
$ -0.94 \pm 0.07 $ &
$ -1.03 \pm 0.07 $ \\
$[13.1, 13.4]$ &
$ -0.90 \pm 0.05 $ &
$ -0.89 \pm 0.07 $ \\
$[13.4, 13.7]$ &
$ -0.86 \pm 0.04 $ &
$ -0.89 \pm 0.05 $ \\
$[13.7, 14.8]$ &
$ -0.81 \pm 0.05 $ &
$ -0.85 \pm 0.05 $ \\

    \end{tabular}
    \label{tab:prof_slopes}
\end{table}

%% file: sections/tngdark_profile.tex
Here we explore the differences between satellite galaxy profiles of groups in
TNG and the equivalents from TNG-Dark runs,
in order to determine the extent to which baryons adjust the shape of the profile.

\subsection{Groups and subhaloes from DMO runs}
\label{sec:tng300D}

We make use of two methods to extract samples from TNG300-1-Dark
to compare to the full-physics run.
Here, as we are just comparing between simulation runs,
we do not apply the group selection to match GAMA.
Instead, we simply select all galaxies with $\mathcal{M}_{\star} \geq 10^9 \Msun$
in groups with $12.0 \leq \log_{10} (\mathcal{M}_{h} / \hMsun) < 14.8$.

The simplest method to generate a sample of TNG300-1-Dark subhaloes 
that correspond to the selected luminous satellites is an abundance matching approach.
We perform this abundance matching using the maximum circular velocity ($V_{\mathrm{max}}$) of each subhalo,
as it is expected this will correlate better than halo mass with galaxy properties \citep[e.g.][]{Zehavi2019}.
In this case we sort the subhaloes in the dark matter-only simulation by their $V_{\mathrm{max}}$,
and select those with the greatest $V_{\mathrm{max}}$ so we have the same number of dark matter subhaloes 
as there are galaxies above the resolution limit in the full-physics run.

The second, more comprehensive, method we use is the subhalo SubLink matches of \cite{RodriquezGomez2015}, 
selecting the matching subhalo from the dark matter-only simulation
for each of our selected TNG300-1 galaxies.
These matches were generated by determining the subhaloes containing the same particles,
and calculating a matching score by 
weighting these particles inversely by their rank ordered binding energy.
For each TNG satellite, the TNG-Dark subhalo with the highest matching score
is taken to be the best match.
This can in some cases lead to multiple TNG galaxies matching to a single TNG-Dark subhalo.
We remove these duplicates from the TNG-Dark run so each subhalo is only included once.

However, these duplicates are important for the TNG run as they 
allow us to split the TNG300-1 satellites into the equivalent
of Type 1 and Type 2 satellites in SAMs.
Type 1 satellites are those contained in dark matter subhaloes,
so all uniquely matched satellites are automatically Type 1s.
Type 2 satellites can then be considered as the unmatched TNG satellites.
A similar application of matching is used by \cite{Renneby2020}.

We use $V_{\mathrm{max}}$ to determine the type for duplicated matches 
at this stage for consistency with the abundance matching method.
The matched TNG300-1 galaxy with the highest maximum circular velocity
is taken to be the Type 1 (or this may be the central Type 0),
while all other matches are allocated as a Type 2 without a matching TNG300-1-Dark subhalo.

These two choices of matching method therefore give us slightly different samples of TNG-Dark subhaloes.
In the first selection we have the same number of objects as TNG,
but they may not be contained in the same environments,
whereas in the second selection the subhaloes we select are known to be comparable to the TNG sample,
but the number of objects differs.

\subsection{Radial profiles in TNG300-1-Dark}
\label{sec:dmo_prof}

\begin{figure*}
    \centering
    \includegraphics[width=\linewidth]{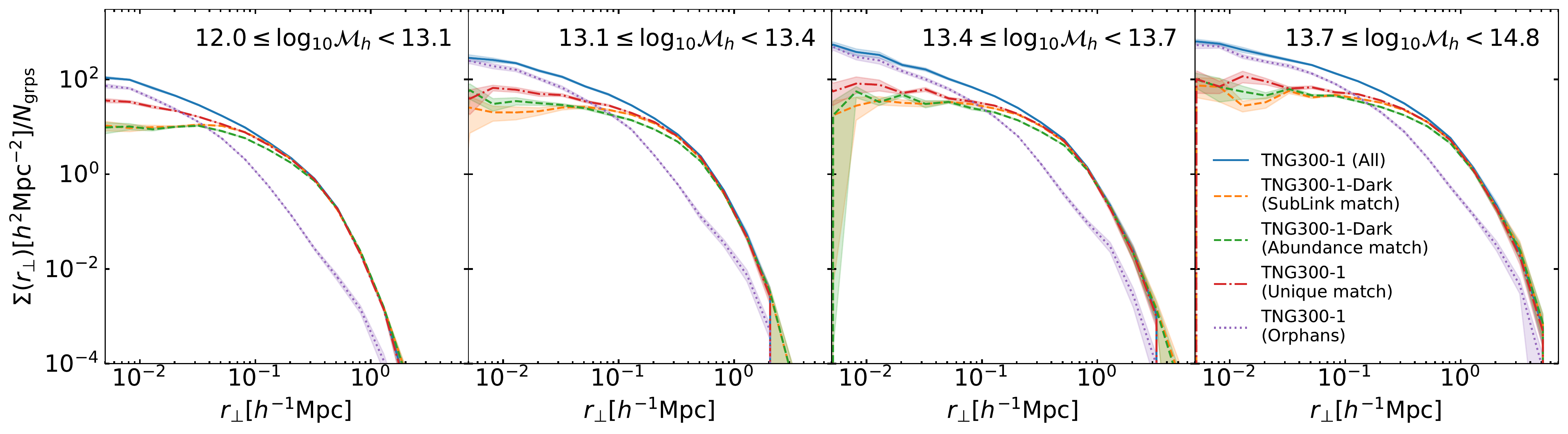}
    \caption{Satellite galaxy projected profiles at $z=0.2$ for TNG300-1 compared to equivalents from the TNG300-1-Dark run.
    The panels show the same mass bins as given in Table \protect\ref{tab:gama_groups}, 
    but now including all galaxies with $\mathcal{M}_{\star} > 10^9 \Msun$ in groups,
    meaning the amplitude of the TNG300-1 profile has increased relative to Fig.~\protect\ref{fig:profiles}.
    Two different methods of selecting matched TNG300-1-Dark subhaloes are used,
    as explained in Section \protect\ref{sec:tng300D}.
    We also show the TNG300-1 galaxies split by those which can be matched to TNG300-1-Dark satellites and those which are unmatched (orphans).}
    \label{fig:tng300_dmo}
\end{figure*}

In Fig.~\ref{fig:tng300_dmo} we compare the profiles of satellites in TNG300-1 against
those from the matched subhaloes in the TNG300-1-Dark simulation at $z=0.2$.
This is the sample used in Fig.~\ref{fig:profiles} but without the group selection method applied.

It is clear that on large scales there is a close agreement between TNG and TNG-Dark. 
However, at small halo-centric distances, 
the density of TNG satellites is enhanced over their matched subhaloes in the TNG-Dark simulation 
(solid blue vs. orange and green dashed curves). 
In particular, the number density profile of the TNG-Dark subhaloes flattens, 
while the TNG profile of luminous satellites continues to rise down to smaller scales, albeit at a reduced rate.
The two options for selecting subhaloes from the TNG-Dark simulation
are seen to be consistent, with the profiles matching within uncertainties,
suggesting this is not just a result of the matching scheme used.

This difference between TNG and TNG-Dark can be attributed to two effects,
which we also show in Fig.~\ref{fig:tng300_dmo}.
Firstly, there is evidence of an inwards displacement in the TNG simulation,
with the directly matched (Type 1) satellites being closer to the centre in TNG.
Secondly, there is a population of galaxies that are not uniquely matched to TNG-Dark subhaloes (Type 2s), 
suggesting that they have been merged or disrupted in the TNG-Dark simulation but not in TNG.
Both of these effects primarily affect scales $r_{\bot} \la 0.1 \hMpc$,
but there is some impact out to at least $r_{\bot} \approx 0.5 \hMpc$ in the largest groups.
Together, these effects suggest baryons enhance both the rate of inwards motion
and the survival time of subhaloes that host galaxies.

We note that in all mass bins the dominant effect on the smallest scales 
is the population of unmatched satellites, 
and that the contribution due to the inwards displacement of matched satellites 
decreases as halo mass increases and the groups become wider.

\subsection{Radial profiles at differing resolutions}
\label{sec:profiles_allres}

\begin{figure}
    \centering
    \includegraphics[width=\linewidth]{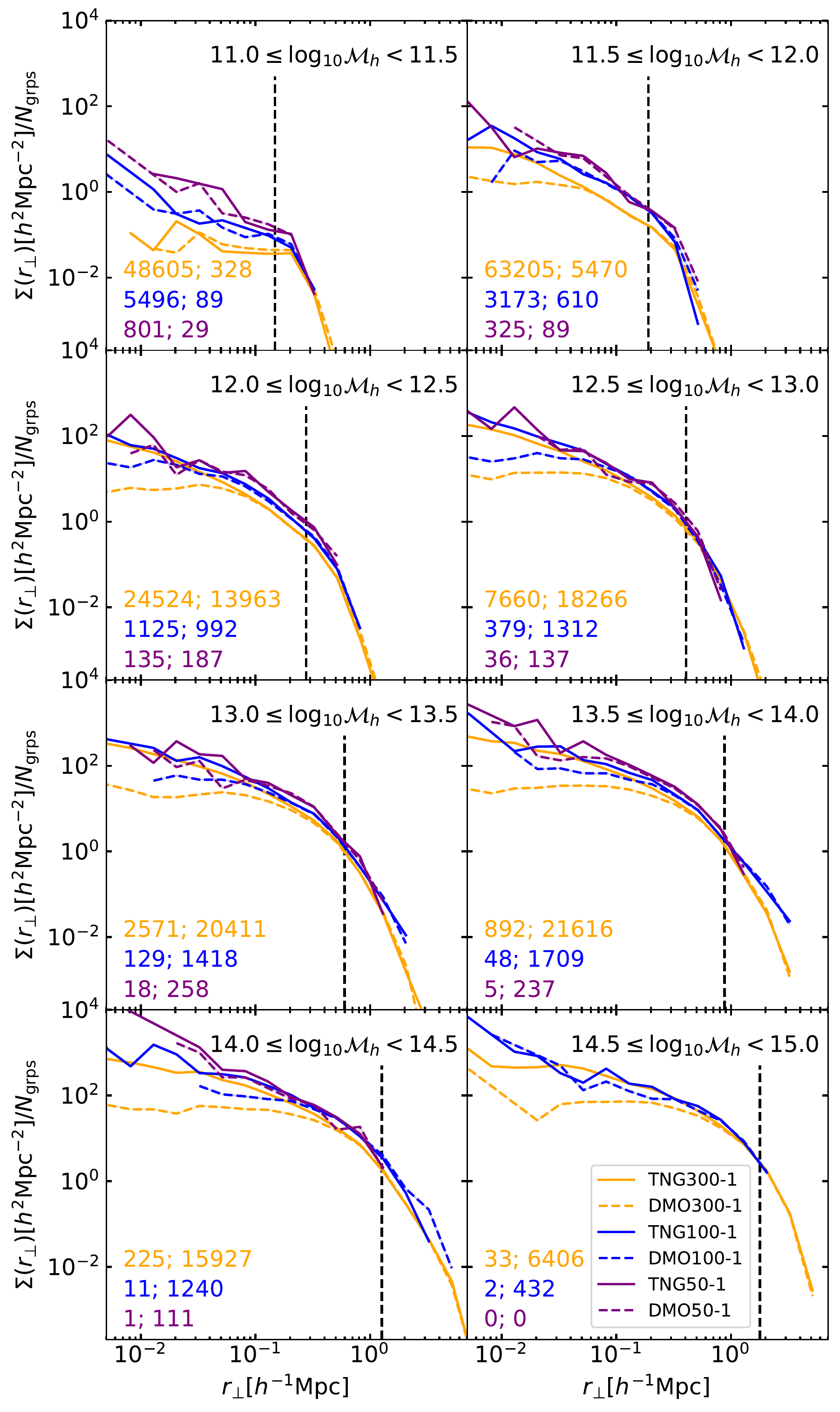}
    \caption{Satellite galaxy projected profiles at $z=0$
        for TNG50-1 (purple), TNG100-1 (blue) and TNG300-1 (orange), 
        with their TNG-Dark equivalents in a range of halo mass bins.
        Galaxies are selected with $\mathcal{M}_{\star} \geq 10^9 \Msun$ 
        and halo mass limits are given in $\hMsun$.
        The numbers in the lower left of each panel give firstly the number
        of groups and then the number of satellites in the TNG run presented in the same colour.
        The vertical dashed lines show the median $R_{200 \rm m}$ of the TNG300-1 groups in that panel.}
    \label{fig:all_tng_profiles}
\end{figure}

To explore the effect of resolution on the profiles in the TNG simulations,
we measure the average group profile in each of the TNG50-1, TNG100-1 and TNG300-1 simulations at $z=0$,
and in their TNG-Dark equivalents.
While we could compare resolutions by using the different runs at identical box size,
we choose to use the largest box available at each resolution to give us a larger galaxy sample,
although we show in Appendix \ref{app:profile_lowres} that the same conclusions are reached 
using different resolutions at the same box size.
To enable comparison between the different simulations we apply the same mass limit
in each case, $\mathcal{M}_{\star} \geq 10^9 \Msun$,
although the resulting low number counts for TNG50-1 make comparisons involving it challenging.
We show in Appendix \ref{app:profile_mstar} that the choice of mass limit only affects the amplitude of the profile,
and that the full-physics runs show very close agreement when normalised.
We have selected TNG-Dark subhaloes which are equivalent to TNG satellites 
using the SubLink matching.

Fig.~\ref{fig:all_tng_profiles} shows the radial profiles from these samples 
in bins of host halo mass.
In group mass bins of increasing mass we see the profile increases in amplitude
and extends to greater radii,
as we observed in GAMA.

It is apparent that there is a reasonable consistency between the different
TNG simulation resolutions in most group mass bins.
The main exception is the least massive bin where the majority of haloes contain no satellites
with stellar mass above $10^9 \Msun$.
We also note that the most massive bins are subject to a high uncertainty due to
containing very few groups.

The agreement between the distributions of well-resolved satellites at differing resolution
matches the conclusions of \cite{Grand2021} with the \textsc{auriga} simulations. 
However, as shown in that work, this consistency is likely to break down
for satellites with very few stellar particles.

Looking at the TNG-Dark results,
we find the same effect noted before of
flatter radial profiles in the centres of groups than in the full baryonic runs, 
at least for subhaloes matched to satellites with a given minimum stellar mass. 
However, such flattening varies across simulations,
with the DMO profiles becoming flatter at small distances for progressively worse numerical resolution.

The implication of this is that the extent of the differences
between TNG and TNG-Dark are affected by the simulation resolution,
and that this is driven by differences with resolution across DMO runs---the full-physics runs 
are in much better agreement across the three resolution levels; see also Fig.~\ref{fig:mstar_prof}.
Instead, the changes in the TNG-Dark profiles for different resolutions are unlikely to be entirely physical
and instead may be the result of the numerical disruption
effects found by \cite{VanDenBosch2018a}.
However, there are still profile differences in TNG50-1, the highest resolution simulation,
(and these differences become more apparent if lower mass satellites are also included)
so there may be a physical effect at work here too.
This could be due to the baryonic feedback, 
which is known to change the shapes of the haloes \citep{Chua2021},
but could also be related to the baryonic core keeping the satellites more bound
and so less prone to disruption (both physical and numerical).
This would match the findings of earlier works such as \cite{Weinberg2008}.
We return to the question of whether the effects we see are
physical or numerical in Section \ref{sec:physical_discussion}.

%% file: sections/fitting_selection.tex
Having established that significant differences exist between satellite
locations in the full-physics and DMO runs,
we now aim to create models to correct for these differences.
To create these models we use the runs with the best resolution,
TNG50-1 and TNG100-1, as these allow us to explore 
smaller scales and lower masses with more confidence.
Following this we use the runs at worse resolution to explore 
the dependence of the required correction on simulation resolution.

\subsection{Splitting Type 1 and Type 2 satellites}

We first split the satellite sample into Type 1s 
(which have a matching TNG-Dark subhalo)
and Type 2s (whose subhalo has disrupted or merged in TNG-Dark),
with a similar method to that which we used for TNG300-1-Dark in Section \ref{sec:tng300D}.

Using the SubLink matches of \cite{RodriquezGomez2015} 
we select the matching TNG-Dark subhalo for each full-physics galaxy.
Uniquely matched satellites become Type 1s.
In the cases where the matches are not unique we assign the best match
as the Type 1, and all others as Type 2s.
Here, we determine the best match by picking the 
subhalo which has the highest matching score, 
as determined by the SubLink matching algorithm.

To clean our sample further,
Type 1 satellites are removed if the central and satellite assignment differs between TNG and TNG-Dark,
leading to 971 subhaloes being excluded in the case of TNG50-1 (about 6\% of the total).
The excluded fraction becomes smaller in the simulations with worse resolution.
There are a few possible reasons the type (central/satellite) can differ between TNG and TNG-Dark: 
either the structure formation has occurred differently,  
the matching scheme is inaccurate, 
the FoF algorithm has combined two close haloes, or 
the subhalo has been accreted earlier in one simulation than the other.
The majority of the subhaloes we remove have a radial separation from the central 
exceeding the host $R_{200 \rm m}$, 
suggesting they have either only just been accreted 
or are part of neighbouring haloes joined by the FoF algorithm.
However, there are a small number closer to the central, 
suggesting different structure formation or incorrect matching. 
We do not attempt to correct these matches, instead just excluding these subhaloes.

We also exclude Type 1 satellites where the host halo mass differs enough to suggest that they
are attached to different groups.
This choice of halo mass difference is somewhat subjective,
but we have determined that excluding cases where
$|\log_{10} (\mathcal{M}_h^{\mathrm{TNG}} / \mathcal{M}_h^{\mathrm{DMO}})| > 0.15$
removes all clearly different hosts, while allowing for some scatter between the simulations.

For TNG50-1 this gives us a sample size of 6915 Type 1s, 781 Type 2s and 8237 central Type 0s
with stellar mass $\mathcal{M}_{\star} \geq 10^7 \Msun$.
This rises in TNG100-1 to 24759 Type 0s, 16842 Type 1s and 2862 Type 2s
with stellar mass $\mathcal{M}_{\star} \geq 10^8 \Msun$.

%% file: sections/type1_model.tex
\subsection{Model for Type 1s}
\label{sec:t1_model}

We first consider the modelling of the Type 1 satellites,
aiming to quantify the expected difference in position between the subhaloes
in the TNG and TNG-Dark runs.
We describe our model here, before giving the parameters for it in Section \ref{sec:t1s_res}.

While the differences in satellite positions between full-physics and DMO runs
may depend on any properties of the subhaloes or host haloes,
we find a simple model adequately describes the differences.
We first present this model, 
then explore the reasons why we are able to exclude other dependencies.

\begin{figure*}
    \centering
    \includegraphics[width=\linewidth]{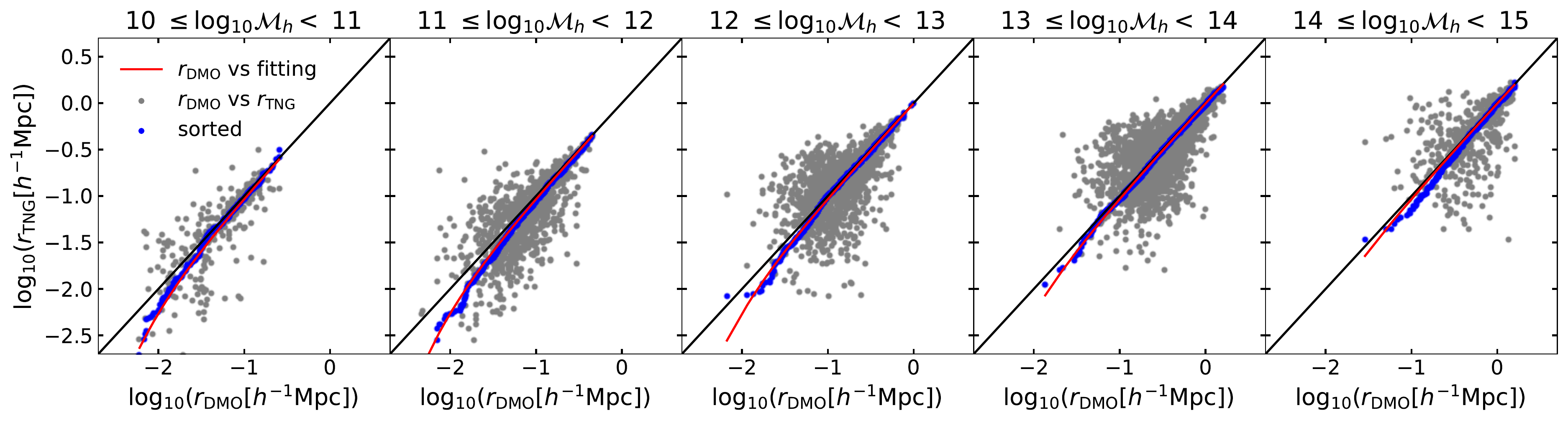}
    \caption{Fitting to the difference between matched satellite positions in TNG50-1 and the TNG50-1-Dark run at $z=0$,
            with the data split into bins of host halo mass, given in $\hMsun$,
            and galaxies shown for $\mathcal{M}_{\star} > 10^7 \Msun$.
            In each panel, the grey background points show the scatter between exactly matched satellites.
            The blue points then show the result of sorting the positions by distance from the centre.
            Finally, the red lines show our fit.}
    \label{fig:tng_t1_fitting}
\end{figure*}

Our model for the correction to Type 1 positions
includes only the {\it comoving} radial distance from the group centre.
We model the correction to the position as a power law,
\begin{equation}
\label{eq:t1_fit}
    \log_{10} (r_{\mathrm{TNG}} / r_{\mathrm{DMO}}) = -(r_{\mathrm{DMO}}/a)^b
\end{equation}
where $r_{\mathrm{TNG}}$ is the radial position in the full-physics run 
and $r_{\mathrm{DMO}}$ is the radial position in the DMO run.

To determine the parameter values in this model
we first sort the positions of the satellites in ascending order
independently for the TNG and TNG-Dark runs,
then fit our model to the sorted positions.
This is done to produce an overall trend in the position difference.

In doing this we are discarding the true associations between
the TNG and TNG-Dark runs, but
without sorting the positions we would potentially fit 
to spurious trends caused by orbital phases.
Objects close to the centre in either simulation will be near pericentre, 
and so a small difference in orbital phase between simulations 
will result in them being further from the centre in the other.
Therefore, without sorting the positions,
we would conclude that objects near the centre should always be moved outwards.
Similarly, this effect matters when considering possible dependencies
on variables which may correlate with radial position.

The position differences between TNG and TNG-Dark satellites,
with the results of applying this model in TNG50-1, are shown in Fig.~\ref{fig:tng_t1_fitting}.
Note that while we have split the sample into halo mass bins for this figure,
we perform the fitting on the whole data sample together
and the fit parameters used are the same in each panel.
The grey points show there is a large scatter between the raw positions in the TNG and TNG-Dark runs,
but the sorted positions in blue show a clear trend for inwards displacement in the full-physics case.
Our fitted results are then shown in red,
demonstrating a good match between our model and the sorted data
across the full range of halo masses.

\subsection{Dependencies on masses}

We now discuss why we are able to exclude other dependencies from our model, 
despite it being anticipated that
the difference between TNG and TNG-Dark may depend both on the properties of the subhalo 
and those of the host halo.
Firstly we note that the aim of our model is to explain the differences between the TNG and TNG-Dark simulations,
while also providing a method that can easily be applied to models such as SAMs and HODs.
For this reason we do not attempt to include all the possible dependencies in our model
(for example dependencies on the star formation rate, colours and gas fraction of galaxies 
may be challenging to incorporate in SAMs).
Additionally, the impact of feedback from active galactic nuclei (AGN) and supernovae on subhaloes
may not be consistent between different hydrodynamic simulations, 
so we do not want to directly consider these effects.

One of the simplest dependencies we expect is with mass,
and halo mass, subhalo total mass and subhalo baryonic mass may all have an impact.

\begin{figure}
    \centering
    \includegraphics[width=\linewidth]{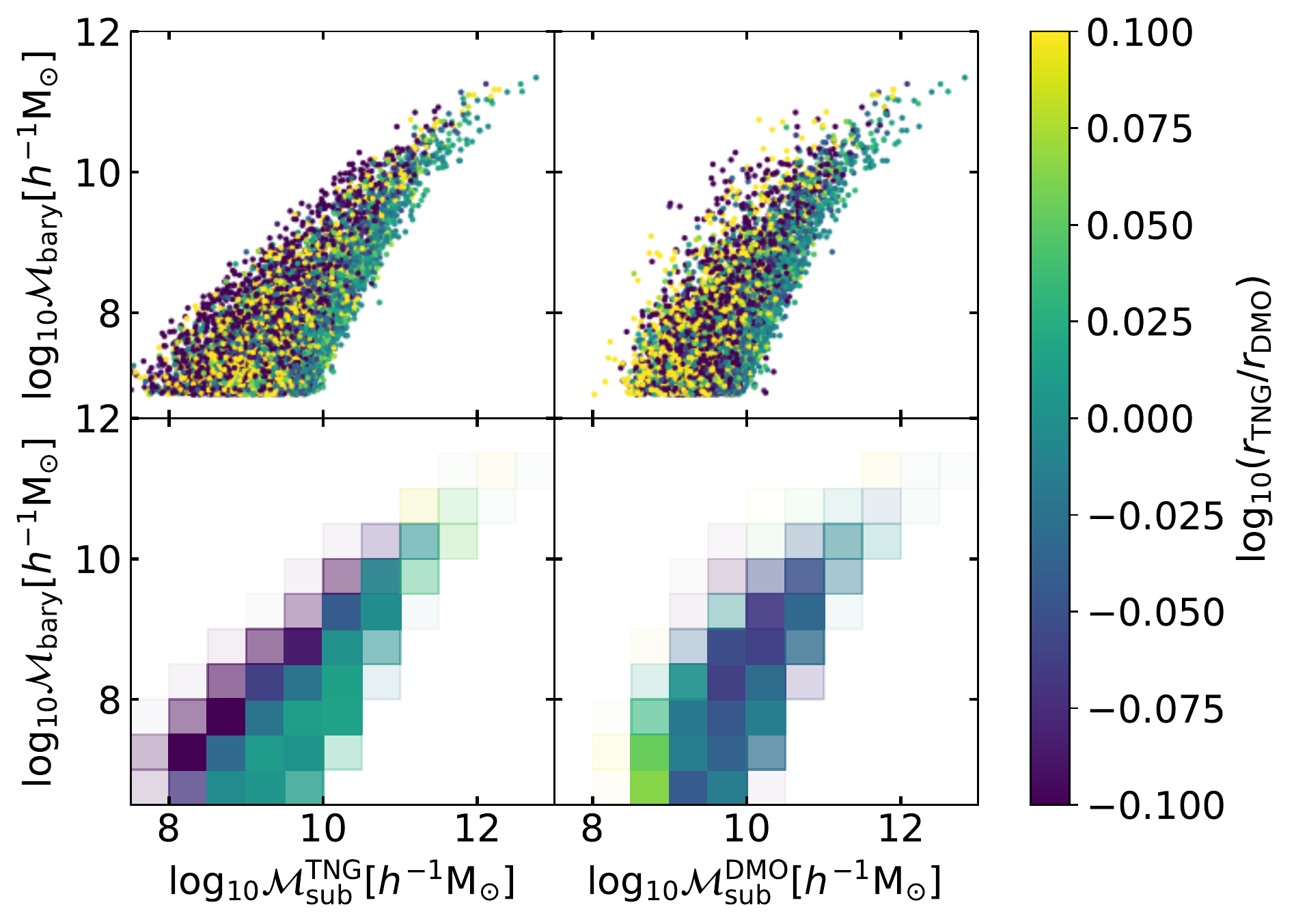}
    \caption{Dependence of the position difference of Type 1s 
        on subhalo total mass and baryonic mass at $z=0$.
        Left panels show the dependence on subhalo mass in TNG50-1, 
        while the right panels show the dependence using the 
        total subhalo masses of matched TNG-Dark subhaloes.
        Upper panels show position differences of individual satellites,
        and the lower panels the binned averages of these.}
    \label{fig:type1_subhalo_mass}
\end{figure}

In Fig.~\ref{fig:type1_subhalo_mass} we show the relation between subhalo baryonic mass and subhalo total mass
in TNG50-1 and for matched TNG-Dark subhaloes.
The colour scaling represents the position difference and shows a large scatter,
and this spread increases at low total subhalo masses.
To account for this, in the lower panels we split the sample into bins of subhalo mass and colour the bins by 
the average position difference.
On the left, using the TNG subhalo masses, 
there is a trend for galaxies which have a high baryonic mass for a given total subhalo mass
to be closer to the group centre in TNG 
(having a lower $\log_{10} (r_{\mathrm{TNG}}/r_{\mathrm{DMO}})$, coloured purple in Fig.~\ref{fig:type1_subhalo_mass}).
However, this trend is not present in the right panels, using TNG-Dark subhalo masses.
This shows that while the baryonic fraction is related to the reduced halo-centric distances
it is likely to be a secondary effect,
such that the closer proximity to the centre causes stripping of some of the subhalo dark matter,
so increasing the baryon fraction.
The secondary nature of this effect, and the lack of this effect in the TNG-Dark panel, 
means this is not an effect which we need to include in our model.

\begin{figure}
    \centering
    \includegraphics[width=\linewidth]{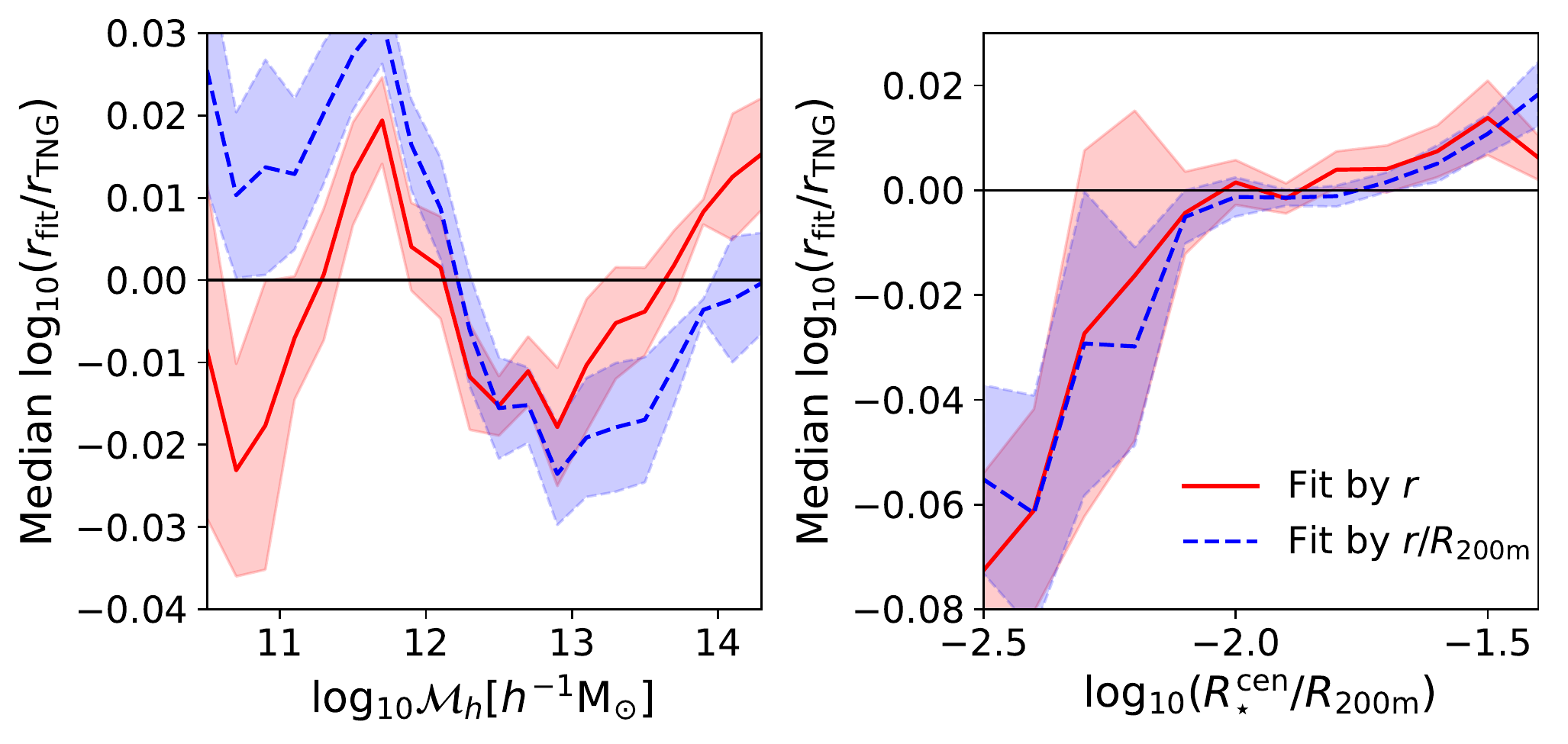}
    \caption{Residuals of the Type 1 position fitting model on TNG50-1 at $z=0$
        for galaxies with $\mathcal{M}_{\star} > 10^7 \Msun$.
        The left panel shows the dependence on halo mass, 
        while the right panel shows the dependence on the relative stellar size of the central galaxy.}
    \label{fig:type1_residuals}
\end{figure}

Fig.~\ref{fig:tng_t1_fitting} has already demonstrated that our model
works across different halo masses,
so we are able to exclude halo mass as an explicit part of our model.
However, there is a residual effect of halo mass on the position differences.
In the left panel of Fig.~\ref{fig:type1_residuals} the median difference 
between the sorted position in the TNG simulation and our model is plotted.
To smooth this we use overlapping mass bins, and the errorbars are calculated using jackkknife.
It is clear that there is some halo mass dependent residual.
Comparing this to figure 4 of \cite{Weinberger2017} shows a very similar trend
to that of the difference in halo masses between TNG and TNG-Dark.
In that work, this is attributed to the effect of stellar and AGN feedback,
and so it is likely our residual is present for the same reasons.
In particular, the drop at $\mathcal{M}_h \approx 10^{12} \hMsun$
is likely due to the onset of feedback from supermassive black holes at this mass scale in TNG.

We also show the comparison here of the outcome of performing our fitting procedure
using {\it normalised} radial distance ($r/R_{200 \rm m}$),
rather than the {\it comoving} radial separation ($r$).
For much of the mass range the discrepancy associated with the two separation options
is comparable.
However, using normalised distances a clear split is seen,
with overestimation in haloes of $\log_{10} (\mathcal{M}_h / \hMsun) \la 12$,
and underestimation in more massive haloes.
This gives a slight advantage to using comoving separations in 
low-mass haloes,
with normalised distances only showing a clear advantage for $\log_{10} (\mathcal{M}_h / \hMsun) \ga 14$.
This motivates our usage of comoving separations in our model.

One further dependence can be seen in 
the right panel of Fig.\ref{fig:type1_residuals} where we show the residual as
a function of the relative stellar size of the central galaxy in the host halo,
defined as the stellar half mass radius divided by the halo radius.
While a similar fitting discrepancy is seen using comoving and normalised radii 
it is likely the stellar size of the central galaxy,
which is proportionally smaller in lower mass haloes \citep[see][]{Pillepich2018},
is also part of the explanation for the residual halo mass dependence seen in the left panel.

\subsection{Model fitting at different resolutions}
\label{sec:t1s_res}

\begin{figure}
    \centering
    \includegraphics[width=\linewidth]{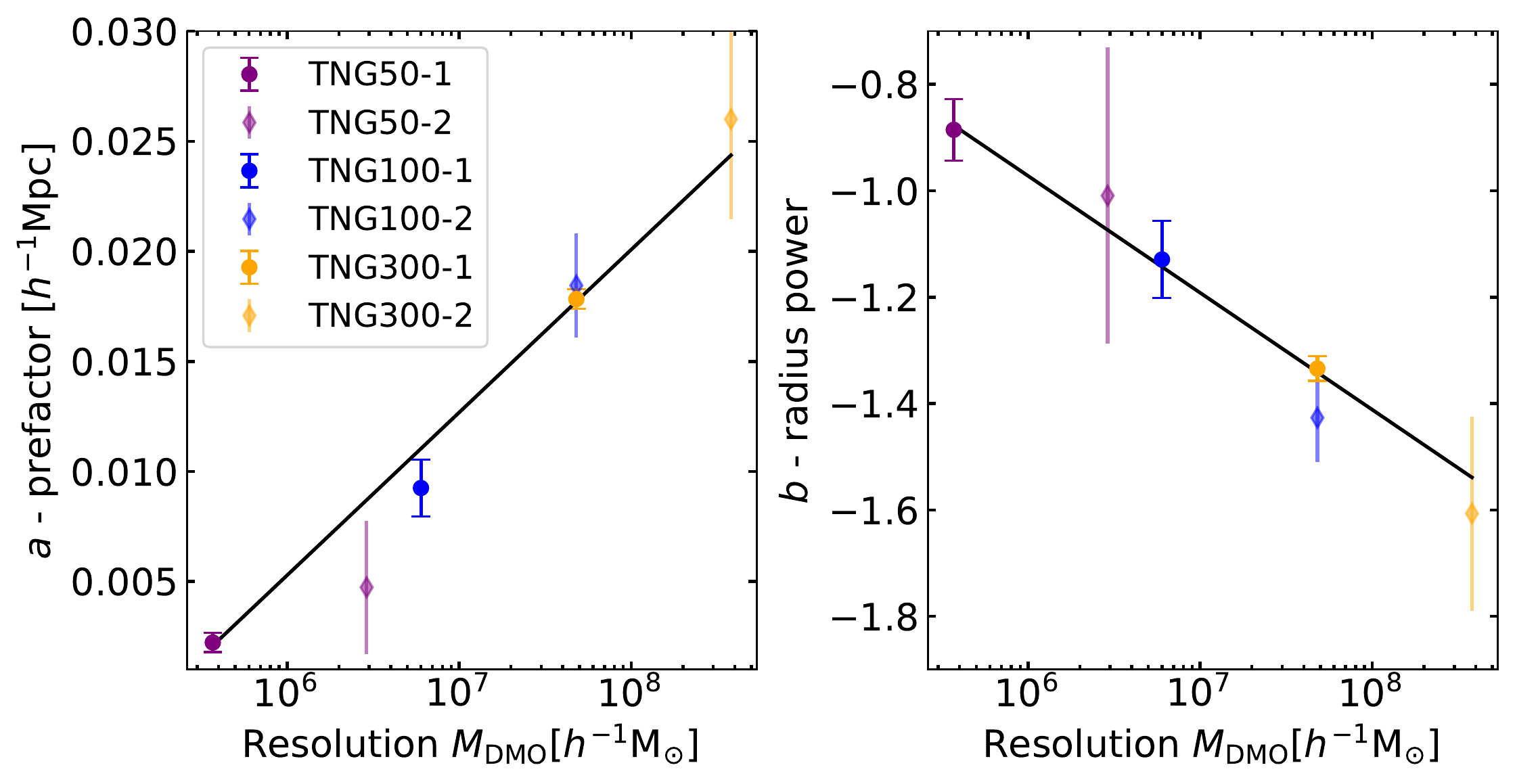}
    \includegraphics[width=0.8\linewidth]{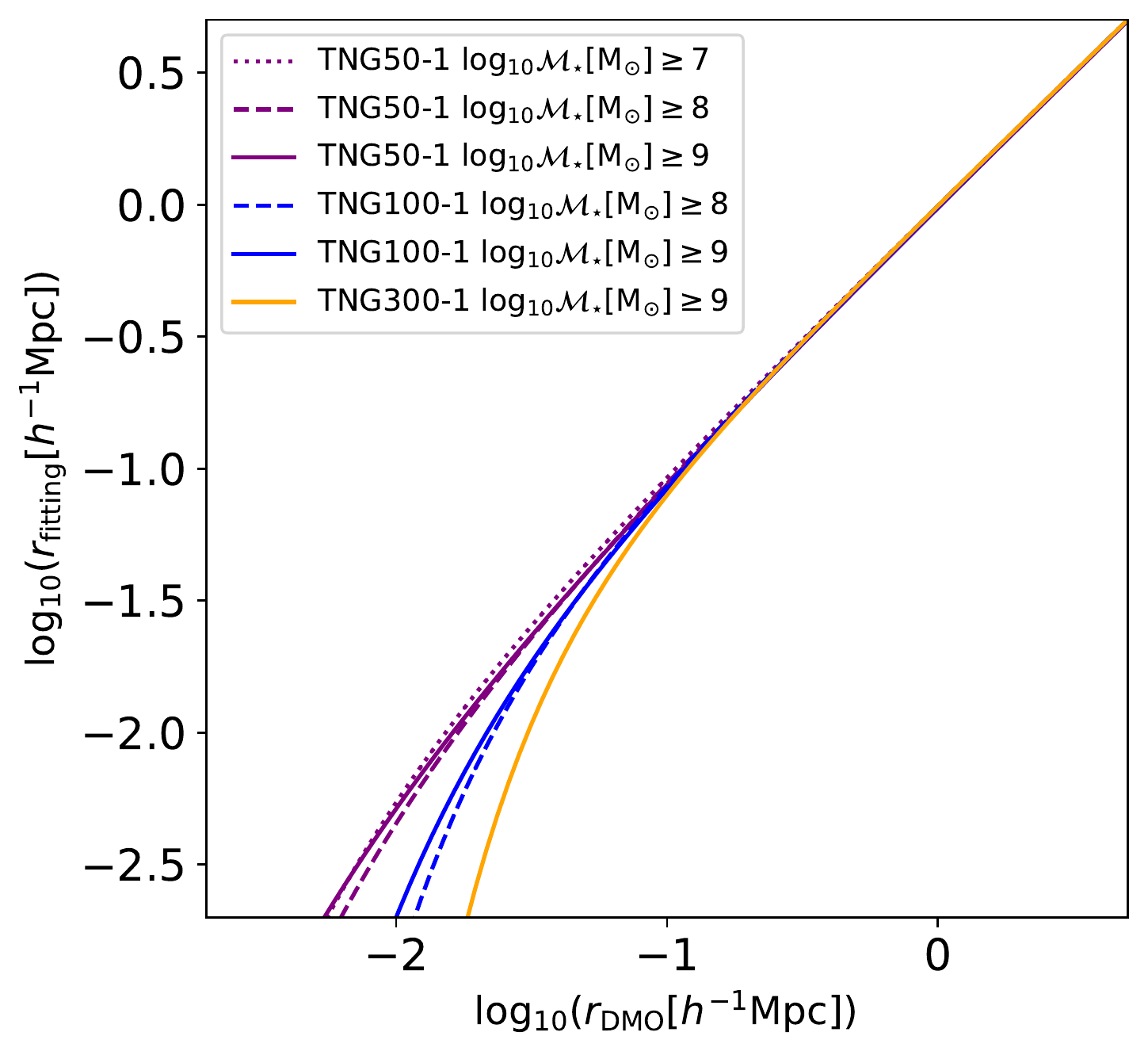}
    \caption{Fitting the radial position change of Type 1 satellites for different resolutions at $z=0$.
    {\bf Upper panels:}
    Fitting parameters from equation \protect\ref{eq:t1_fit}
    for the Type 1s in different resolution TNG simulations, 
    as a function of the dark matter particle mass used in each TNG-Dark simulation.
    Galaxies with $\ga 100$ stellar particles are selected from each simulation.
    {\bf Lower panel:}
    The radial position of Type 1 satellites in the
    full-physics simulations as a function of radial position in the DMO simulations 
    for different resolutions and stellar mass limits.
    }
    \label{fig:params_resolution}
\end{figure}

We then repeat the fitting procedure in different simulations to investigate the effect of resolution.
The upper panels of Fig.~\ref{fig:params_resolution} show the parameters as a function of 
the dark matter particle mass in the TNG-Dark simulation $M_{\mathrm{DMO}}$,
when all resolved galaxies are used in each case.
The uncertainties shown are calculated by jackknife between sub-cubes of each simulation.

It is seen that the pivot radius, $a$, increases, while the power scaling, $b$, decreases.
A linear function of $\log_{10} M_{\mathrm{DMO}}$ is a reasonable fit to both of these parameters.
Applying a linear fit we find
\begin{equation}
    a = -0.039 + 0.0074 \log_{10} M_{\mathrm{DMO}}
\end{equation}
and
\begin{equation}
    b = 0.35 - 0.22 \log_{10} M_{\mathrm{DMO}}.
\end{equation}

The overall correction required is enhanced at worse resolution, as seen in the lower panel of Fig.~\ref{fig:params_resolution}.
It may be hypothesised that this is due to including haloes and subhaloes of differing masses in each
simulation selection.
However, the lower panel of Fig.~\ref{fig:params_resolution} also shows the fit does not shift substantially if
the better resolution simulations are restricted to only use the most massive galaxies,
and therefore this is a true effect of the resolution.

\subsection{Redshift dependence}

\begin{figure}
    \centering
    \includegraphics[width=\linewidth]{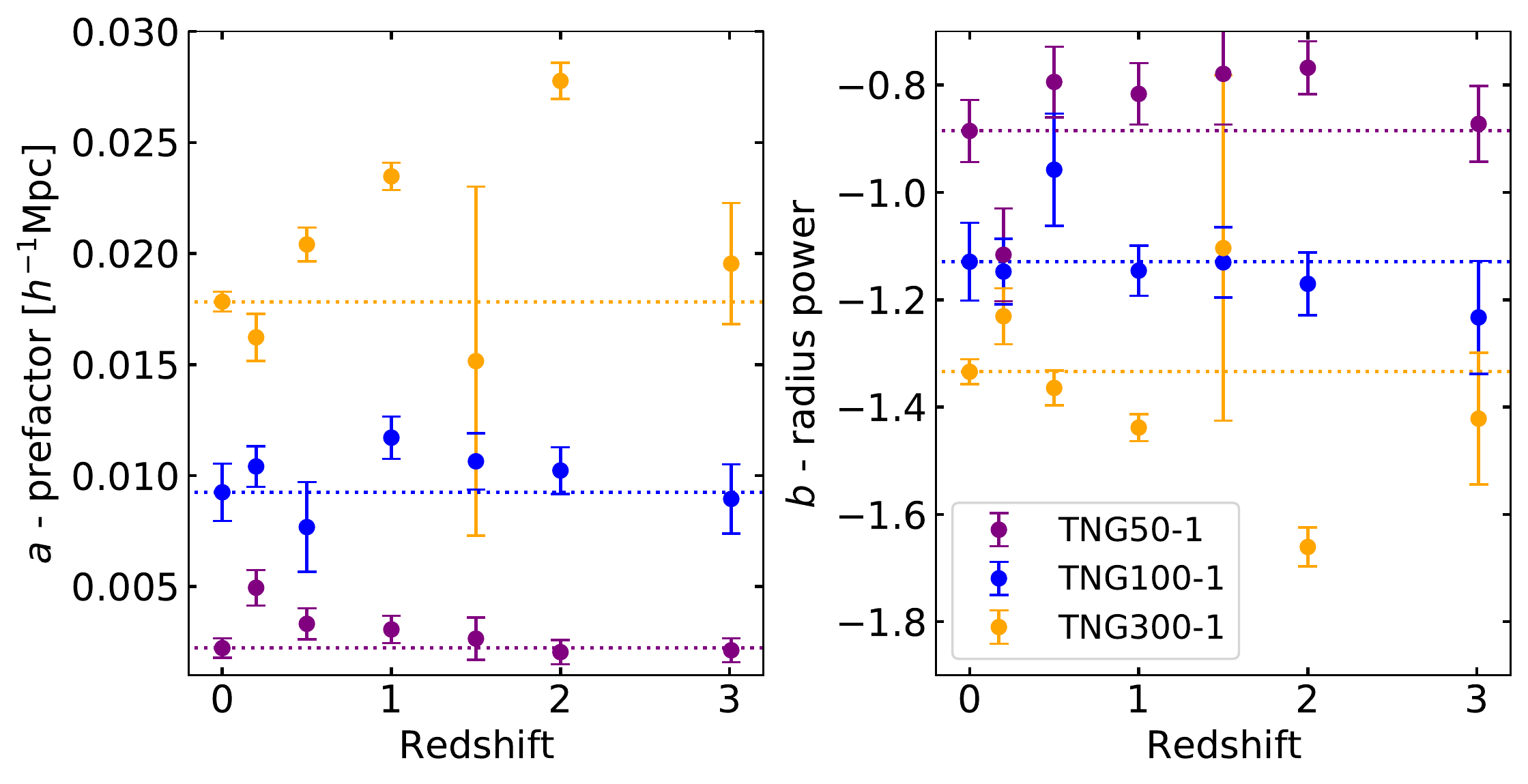}
    \caption{The fitting parameters from equation \protect\ref{eq:t1_fit} for Type 1 satellites 
    as a function of redshift for the run with the best resolution of each box size.}
    \label{fig:type1s_redshifts}
\end{figure}

Finally, we examine whether our model depends on the redshift at which it is applied.
We repeat the fitting of equation \ref{eq:t1_fit} at a series of snapshots
in the run with the best resolution of each box size,
and we show the results of this fitting in Fig.~\ref{fig:type1s_redshifts}.

In TNG50-1 and TNG100-1 there is no systematic trend visble 
in the parameters at different redshifts,
with the parameters consistent with the redshift zero result in most instances.
In TNG300-1 there is a trend for the pivot radius to increase 
and the power scaling to decrease with redshift, 
which is largely attributable to the degeneracy in the fitting of the two parameters.
Overall, we therefore expect that the fitting parameters we found
at redshift zero will be sufficient for any applications at higher redshifts.

\subsection{Subhalo mass differences}

Our result in Fig.~\ref{fig:type1_subhalo_mass} that the radius change is related to the 
subhalo mass from the full-physics simulation but not in DMO
suggests a systematic difference in the masses as well as the positions of satellites,
as previously found by comparisons of full-physics and DMO simulations by \cite{Sawala2013}.
Following on from this, we briefly consider here what correction would be required for the masses.

\begin{figure}
    \centering
    \includegraphics[width=\linewidth]{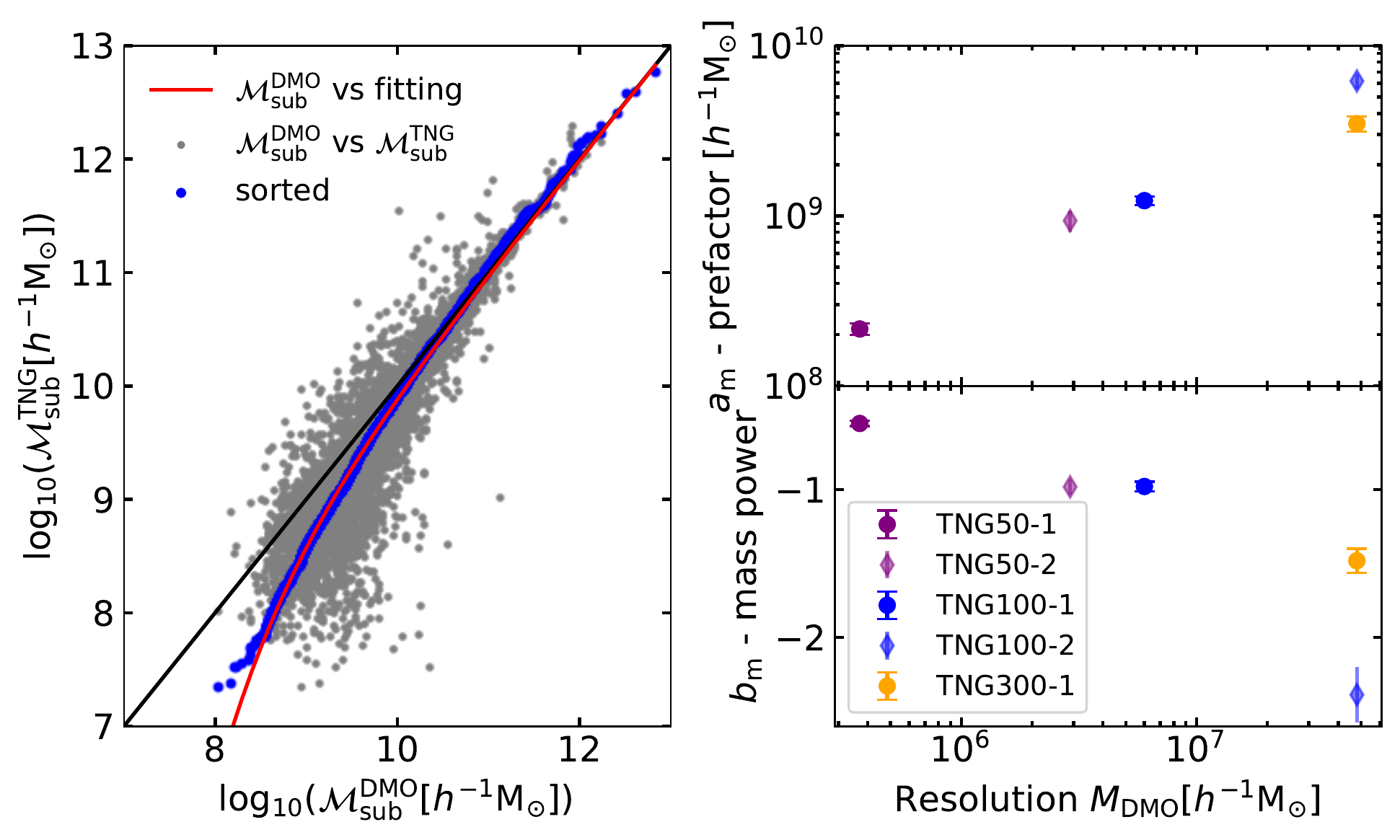}
    \caption{Fitting the mass change of Type 1 satellites between the TNG and TNG-Dark runs at $z=0$.
    \textbf{Left panel:} The difference in mass of satellites of mass $\mathcal{M}_{\star} > 10^7 \Msun$ in TNG50-1 and TNG50-1-Dark. 
    The grey background points show the scatter between exactly matched satellites, 
    blue points show the result of sorting the masses, and the red line the fit.
    \textbf{Right panels:} The dependence of the mass fitting parameters on simulation resolution,
    each simulation using galaxies with $\ga 100$ stellar particles.}
    \label{fig:type1s_mass_fit}
\end{figure}

We see in the left panel of Fig.~\ref{fig:type1s_mass_fit} 
that in TNG50 the total mass identified by Subfind \citep{Springel2001} 
as belonging to the subhalo is reduced.
We speculate that this mass difference may be partly a physical effect due to ram-pressure stripping
\citep[e.g.][]{Ayromlou2019, Ayromlou2021}, 
but also a numerical effect due to the ability of Subfind to distinguish the structures \citep[e.g.][]{Onions2012}.

We apply a similar fitting for mass change to that which we used for radial position change,
\begin{equation}
    \log_{10} (\mathcal{M}^{\mathrm{TNG}}_{\mathrm{sub}} / \mathcal{M}^{\mathrm{DMO}}_{\mathrm{sub}}) = -(\mathcal{M}^{\mathrm{DMO}}_{\mathrm{sub}}/a_{\rm m})^{b_{\rm m}},
    \label{eq:t1_mass_change}
\end{equation}
and follows a power law about a pivot mass $a_{\rm m}$.
The red line in the left panel of Fig.~\ref{fig:type1s_mass_fit} shows the outcome
of fitting this function, successfully reproducing the typical mass difference.

Similarly to the radius change, the mass change could depend on a number of the properties
of the subhalo and host halo, as well as simulation resolution.
Additionally, we expect a covariance between the mass and radius change.
However, to be consistent with the corrections we provide for radius change
we again apply only a one-dimensional fitting.
This gives us $a_{\rm m} = 2.2 \times 10^8 \hMsun$, $b_{\rm m} = -0.55$ for TNG50-1,
and $a_{\rm m} = 1.2 \times 10^9 \hMsun$, $b_{\rm m} = -0.98$ for TNG100-1.

The resolution dependence is somewhat more complicated than it was for radii.
The right panels of Fig.~\ref{fig:type1s_mass_fit} show the fitting parameters
in different runs.
In the runs at the lower end of the $M_{\mathrm{DMO}}$ range, 
a trend is seen for pivot mass $a_{\rm m}$ to increase
and power $b_{\rm m}$ to decrease as $M_{\mathrm{DMO}}$ increases.
However, in the runs with worse resolution (higher $M_{\mathrm{DMO}}$) the pivot mass 
and minimum resolved satellite mass converge, and the fitting method breaks down.
For this reason, our results from TNG100-2 and TNG300-1 are not in agreement,
and we are unable to fit to TNG300-2.

Consequently, while we note that satellite masses are reduced 
in the TNG simulation relative to TNG-Dark,
and that this change can be approximated by equation \ref{eq:t1_mass_change},
we do not provide fits for simulation resolution.

%% file: sections/type2_profile.tex
For the unmatched Type 2 satellites, 
we want to know their radial locations
after they are no longer found in the DMO simulation.
Our sample here consists of the residual satellites from the cases where
multiple TNG galaxies match to one TNG-Dark subhalo.
Note that in our fiducial matching algorithm all TNG galaxies map to a TNG-Dark subhalo, 
so if the corresponding subhalo in TNG-Dark has already merged into a central, 
then the mapping will be to that central. 
As a result, all TNG satellites are included in either the Type 1 or Type 2 sample,
except for the small number rejected earlier due to differences in the host halo of the matched subhaloes.

We take two approaches to explore their radial distribution.
Firstly, we consider the radial profile of the Type 2s 
at a single snapshot.
Secondly, we look at the radial motion between snapshots.

\subsection{Radial profiles of unmatched satellites}
\label{sec:t2_prof}

Positions of Type 2 satellites at a single snapshot 
can be selected by using fits to their radial distribution.
As shown in Fig.~\ref{fig:tng300_dmo}, 
the Type 2s are generally distributed much closer to the central galaxy than the Type 1s,
and this gives a different profile shape.

Rather than fitting to these profiles directly,
we fit the cumulative distribution of the number of satellites as a
function of distance from the centre.
This directly provides a distribution from which satellite positions can be drawn.
We examine distances and profiles in three dimensions as we are only considering simulated galaxies.

Desiring a profile from which we can readily draw samples,
we find that the cumulative distribution is well fit by assuming 
the galaxy number counts of Type 2s follow a log-normal distribution
\begin{multline}
    N(r/R_{200 \rm m})\\
    = N_{\mathrm{sats}} \mathrm{exp} \left( { \frac{ -(\log_{10} (r/R_{200 \rm m}) - \log_{10} (r_s/R_{200 \rm m}) )^2 }{2 \sigma^2} } \right).
\label{eq:lognorm_prof_num}
\end{multline}
This implies the number density profile of Type 2s can be determined from this using
\begin{equation}
    n(r/R_{200 \rm m}) = \frac{ N(r/R_{200 \rm m}) }
    { \sqrt{2 \pi^3} 4 \sigma (r/R_{200 \rm m})^3 \ln (10)} ,
\label{eq:lognorm_prof_den}
\end{equation}
where $N_{\mathrm{sats}}$ is the total number of satellite galaxies, $r$ is the radial position of the satellite,
$r_s$ is a scale radius and $\sigma$ is the distribution width.
We note that an Einasto profile \citep{Einasto1965} is also able to fit the data, 
but that we select the log-normal approach 
due to the comparative ease of drawing random samples from it.

Applying this fit we find the average parameters for the three simulations are
$r_s / R_{200 \rm m} = 0.18$ and $\sigma = 0.43$.
We note that due to fitting in terms of $r/R_{200 \rm m}$ this depends on the halo mass estimate used.
If, instead of using an overdensity of 200 times the mean density,
we use 200 times the critical density then $r_s /R_{200 \rm (c)}$ increases to 0.30 but $\sigma$ does not change.

\begin{figure}
    \centering
    \includegraphics[width=\linewidth]{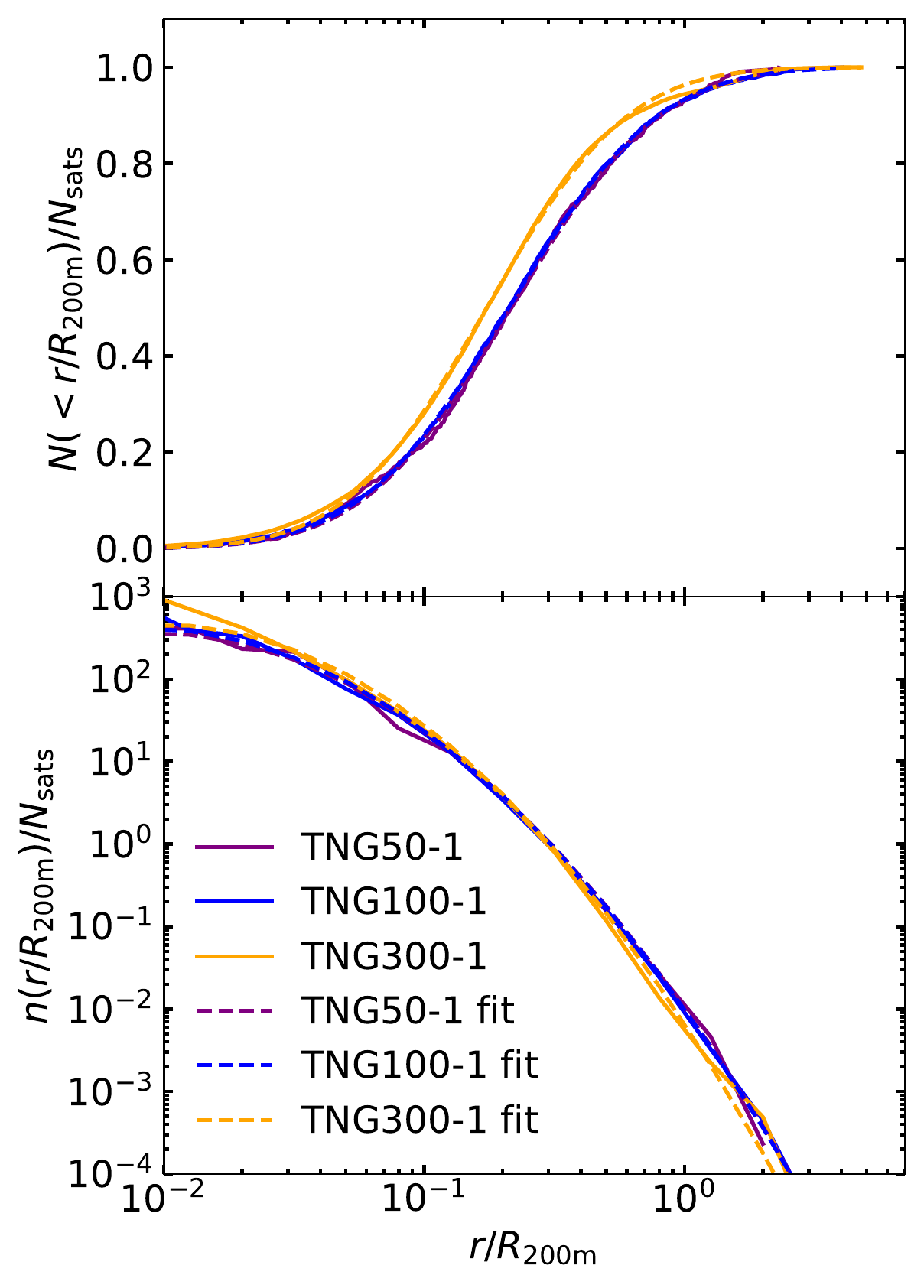}
    \caption{Fitting to the distribution of Type 2 satellite galaxies in TNG,
    as a function of distance normalised by halo radius at $z=0$.
    The upper panel shows the cumulative distribution and the lower panel
    the number density profile. 
    Solid lines show the Type 2 satellites of TNG, 
    and dashed lines show the fits using equation \protect\ref{eq:lognorm_prof_den}, 
    which assumes the number counts follow a log-normal distribution.}
    \label{fig:T2_prof_fit}
\end{figure}

In Fig.~\ref{fig:T2_prof_fit} we show the outcome of this model fit.
In the upper panel we show the cumulative number of satellite galaxies in TNG50-1, TNG100-1 and TNG300-1,
and in the lower panel we show the number density profile.
We have not set matching mass limits in this case, 
instead selecting all Type 2s above the stellar mass limit for each simulation 
and then normalising by the total number and the halo $R_{200 \rm m}$.
It is immediately apparent that there is a very similar distribution of Type 2 satellites 
in the different resolution runs,
and these agree well with fits given by equation \ref{eq:lognorm_prof_den}.
Slight discrepancies in the fits are visible on the smallest and largest scales,
particularly in TNG300-1 where the tails are underestimated due to the 
profile shape differing slightly from that of TNG50-1 and TNG100-1.
However, in the range $0.02 \la r/R_{200 \rm m} \la 1$ our fitting is seen to work well
for all the simulations.

\begin{figure}
    \centering
    \includegraphics[width=\linewidth]{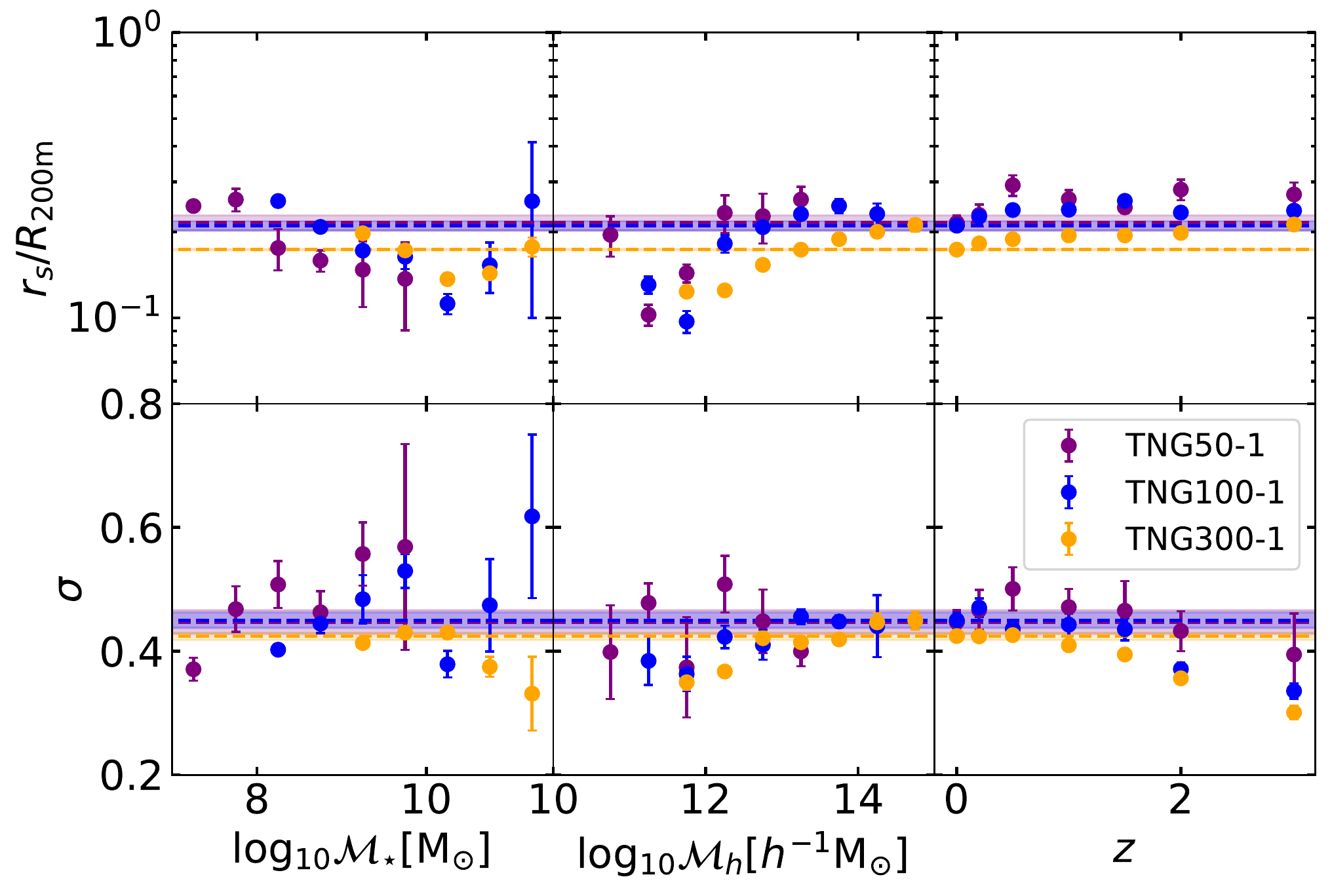}
    \caption{The dependency of the parameters of the Type 2 number density profile fits
    on halo mass, stellar mass and redshift.
    The $r_s / R_{200 \rm m}$ (upper panels) and $\sigma$ (lower panels)
    parameters of equation \protect\ref{eq:lognorm_prof_den} are
    shown in bins of stellar mass (left panels), halo mass (centre panels) and redshift (right panels) 
    for TNG50-1, TNG100-1 and TNG300-1.
    The dashed lines show the respective overall fits, as plotted in Fig.~\protect\ref{fig:T2_prof_fit}.
    For clarity we only show bins containing satellites of at least 10 groups.}
    \label{fig:T2_prof_pars}
\end{figure}

In Fig.~\ref{fig:T2_prof_pars} we then examine whether the fits depend
on halo or stellar mass.
We select Type 2 satellites in evenly spaced bins of halo and stellar mass
and recalculate the profile fits in each bin.
The changes in the parameters are all relatively small,
with a slight reduction in $r_s / R_{200 \rm m}$ in low mass haloes,
and for the highest mass galaxies.
This consistency is expected from our earlier conclusions 
that the overall normalised profiles do not depend on halo mass.
We do not consider the dependence on subhalo total mass,
as this would have no equivalent in the case of SAMs,
where these satellites are not contained in subhaloes.
Additionally, we show the redshift dependence to this fitting,
which leads to a slight reduction of $\sigma$ at higher $z$.

In SAMs, the number of Type 2 satellites is known.
However, in some simpler empirical models
the number of Type 2 satellites would need to be input 
in order to apply these profiles to DMO simulations.
Despite the over-simplifications of halo occupation distribution models \citep[HODs, see e.g.][]{Hadzhiyska2020},
we fit the number of Type 2s per group, $\frac{N_{\mathrm{T}2}}{N_{\mathrm{grp}}}$,
with a simple three parameter model,
\begin{equation}
    \frac{N_{\mathrm{T}2}}{N_{\mathrm{grp}}} = \left(\frac{\mathcal{M}_{h} / \hMsun - M_0 }{M_1}\right)^{\alpha}.
    \label{eq:hod_unfit}
\end{equation}

\begin{figure}
    \centering
    \includegraphics[width=\linewidth]{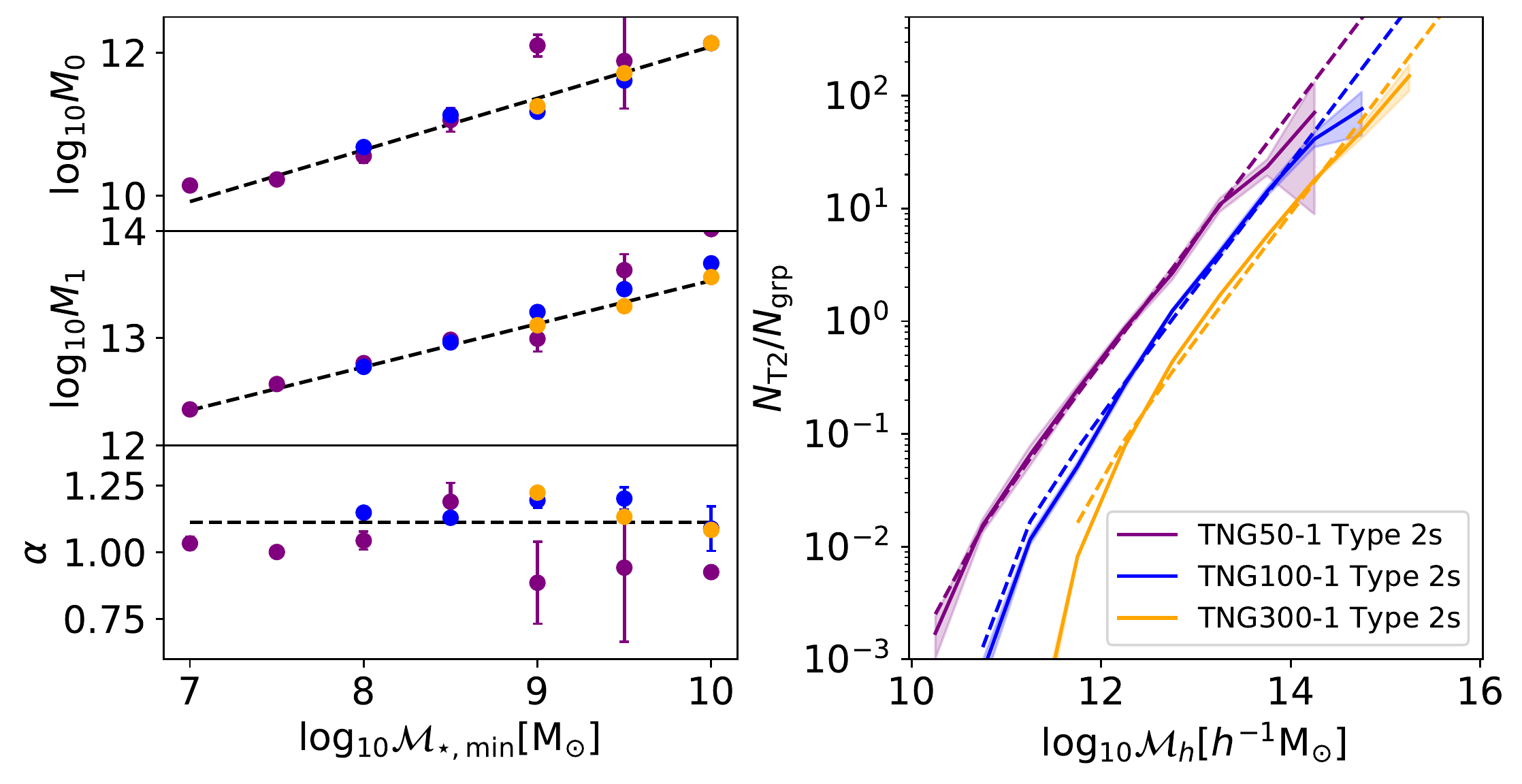}
    \caption{The number counts of Type 2 satellites in the TNG simulations at $z=0$.
    Left panels show the parameters of equation \protect\ref{eq:hod_unfit} 
    as a function of the stellar mass limit used, 
    fit with the black dashed lines, which are given by equation \protect\ref{eq:hod_pars}.
    The right panel then shows the number counts in the TNG simulations
    as solid lines, and the model results as dashed lines.
    Different mass limits are used for each resolution,
    giving different numbers of Type 2s:
    TNG50-1 is shown for $\log_{10} (\mathcal{M}_{\star} / \Msun) \geq 7$,
    TNG100-1 is shown for $\log_{10} (\mathcal{M}_{\star} / \Msun) \geq 8$
    and TNG300-1 is shown for $\log_{10} (\mathcal{M}_{\star} / \Msun) \geq 9$.}
    \label{fig:HOD_counts}
\end{figure}

This model is illustrated in Fig.~\ref{fig:HOD_counts}.
The left panels show the three parameters of equation \ref{eq:hod_unfit}
as a function of the stellar mass cut applied to the galaxy sample, $\mathcal{M}_{\star, \mathrm{min}}$,
for TNG50-1, TNG100-1 and TNG300-1.
It can be seen that the fits are relatively insensitive to the simulation choice, 
but there is a dependence on the stellar mass of selected galaxies,
which we have fit with the black dashed lines,
given by
\begin{equation}
\begin{split}
    M_0 & = 10^{4.84} (\mathcal{M}_{\star, \mathrm{min}} / \Msun)^{0.724}, \\
    M_1 & = 10^{9.50} (\mathcal{M}_{\star, \mathrm{min}} / \Msun)^{0.404}, \\
    \alpha & = 1.11. \\
\end{split}
    \label{eq:hod_pars}
\end{equation}
This stellar mass dependence is a result of an increased number of satellites
per group and the inclusion of satellites in lower mass groups,
both resulting from a lower minimum mass threshold.
However, the lack of dependence on the simulation resolution is more surprising,
as it means that even with improved resolution we are still finding some 
massive galaxies lose their subhalo to become `orphaned'.

The right panel of Fig.~\ref{fig:HOD_counts} then shows
the number of Type 2 satellites per group,
and the fits resulting from equations \ref{eq:hod_unfit} and \ref{eq:hod_pars}.
While the number of satellites is reproduced for intermediate halo masses
$\mathcal{M}_h \approx 10^{12.5} \hMsun$,
the fitting is less accurate at either end of the mass scale,
particularly the most massive haloes have the number of Type 2s over-estimated.
As we are only interested in knowing the approximate number of Type 2s,
we do not attempt to correct this further.

%% file: sections/type2_model.tex
\subsection{Modelling Type 2 radial motion}
\label{sec:t2_infall}

The second approach we consider for determining the location of Type 2 satellites
is to trace and model their radial motion.
To do this we take all the galaxies we have assigned as Type 2s at redshift zero,
and find the difference in position from earlier snapshots.
We restrict ourselves to galaxies that remain satellites at earlier snapshots,
and which reside in haloes which differ in mass by less than 0.15 dex between snapshots.

Fitting a relationship between successive snapshots and propagating this over time
invites an increasingly large error on each iteration.
Instead, we look at the change in radial separation of Type 2s from their host
across a range of time steps.
Given an initial radial distance at an earlier time, 
this can then be applied to generate radial distances at later times.
By implication, the application of this means the positions of satellites
at successive snapshots are not directly related,
but instead they are both related to 
the radial separation at the starting time.

Strictly, we are then concerned only with the radial change since 
a certain starting time, that at which the satellite was last identified as a Type 1.
However, this is very restrictive on the number of satellites available at each snapshot,
and has a strong dependence on the criteria used to identify the galaxy type.
Instead, we consider the radial change from all snapshots at which the galaxy
remains a satellite (Type 1 or Type 2).
Regarding the number which remain Type 2s on tracing back from redshift zero,
about 90\% of the Type 2s are still Type 2s after a single timestep,
dropping to 50\% after slightly over 2 Gyr (15 snapshots),
and decreasing slowly for greater times.

Having found the historical locations of the satellites,
we can then consider statistically the distribution of 
possible radial movements of a galaxy at a given initial position.
This allows us to estimate the positions of Type 2s over time 
by drawing randomly from this distribution.

To find this distribution we first calculate for each galaxy
the probability, $\lambda$, that a galaxy at the same initial radial distance
has experienced more radial motion towards the centre of the group.
We calculate this by examining all galaxies starting in 
a bin of log radial location centred on the selected galaxy with width 0.1 dex,
and determining the fraction that move inwards by the same or a greater proportion
(equal or lower value of $\log_{10} ( r_{\mathrm{end}} / r_{\mathrm{start}}  )$),
giving a $\lambda$ value in the range $0 < \lambda \leq 1$.

\begin{figure}
    \centering
    \includegraphics[width=\linewidth]{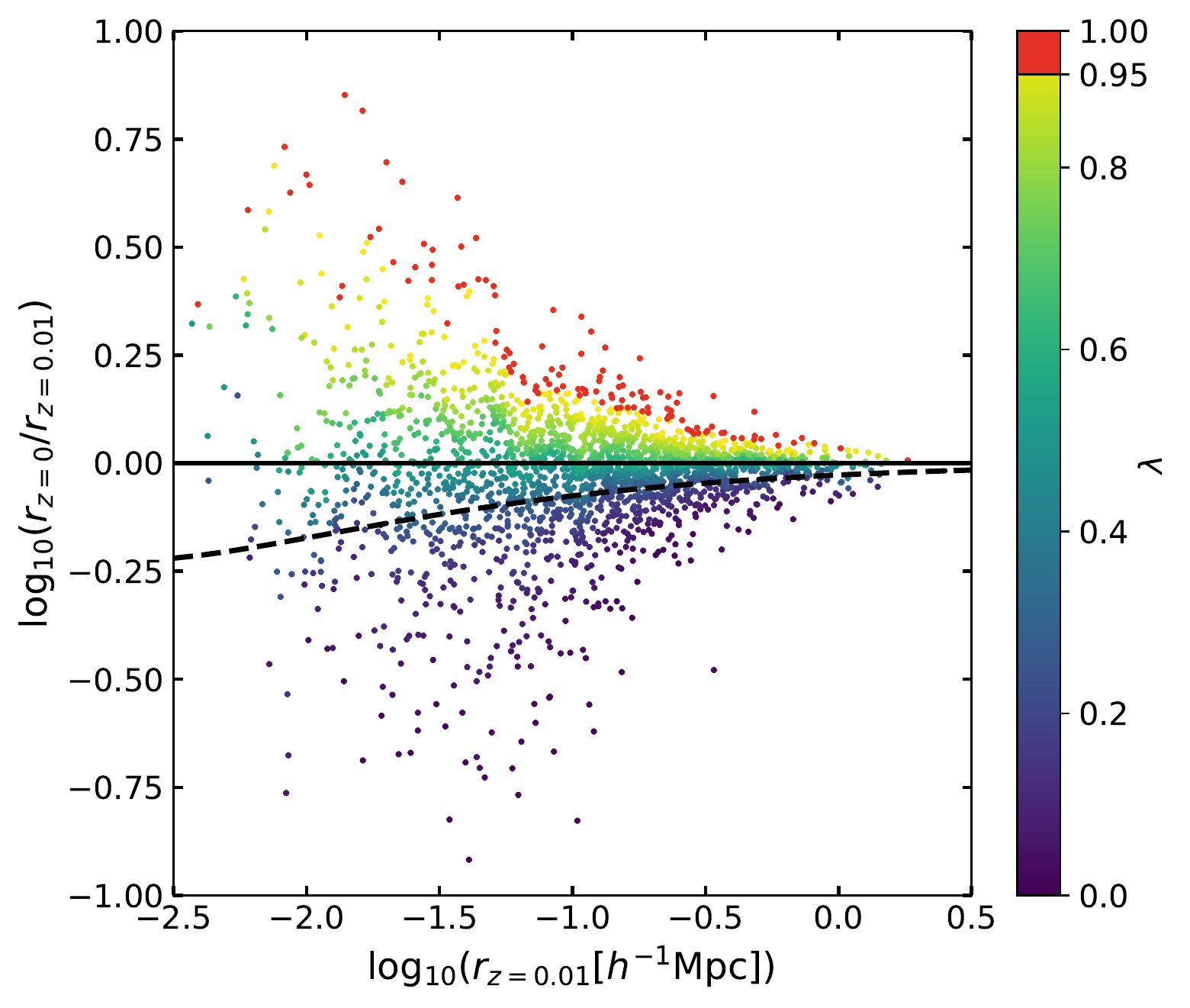}
    \caption{Movement of Type 2s between the final two snapshots of the TNG100-1 simulation,
    $z=0.01$ and $z=0$,
    plotted against their starting location at $z=0.01$.
    Galaxies are shown for masses $\mathcal{M}_{\star} > 10^8 \Msun$
    and colour-coded by the probability of a galaxy at
    a similar starting location moving further inwards.
    We highlight in red the ones which move furthest outwards.
    The black dashed line then shows the prediction of the
    model given in Table \protect\ref{tab:type2_fit_table} for $\lambda=0.5$.}
    \label{fig:snap98_type2s}
\end{figure}

Due to the small number of Type 2 satellites in TNG50-1,
we instead focus on TNG100-1 when developing our model.
In Fig.~\ref{fig:snap98_type2s} we show the radial movement of Type 2 satellites 
in TNG100-1 between $z=0.01$ and $z=0$.
We colour the points by their $\lambda$ value, 
and highlight in red the ones with $\lambda \geq 0.95$ 
which we use to fit the upper power law described below.
We also show the prediction of the model given below for $\lambda=0.5$,
showing the satellites tend to move slightly inwards on average between these times.

Typically we see that Type 2 satellites gradually move towards the halo centre over time,
but that there is a chance that they move away from the centre.
In particular, those which begun close to the centre
(and so close to the pericentre of their orbit)
are more likely to move outwards.

We model this distribution as a sum of power laws.
This provides a relatively simple model which approximately visually matches 
the shape of the contours of equal $\lambda$,
and allows for the possibilities that satellites move outwards or inwards.
Alternative models can be proposed
(and this could perhaps be done using the machine learning methods of \citealt{KroneMartins2014}), but
fitting in bins of radius would require more parameters,
and fitting an overall trend would ignore 
the spread of satellite orbits followed.

Our power law sum consists of two terms:
an upper power law which describes the maximum outwards movement possible,
and a lower power law describing the inwards motion.
We express this in the form
\begin{equation}
    \log_{10} ( r_{\mathrm{end}} / r_{\mathrm{start}}  ) = u(t) r_{\mathrm{start}}^{v(t)} - C(\lambda, t) r_{\mathrm{start}}^{D(\lambda, t)},
    \label{eq:t2_2powers}
\end{equation}
where $r_{\mathrm{start}}$ is the initial radial location,
$r_{\mathrm{end}}$ is the final radial location and $t$ is the time between the snapshots.
The first term in this equation is our upper power law,
and the second term the lower power law.

Our procedure for fitting this is as follows.
We fit the galaxies with $\lambda \geq 0.95$ with a single power law $u r^v$.
Then we fit the lower power law $C r^D$ as a function of $\lambda$ in bins of width 0.05.
For both of these fittings we use only galaxies at $r_{\mathrm{start}} > 0.01 \hMpc$,
to avoid biasing the fit with the very few on the smallest scales
which may be affected by the spatial resolution of the simulation. 

\begin{figure*}
    \centering
    \includegraphics[width=\linewidth]{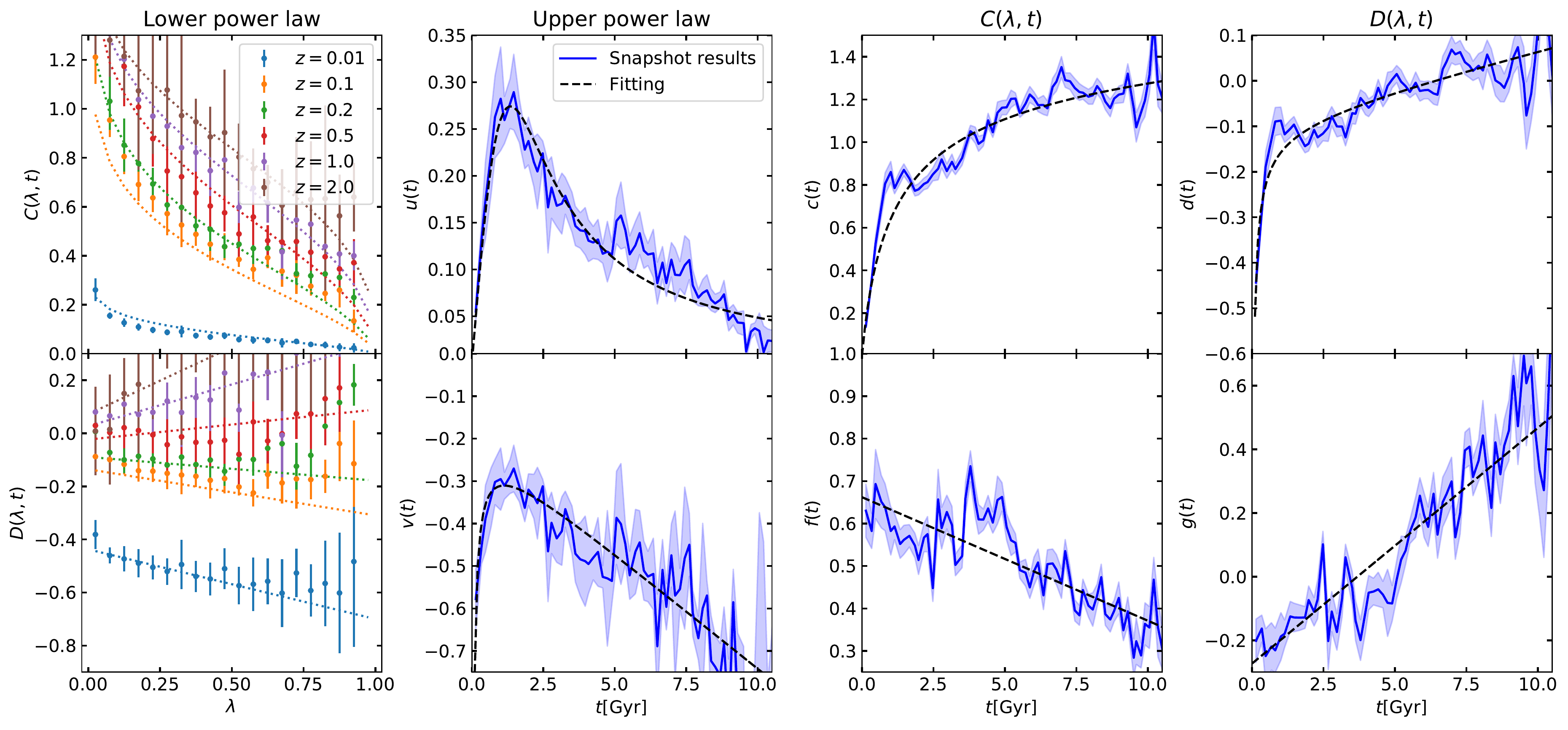}
    \caption{Parameters of the model given in Table \protect\ref{tab:type2_fit_table}
    used to fit the movement of $z=0$ Type 2 satellites between snapshots in TNG100-1.
    The left panels show the lower power law parameters as a function of 
    the distribution location $\lambda$,
    with different colours showing different starting redshifts
    and dashed lines showing fittings from Table \protect\ref{tab:type2_fit_table}.
    The other panels show the parameters as a function of time between snapshots,
    with the fittings overplotted as dashed lines.}
    \label{fig:type2_fit_pars}
\end{figure*}

In the left panels of Fig.~\ref{fig:type2_fit_pars} we show 
the dependence of the lower power law parameters $C(\lambda)$ and $D(\lambda)$ 
on the distribution percentile ($\lambda$) across different time periods.
The two components of the lower power law can be fit as
$ C(\lambda, t) = c(t) (-\log_{10} \lambda)^{f(t)} $
and
$ D(\lambda, t) = d(t) + g(t) \lambda $.

Finally, we fit this as a function of the time between snapshots,
for $0.136 < t/{\mathrm{Gyr}} < 10$.
The right hand 6 panels of Fig.~\ref{fig:type2_fit_pars} show
the dependencies of the parameters on time between snapshots.
In considering the time dependence, our primary requirement is that
parameters $u$ and $c$ tend towards zero at small times,
to give no instantaneous satellite movement
(although for times $t < 0.136$~Gyr,
shorter than the minimum this model is fit for,
it would be more appropriate to just set zero movement).
We include the time dependence of the parameters with a summary of the model
in Table \ref{tab:type2_fit_table}.

\begin{table*}
    \centering
    \caption{Details of the parameters of the model given in equation \protect\ref{eq:t2_2powers},
    used to fit the movement of Type 2 satellites between snapshots in TNG100-1
    with time steps $0.136 < t/{\mathrm{Gyr}} < 10$.
    The top half of the table gives the time dependence of the upper power law,
    and the lower half of the table gives the dependence on time and distribution position $\lambda$
    of the lower power law.}
    \begin{tabular}{ccccc}
        \hline
Section & Model & \multicolumn{3}{c}{Parameters} \\ 
\hline 
Upper power law & $u(t) r^{v(t)}$ & & & \\ 

& $u(t) = \frac{t}{u_1 + u_2 t^{u_3}}$ & $u_1$= 2.7 & $u_2$= 1.1 & $u_3$= 2.3  \\ 
& $v(t) = (v_1 + v_2 t)(1 + t^{v_3})$ & $v_1$= -0.10 & $v_2$= -0.053 & $v_3$= -0.78  \\ 

\hline 
Lower power law & $C(\lambda, t) r^{D(\lambda, t)}$ & & & \\ 

\multirow{2}{*}{$ C(\lambda, t) = c(t) (-\log \lambda)^{f(t)} $} & $c(t) = \frac{t}{c_1 + c_2 t^{c_3}}$ & $c_1$= 0.66 & $c_2$= 0.91 & $c_3$= 0.90  \\ 
& $f(t) = f_1 + f_2 t$ & $f_1$= 0.66 & $f_2$= -0.029 &  \\ 
\multirow{2}{*}{$ D(\lambda, t) = d(t) + g(t) \lambda $} & $d(t) = (d_1 + d_2 t)(1 + t^{d_3})$ & $d_1$= -0.094 & $d_2$= 0.015 & $d_3$= -0.67  \\ 
& $g(t) = g_1 + g_2 t$ & $g_1$= -0.27 & $g_2$= 0.074 &  \\ 

\hline
    \end{tabular}
    \label{tab:type2_fit_table}
\end{table*}

\subsection{Interpreting the fitting}

Our model has been designed to match the statistical distribution of satellite radial positions.
This means that for any individual satellite we are treating the orbital phase as a random variable,
and so the motion will not be accurately predicted from the initial location of the satellite,
but for the whole population the distribution should be reproduced.
It also means that our model parameters have no direct physical meanings,
but we can still infer some information from them.

Firstly, considering the parameters at small timesteps,
the shape of $C(\lambda, t)$, which has sharp upturn at 
lower  end of the $\lambda$ range,
demonstrates that most satellites do not move far,
but that the distribution has a large tail
of satellites with substantially greater radial movement,
perhaps those on first infall with radial orbits.

Looking at the time dependence,
the strengths of the power laws, given by $u(t)$ and $c(t)$,
inform us of the relative probabilities of a satellite moving 
towards or away from the group centre.
Across a few snapshots, both $u(t)$ and $c(t)$ increase rapidly,
showing the satellites can have large radial movements on their orbits,
but the overall population does not have a significant inwards or outwards movement.
At greater timesteps, $u(t)$ and $c(t)$ both become smoother,
with a gradual decrease in $u(t)$ and an increase in $c(t)$.
This shows a transition from the scatter associated with the orbital motion
to an average inwards motion for the satellite population.

This switch to an overall infall is also visible in $d(t)$,
which tends towards zero at large times, 
showing that some of the radial dependence is washed out by the overall infall.
However, there is still some radial dependence,
with $g(t)$ changing sign at large times.
This sign change, and the growth of $v(t)$,
is indicative of a continued tendency for those satellites which 
began close to the central to move outwards on average.
This is to be expected, as any satellites which began close to the central
and moved inwards will have merged into the central,
and so not be included in our analysis.

These interpretations show that our model has encapsulated
much of the expected satellite motion,
and should provide a practical method to predict the overall movement of satellite populations.

%% file: sections/model_tests.tex
We further explore our results and models here,
firstly via some tests of the application of our models
and then by discussing the interpretation and caveats of this work.

\subsection{Testing the Type 2 model for TNG subhaloes}

\begin{figure*}
    \centering
    \includegraphics[width=\linewidth]{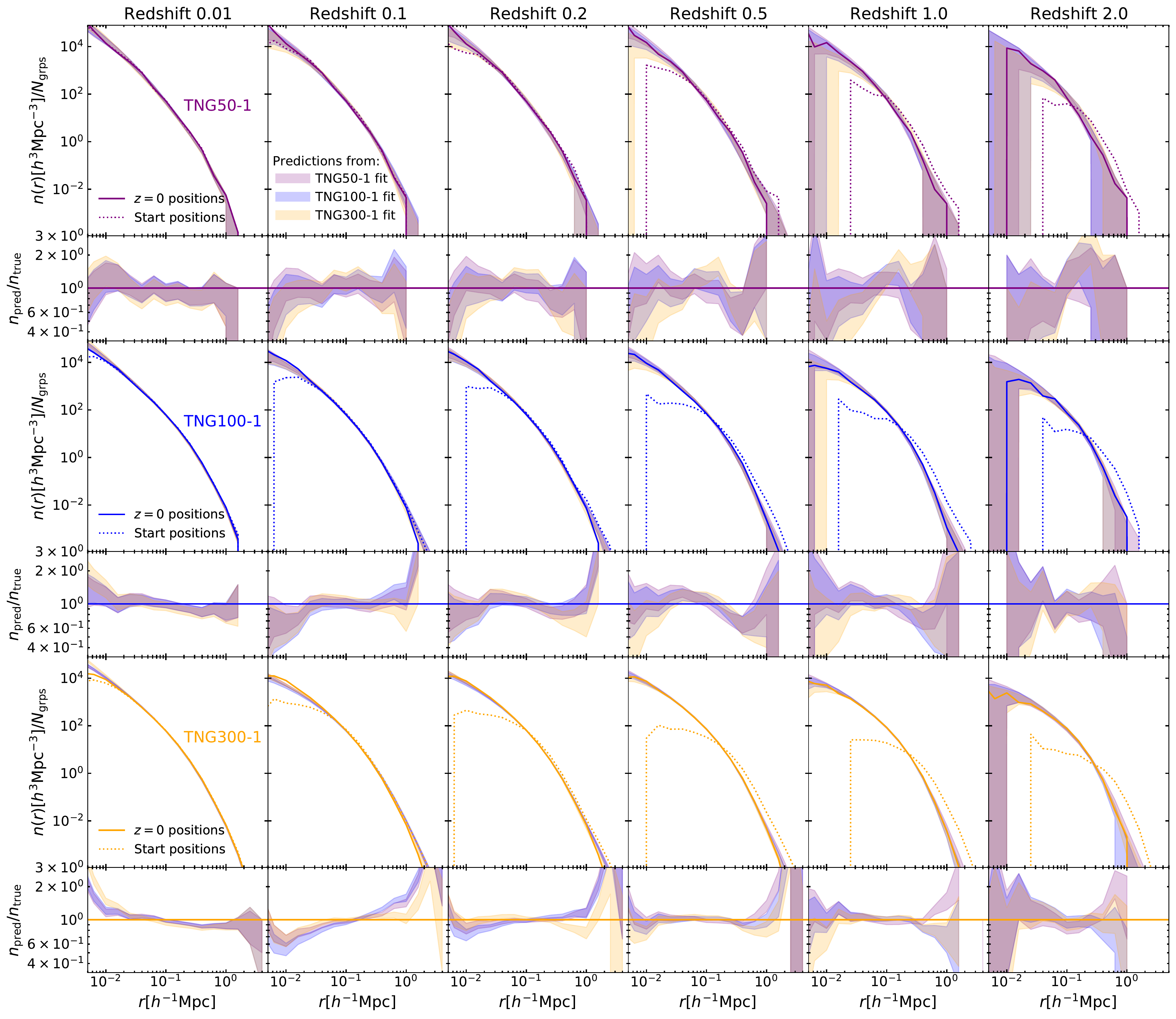}
    \caption{  
    Radial distributions of Type 2 satellites in the TNG simulations,
    before and after applying our model for their radial movement given in Table \protect\ref{tab:type2_fit_table}.
    The top two rows show TNG50-1 profiles, middle two rows TNG100-1 profiles and lower two rows TNG300-1 profiles.
    From left to right the panels show satellites tracked to higher redshifts.
    The larger panels show the radial profiles, 
    while the smaller panels show the ratio of the predicted profiles to the true profile.
    We include resolved galaxies in all groups, but show comoving distances as those are the input to our model.
    In each of the larger panels the dotted line shows the distribution of satellites at the redshift of the column
    and the solid line shows the distribution of the satellites at redshift zero.
    The shaded bands in all panels show the 95\% region for 1000 applications
    of the model predictions at redshift zero.
    The model predictions are calculated for the satellite locations shown in the dotted lines, 
    with random values of $\lambda$,
    and the different colours show the prediction of the model when fit to the different simulations.
    }
    \label{fig:type2_scatter_prof}
\end{figure*}

The primary test of the model from Section \ref{sec:t2_infall} is the application of it
to the traced locations of the satellites over time.
We show in Fig.~\ref{fig:type2_scatter_prof} the profiles of satellites at
redshift zero in TNG50-1, TNG100-1 and TNG300-1 as solid lines in each panel,
selected with $\log_{10} (\mathcal{M}_{\star} / \Msun) \geq$ 
7 (TNG50-1), 8 (TNG100-1) or 9 (TNG300-1).
The dotted lines then show the radial distribution of these same satellites
traced back to the redshift of the column.
If successful, our model should take the radial positions shown by the dotted line
in each panel, and reproduce the solid lines.

The blue shaded region in each panel shows the result of the application
of our model specified in Table \ref{tab:type2_fit_table}.
The model was applied 1000 times,
with a different set of random $\lambda$ values each time,
and the shaded regions show the 95\% region of the spread of these results.
In most cases, it can be seen that our model is successfully generating 
the distribution of satellites at redshift zero.
Discrepancies in our model are most apparent on small scales when it is applied to TNG300-1,
likely due to the different halo masses sampled by it.
More generally, there is a small tendency 
to move satellites too close to the centre when starting at higher redshifts.

Fitting our model on Type 2 satellites in TNG50-1, TNG100-1 and TNG300-1 leads to 
slightly different parameterisations,
although the overall trends are similar between them.
These different fits are shown in Appendix \ref{app:t2_resolutions}.
We show using the purple and orange shaded regions the results
of alternatively applying the model as fit on TNG50-1 or TNG300-1,
demonstrating the comparable results of each.

\begin{figure}
    \centering
    \includegraphics[width=\linewidth]{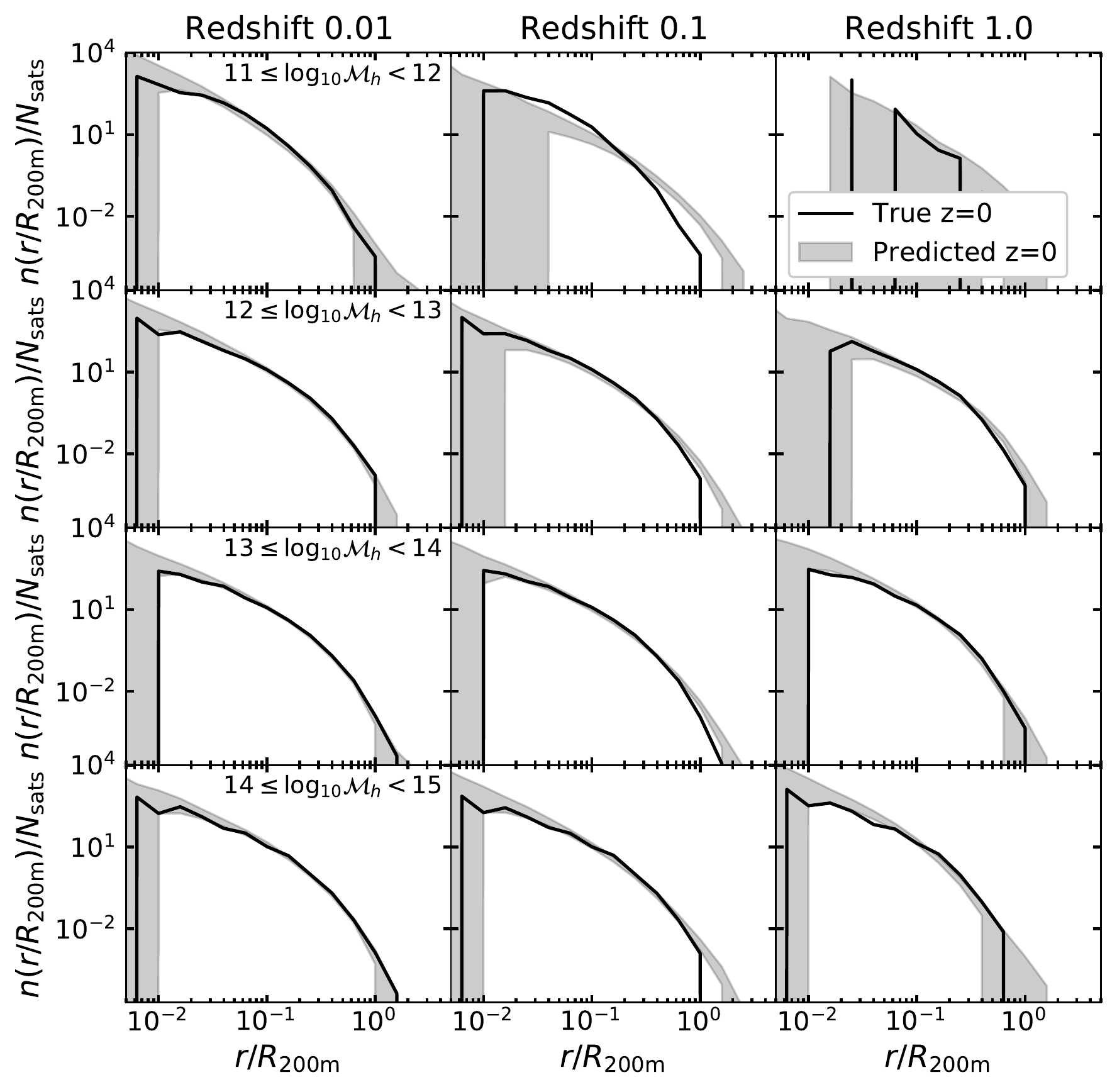}
    \caption{The Type 2 profile fits in different group mass bins for TNG100-1.
    The rows each show a different group mass selection, 
    while the columns show satellites traced back to different redshifts.
    The solid lines show the redshift zero positions of satellites
    starting at the specified redshift, 
    and the shaded regions show the 95\% spread of the positions predicted by our model over 1000 applications.}
    \label{fig:T2_fit_grps}
\end{figure}

We may anticipate some halo or stellar mass dependence to these fits,
as dynamical friction is a function of both of these \citep[e.g.][]{Binney1987}.
We show in Fig.~\ref{fig:T2_fit_grps} the radial profile in four halo mass bins,
starting at three different redshifts.
It is apparent that our model achieves reasonable success in every case,
although there are some minor discrepancies.
In particular the model performs less well for halo masses below $10^{12} \hMsun$,
which is unsurprising given we have fewer Type 2s to fit in those haloes.

Some of the differences are attributable to variation in the distribution locations as a function of halo mass.
We find that Type 2s in lower mass haloes are assigned $\lambda$ values 
which are on average less than 0.5, while the opposite applies to high mass haloes.

A similar picture emerges for stellar mass, 
with our model working well for the lower mass satellites which are the most frequent,
and slightly less well for higher masses.
Therefore we conclude that, while there are mass dependencies,
these are small and so our model is able to perform adequately without 
these extra dependencies.

\subsection{Testing the full model on TNG300-Dark}

The accuracy of the power law model for the inwards displacement of Type 1s given in Section \ref{sec:t1_model} 
plus the distributions of Type 2 satellites given in Section \ref{sec:t2_prof}
can be tested simply by application to the positions of TNG300-1-Dark subhaloes
we selected in Section \ref{sec:dmo_prof}.
To do this we move the Type 1s radially inwards along the vector separating them from the central,
and add Type 2s randomly distributed in a sphere around the central 
with the log-normal radial distribution given in Section \ref{sec:t2_prof}.

\begin{figure*}
    \centering
    \includegraphics[width=\linewidth]{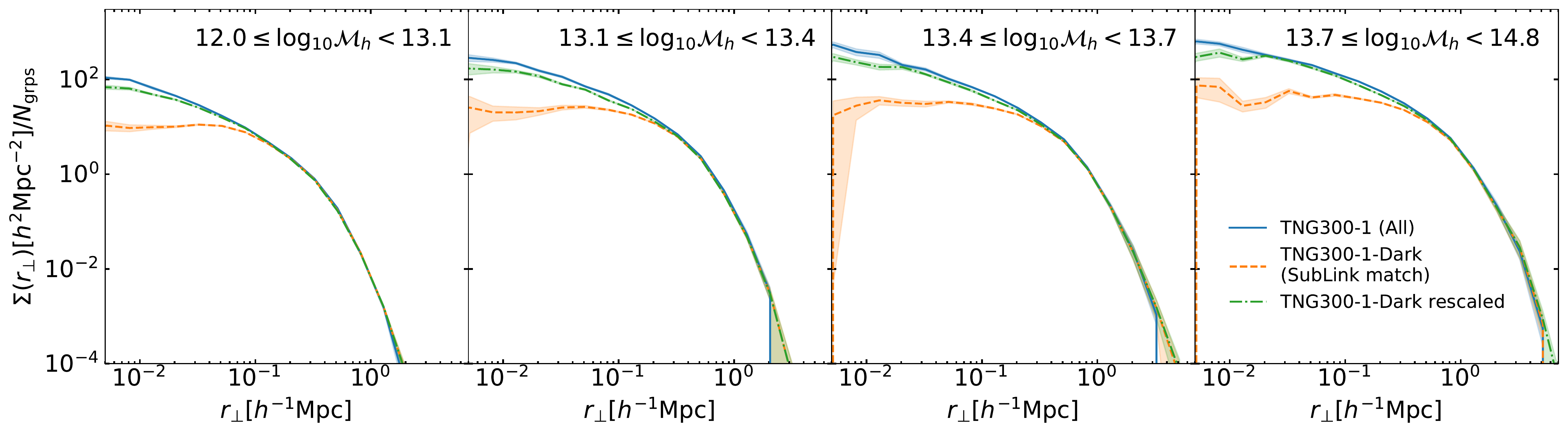}
    \caption{The outcome of the application of our model for Type 1 satellites on the TNG300-1-Dark profile at $z=0.2$,
    with the addition of the profile for Type 2 satellites,
    shown as green dot dash lines,
    compared to the TNG300-1 profile (blue solid lines) and the original TNG300-1-Dark profile (orange dashed lines)
    from Fig.~\protect\ref{fig:tng300_dmo}.}
    \label{fig:dmo_rescaled}
\end{figure*}

In Fig.~\ref{fig:dmo_rescaled} we show the outcome of this test
compared to the TNG and TNG-Dark profiles of Section \ref{sec:dmo_prof}.
On scales $r_{\bot} \geq 0.02 \hMpc$ our model accurately modifies the TNG-Dark simulation
to give it the same profile as TNG.
On the smallest scales we see a slight underestimation of the number of satellites,
which is related to the slightly different small-scale
profiles seen amongst the TNG simulations in Fig.~\ref{fig:T2_prof_fit}.
While we underestimate the profile for TNG300-1 on small scales in
both Fig.~\ref{fig:T2_prof_fit} and Fig.~\ref{fig:dmo_rescaled},
we expect the simulations with better resolution to be more accurate on
small scales---and these were well reproduced in Fig.~\ref{fig:T2_prof_fit}---so 
we do not try and correct the discrepancy remaining here any further.

Overall the Type 1 model and the model for the Type 2 profiles is seen to accurately 
reproduce the profile from the full-physics simulation.
In future work we will test these further,
and also evaluate our model for Type 2 radial motion (Section \ref{sec:t2_infall}), 
based on tracing subhaloes across snapshots, 
by application directly to a semi-analytic model for galaxy formation.

%% file: sections/discussion.tex
\subsection{Physical interpretations}
\label{sec:physical_discussion}

By comparing the TNG and TNG-Dark simulations,
we showed that there are two primary effects of baryons on the 
radial distribution of satellites (and subhaloes) in groups.

Firstly, the comparison between satellites and their matched subhaloes in the DMO runs 
shows that satellites in the full-physics simulations are located at smaller halo-centric distances 
at the time of inspection than their surviving analogue subhaloes in the DMO simulations. 
Secondly, the existence of a population of satellites with no DMO matches suggests 
an increased survival time of satellites in full-physics simulations. 
These effects are connected, as satellites that spent more time in their current hosts 
are typically found closer to their host centres \citep{Rhee2017}.

The greater survival time of satellites in TNG can be explained by 
the inclusion of a baryonic core to the satellites.
Many studies \citep[e.g.][]{Smith2016, Joshi2019, Lokas2020, Engler2021a} 
show that tidal stripping acts primarily on the dark matter
component of subhaloes, and that the baryonic component is not extensively stripped.
This central component can thus be postulated to keep the satellite
bound beyond the point it is disrupted in a DMO simulation,
in agreement with \cite{Nagai2005}.

This would seemingly be in contrast to the \cite{Chua2017}
result that the addition of baryons reduces the survival time,
or the conclusion of \cite{Bahe2019} that baryons make little difference to survival times.
However, our findings are not necessarily in tension with such results. 
Importantly, throughout this work we have focused on satellite galaxies above a certain minimum stellar mass 
and on their analogues in the DMO simulations, whereas \cite{Chua2017}, for example, 
analyse the entire population of subhaloes, whether luminous or not, and also include lower-mass ones. 
Secondly, our orphan population, i.e. the satellites with no DMO surviving counterparts, 
is only a small proportion of the total group--galaxy population and is biased towards the centre. 
Instead, our results therefore seem to suggest that there is a 
strong radial dependence to the effect of baryons on the survival of satellites. 
We cannot exclude, but do not think it the case, 
that some differences across works may be due to different simulations using different astrophysical feedback mechanisms.

Different survival times between works could also be related to
the opposite effect to that considered in this work: disruption caused by baryons.
A suppression in the number of substructures is known to occur 
due to the destruction of satellites by baryonic discs \citep[e.g.][]{GarrisonKimmel2017, Kelley2019}.
Our choice to select only galaxies from the full-physics simulation 
and then determine their DMO analogues means we do not account for this,
but it will affect the relative survival times of full-physics and DMO substructures.

An alternative explanation for the greater survival time we see
is provided by \cite{Haggar2021},
who argue that the baryonic material in the centre of the subhaloes
causes a contraction of the surrounding dark matter distribution,
as seen in other works \citep[e.g.][]{Dolag2009, Adhikari2020}.
This leads to a more pronounced density contrast between the subhalo and the host halo,
making it easier for the halo finder to detect the subhalo.
If the differences are indeed due to the subhalo detection and tracking,
then this might in future be resolved by 
more advanced structure finders such as those of \cite{Elahi2019a} and \cite{Springel2021},
and alternative methods such as the merger graphs of \cite{Roper2020}.

A similar contraction argument can be used to explain the inwards displacement of full-physics satellites.
Baryons change both the concentration \citep[e.g.][]{Bryan2013, Lovell2018, Chua2019}
and shape \citep[e.g.][]{Rasia2004, Lin2006} of haloes,
which can change the location of the satellite galaxies in the potential of the host.
Contraction of the halo can then be suggested
to lead to the satellite being further out in the potential,
and then falling inwards towards the halo centre to balance this.
Alternatively, it is possible that the baryons are increasing the drag force
experienced by the satellites, causing the orbits to reduce in size \citep[e.g.][]{Gu2016}.

These explanations do not account for the resolution dependence to the position differences.
Instead, the resolution dependence of the DMO results implies that the inwards displacement is
at least partly a numerical effect of the simulations,
perhaps due to gravitational changes associated with 
the reduced sampling of the distribution of mass in the halo by the particles at poorer resolution.

Such degeneracies in the explanations should be remembered throughout.
Overall, when thinking about physical interpretations,
we cannot definitively distinguish the physical effects of adding baryons
from numerical effects. 
While we have provided some speculation for the reasons behind differences between full-physics and DMO results,
detailed explanations of the causes are beyond the scope of this work
and do not affect the empirical correction models we have presented.

\subsection{Caveats}

There are a number of assumptions and resulting caveats in the results and models
we have presented in this work.
We discuss a few of the more important ones here.

\subsubsection{Matching scheme}

One of the primary sources of potential uncertainty in our work
lies in the matching between TNG and TNG-Dark satellites,
and the distinction between Type 1 and Type 2 satellites.
Due to the differences in structure formation between the full-physics and DMO cases,
it is not necessarily clear that matched satellites represent the same structures.

One way of exploring this is by using an alternative matching scheme,
and one exists using the LHaloTree method of \cite{Nelson2015}.
The matches given by this method are bijective,
only matching objects where the object with the most matching particles
is the same for the TNG-Dark to TNG direction as for the TNG to TNG-Dark direction.
This provides a stricter criteria for the matching 
and leads to a reduced number of matched satellites (Type 1s),
particularly near the centre of haloes.
This eliminates the need to apply a correction to the locations of Type 1 satellites,
but enhances the need for Type 2s.
This, together with the abundance matching method we explored earlier,
shows that the balance between satellite types can be adjusted,
but the differences between the TNG and TNG-Dark profiles remain.
For our purposes, as we are interested in the expected positions of satellites
placed in DMO subhaloes by a SAM, which is a one-way matching,
it is most appropriate for us to use the one-way SubLink matches we have used throughout
to select the types.
In future, the effect of the matching scheme could be further explored by 
also comparing to results from the Lagrangian matching scheme of \cite{Lovell2018}.

One further comment on the matching is that
we found earlier that up to around 6\% of galaxies in each simulation
are identified as a satellite in TNG or TNG-Dark but as a central in the other.
Rather than attempt to correct for this, we have simply excluded these galaxies.
We have not, however, removed any other satellites which may be in these groups.
Most of these were at large distances from the centre when a satellite,
and our analysis is not affected by these objects.
This does, however, suggest some differences in either the structure formation
or the numerical methods used, particularly the matching scheme and group finder.

\subsubsection{Other physical effects}

While we have attempted to account for the most relevant physical dependencies and processes
in our analysis, there are others which we have not included.

For example, while we have considered dependencies on the masses of the hosts and satellites,
we have not included additional parameters such as those known to be secondary parameters in
assembly bias \citep[e.g.][]{Sheth2004, Xu2021}.
These may include local environment, halo shape and halo maximum circular velocity.
Any of these may impact the motion of satellites,
but, aiming for simplicity in our models, we choose not to pursue these secondary effects.

Finally, we note that throughout this work we have assumed that
all the satellites are directly associated with the central,
and that they do not interact with other satellites.
This simplification ignores effects known to exist in simulations,
including mergers between satellites \citep{Shi2020},
the accretion of groups onto clusters \citep{Haggar2021},
and more generally the pre-processing of satellites in other environments \citep{Donnari2021}.
We also note that we have not included any exclusion principle for the satellites,
and satellites could therefore lie arbitrarily close to each other
when implementing our models.

%% file: sections/conclusions.tex
In this work we have explored the radial distributions of satellite
galaxies in groups in the GAMA survey and in the IllustrisTNG simulations.
We have then compared the distributions of satellites between full-physics 
and dark matter-only (DMO) simulations, 
and developed models to characterise the differences.

For the GAMA survey, we showed the number density profile of all visible satellites
in groups of mass $12.0 \leq \log_{10} (\mathcal{M}_h / \hMsun) < 14.8$ at $z < 0.267$.
We saw that an increasing group mass leads to a greater number of satellites
and more extended radial distributions.
However, normalising by the number of satellites and the group radius 
showed that there is no mass dependence to the shape of the satellite radial profiles.
By comparison to mock catalogues constructed from DMO simulations,
we identified that GAMA group profiles are expected to be accurate for small scales,
but that satellites on the edges of the groups ($r_{\bot} \ga 1 \hMpc$),
are missed by the group finding algorithm and so the profiles are underestimated.

We selected galaxies and groups from the TNG300-1 simulation 
to replicate the GAMA sample
and showed that the profiles derived from these agree well with GAMA.
This agreement demonstrates the accuracy of the satellite population in TNG,
and so we can be confident that our subsequent modelling is performed on a realistic sample.

Comparing the full sample of group satellites above fixed stellar mass limits from the TNG simulations to
matched subhaloes from the equivalent TNG-Dark runs
showed that the satellite profiles are much flatter in the DMO case.
We attribute this to two connected effects;
an inwards displacement and a longer survival time
of the satellites in the full-physics case.

Following this, we developed empirical models to account for these effects.
We showed that the reduced halo-centric distances of matched satellites 
can be accounted for with a simple power-law model,
and that a similar model can also reproduce the mass loss of these satellites.
We fit the unmatched satellites which have endured longer in the full-physics run via two methods.
Firstly, we considered the shape of the radial profile,
finding it can be fit by a model of log-normal number counts.
Secondly, we considered the radial motion of unmatched satellites over time.

In future work, we intend to apply our models to semi-analytic galaxy formation models,
with the aim of improving their predictions of galaxy clustering.
From a simulation perspective, an expansion to this work
would be to apply the same methods to other simulations, such as EAGLE \citep{Crain2015, Schaye2015},
and the use of alternative methods to find, track and match subhaloes.
Observationally, more reliable profiles of galaxy groups 
will be produced in future from the Wide Area VISTA Extragalactic Survey \citep{Driver2019}.
The use of different group finding algorithms in observational data will also provide improvements,
particularly around the edges of groups.

%% file: sections/appendix.tex
\section{Group mass function}
\label{app:GMF}

\begin{figure}
    \centering
    \includegraphics[width=\linewidth]{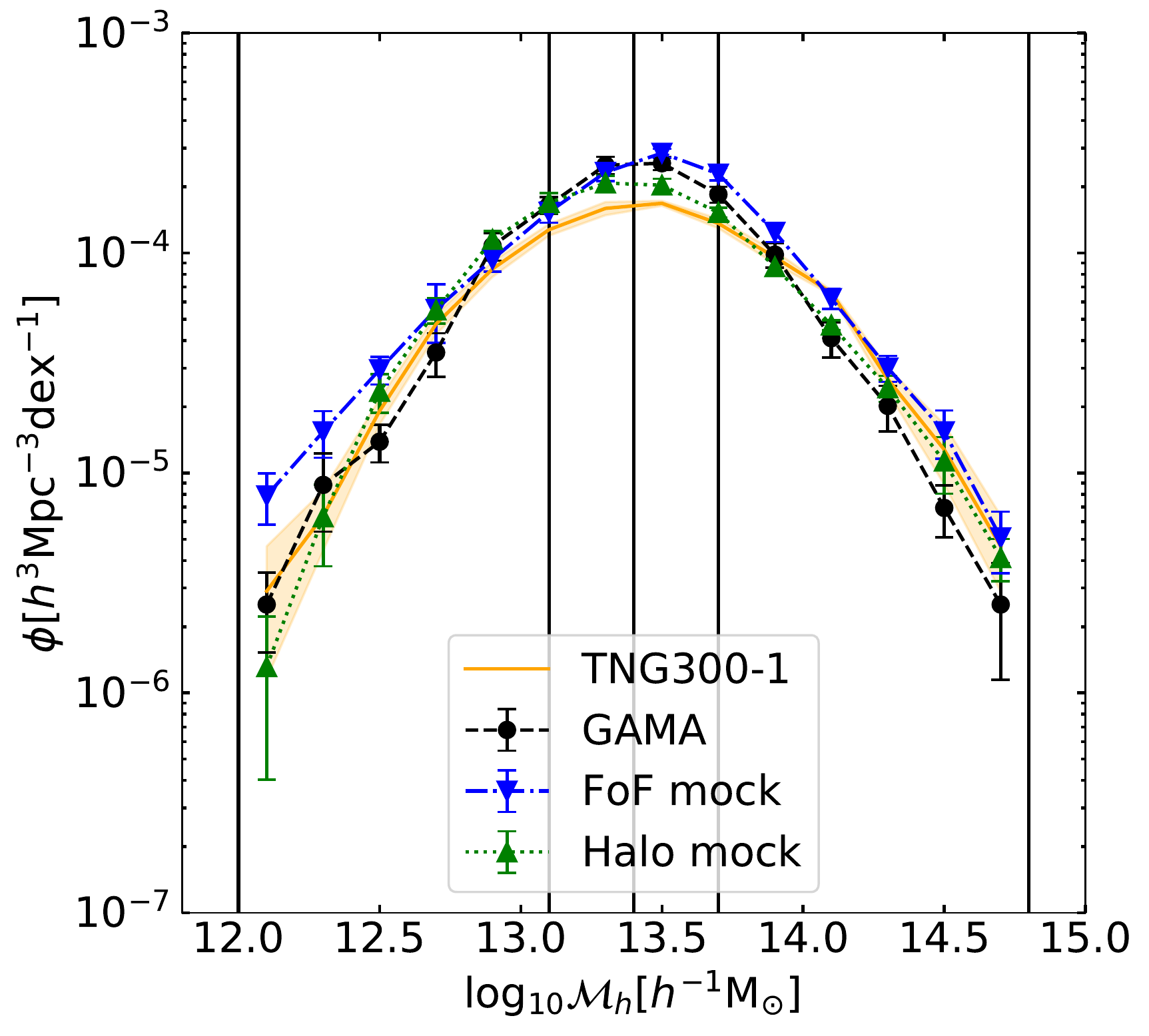}
    \caption{Mass distribution of selected groups in GAMA and the mock catalogues,
    together with a sample from TNG300-1 designed to approximately match the GAMA selection criteria. 
    Vertical lines show the mass bins we use.}
    \label{fig:group_masses}
\end{figure}

We show in Fig.~\ref{fig:group_masses}
the mass distribution of selected groups from GAMA, the mocks and TNG300-1.
Groups in TNG300-1 are selected with the method given in Section \ref{sec:group_sel}.

The primary effect of this selection method is to reduce the number of low-mass groups,
and so the comparable shapes of the mass distributions of GAMA and TNG300-1
demonstrates the success of our selection method for TNG300-1 groups.
Differences in the mass distribution are visible between GAMA and TNG300-1,
but these are mostly at masses above the peak,
where the selection function has less impact,
and so this is more likely related to differences in the underlying
group and galaxy populations \citep[see e.g.][]{VazquezMata2020}.
Additionally, our earlier result that the halo mass 
does not affect the profile shape suggests that 
the differences seen here are unimportant.

\section{Resolution dependence of TNG and TNG-Dark distributions}
\label{app:profile_lowres}

We show here that the results of Section \ref{sec:profiles_allres}
still apply if we instead consider different resolutions with the same box size.
This minimises the impact of different environments on our results,
demonstrating the outcomes are not simply an effect of cosmic variance.

\begin{figure}
    \centering
    \includegraphics[width=\linewidth]{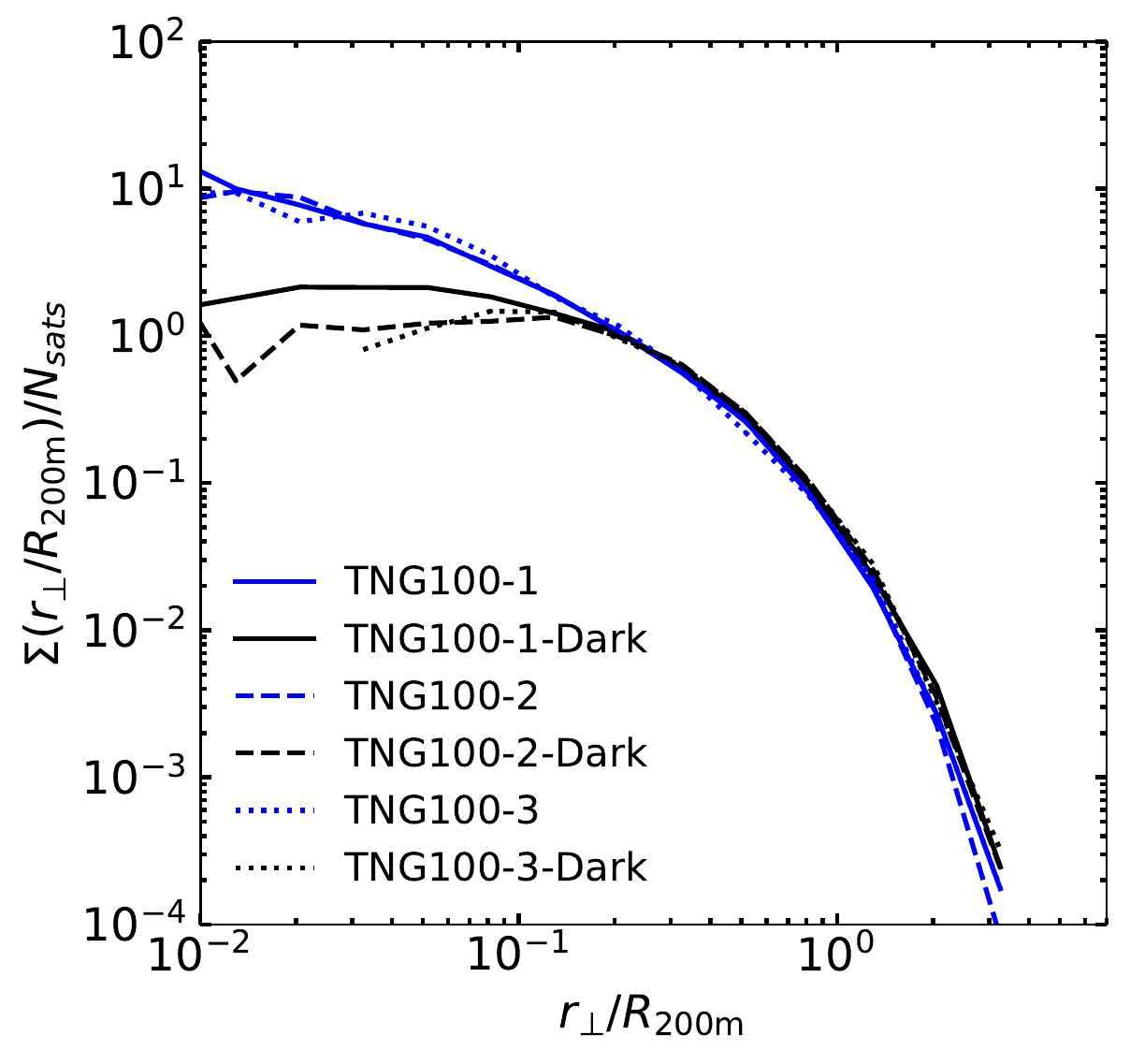}
    \caption{Normalised satellite profile of groups of mass $11 \leq \log_{10}\mathcal{M}_h < 15$ 
    and galaxies with $\mathcal{M}_{\star} \geq 10^9 \Msun$ at $z=0$
    in TNG100-1, TNG100-2 and TNG100-3.}
    \label{fig:lowres_prof}
\end{figure}

In Fig.~\ref{fig:lowres_prof} we show normalised profiles 
of satellites with $\mathcal{M}_{\star} \geq 10^9 \Msun$ for TNG100-1, TNG100-2 and TNG100-3,
each compared against subhaloes from the equivalent TNG-Dark run,
matched using SubLink.
We see the same results as in Section \ref{sec:profiles_allres},
i.e. that resolution does not affect the distribution of full-physics satellites, 
but improved resolution changes the distribution of the matched TNG-Dark satellites.
While the results from the worse resolution TNG-Dark runs are noisy,
they flatten at larger radii than TNG100-1-Dark,
and cut off at larger scales.

\section{Stellar mass dependence of TNG radial distribution}
\label{app:profile_mstar}

\begin{figure}
    \centering
    \includegraphics[width=\linewidth]{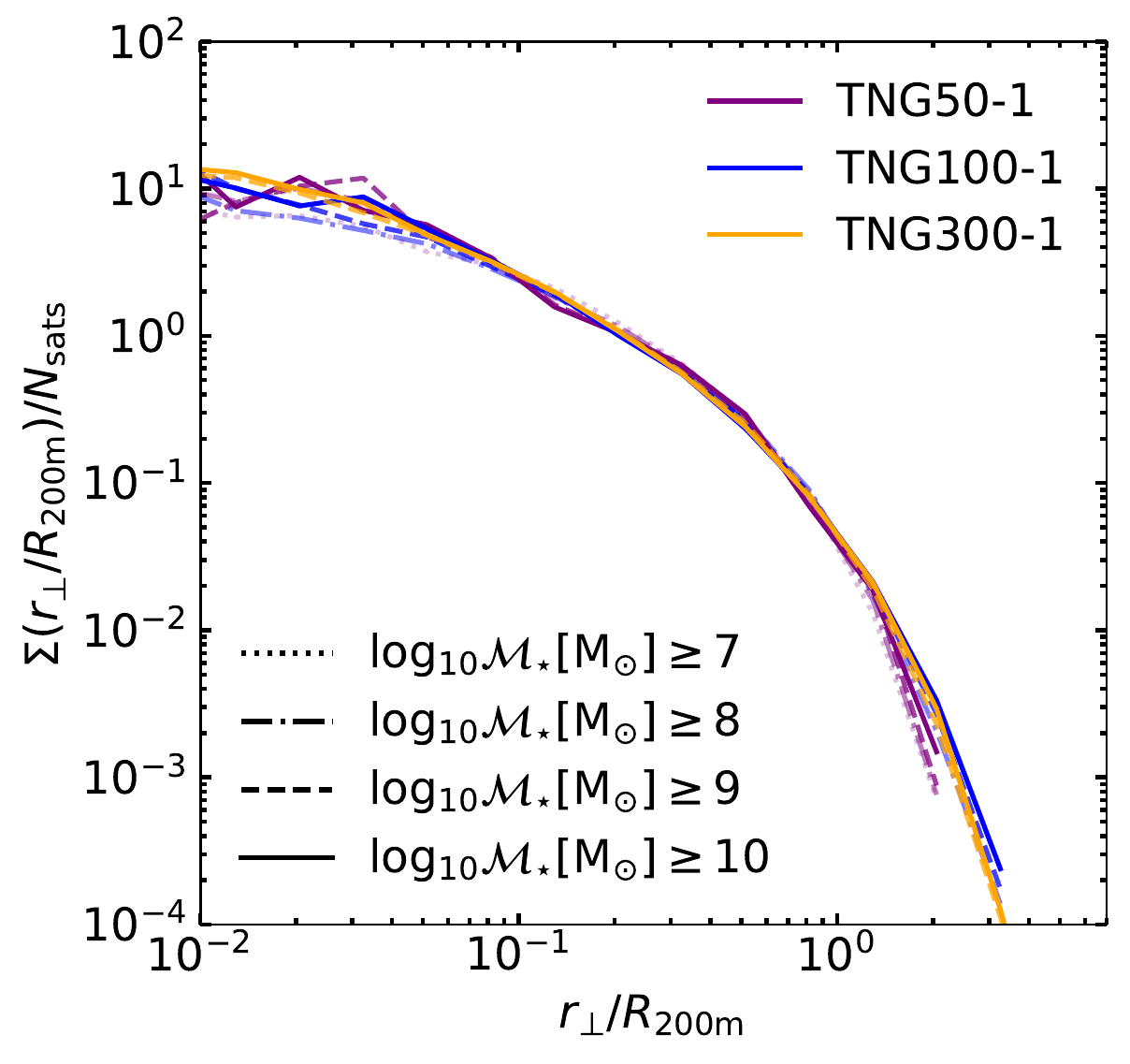}
    \caption{Normalised satellite profile of groups of mass $11 \leq \log_{10}\mathcal{M}_h < 15$ at $z=0$
    in TNG50-1, TNG100-1 and TNG300-1, with different stellar mass cuts on the galaxies included.}
    \label{fig:mstar_prof}
\end{figure}

In Fig.~\ref{fig:mstar_prof} we show that the normalised satellite profiles
in TNG do not depend on the simulation resolution or the stellar mass limit applied.
We include TNG50-1, TNG100-1 and TNG300-1, 
each with a series of increasing minimum satellite masses.
No change is seen in the shape of these normalised profiles 
when these different cuts are applied.

This shows that while the inclusion of lower-mass satellites 
increases the number of satellites, 
and so the amplitude of the average group profile,
these additional satellites are distributed
in the same way as the most massive satellites.

\section{Type 2 model fitting at different resolutions}
\label{app:t2_resolutions}

\begin{figure}
    \centering
    \includegraphics[width=\linewidth]{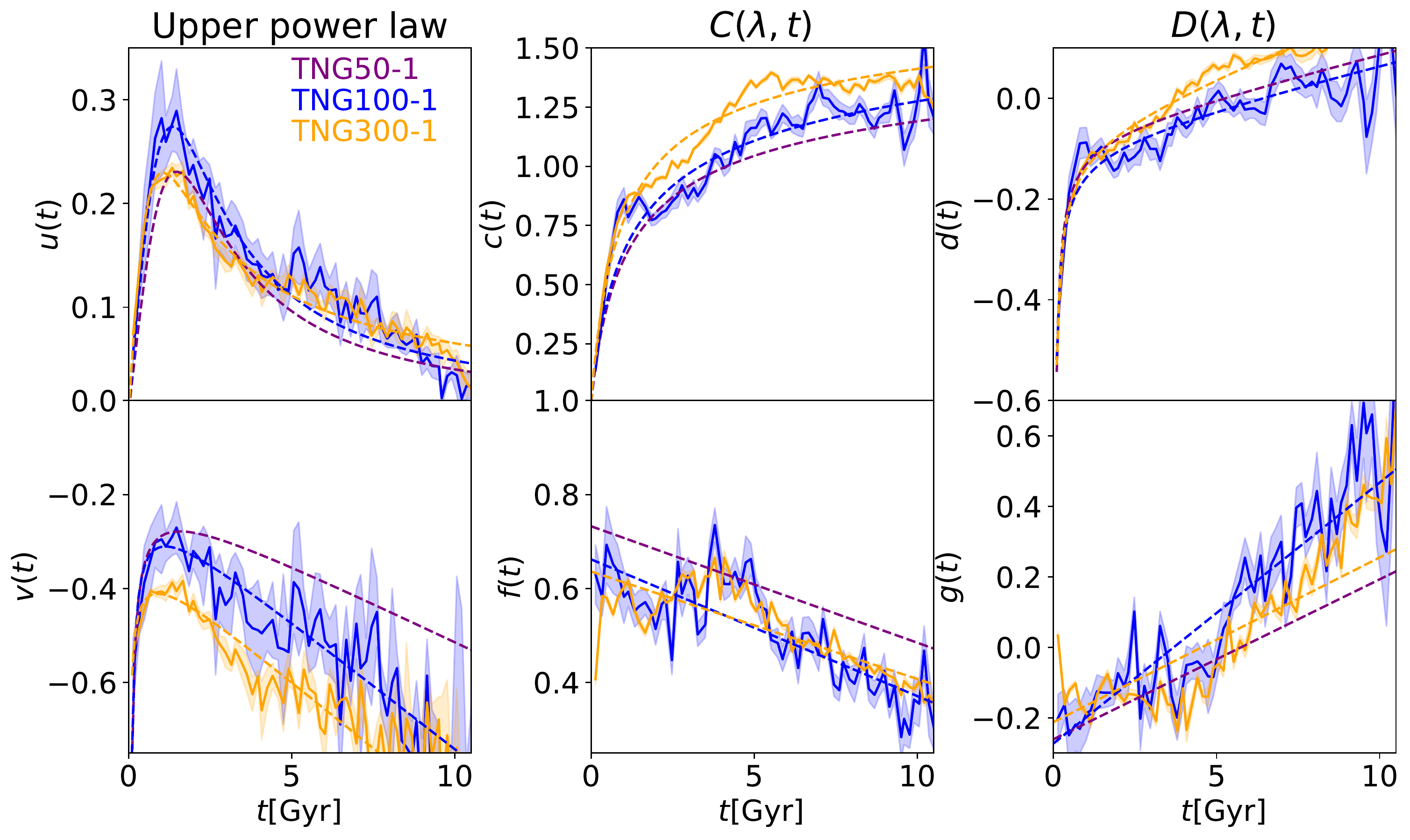}
    \caption{Parameters of the Type 2 model as a function of time,
    for TNG50-1 (purple), TNG100-1 (blue) and TNG300-1 (orange).
    The solid lines with error bands show the fits for each snapshot,
    and the dashed lines the fits as a function of time.
    We do not show the individual snapshot fits for TNG50-1 for clarity, 
    as they have large scatter and uncertainty.}
    \label{fig:t2_pars_all}
\end{figure}

Here we show the parameterisations of the Type 2 model given in Section \ref{sec:t2_infall}
for TNG50-1, TNG100-1 and TNG300-1.
For TNG100-1 and TNG300-1 we show in Fig.~\ref{fig:t2_pars_all} the fits at each snapshot
as solid lines with errorbars, and the overall relation with a dashed line of the same colour.
With TNG50-1 we only show the overall relation,
as the scatter and uncertainties across individual snapshots are large.

It can be seen that the overall trends in the parameters as a function of time
are the same for each resolution.
However, the exact values vary, particularly at larger timesteps
for $v(t)$ and $c(t)$.
This is likely to show the covariances between our parameters,
and also perhaps an effect of different halo and stellar mass selections in the simulations.

%% file: GroupProfiles.bbl
\begin{thebibliography}{}
\makeatletter
\relax
\def\mn@urlcharsother{\let\do\@makeother \do\$\do\&\do\#\do\^\do\_\do\%\do\~}
\def\mn@doi{\begingroup\mn@urlcharsother \@ifnextchar [ {\mn@doi@}
  {\mn@doi@[]}}
\def\mn@doi@[#1]#2{\def\@tempa{#1}\ifx\@tempa\@empty \href
  {http://dx.doi.org/#2} {doi:#2}\else \href {http://dx.doi.org/#2} {#1}\fi
  \endgroup}
\def\mn@eprint#1#2{\mn@eprint@#1:#2::\@nil}
\def\mn@eprint@arXiv#1{\href {http://arxiv.org/abs/#1} {{\tt arXiv:#1}}}
\def\mn@eprint@dblp#1{\href {http://dblp.uni-trier.de/rec/bibtex/#1.xml}
  {dblp:#1}}
\def\mn@eprint@#1:#2:#3:#4\@nil{\def\@tempa {#1}\def\@tempb {#2}\def\@tempc
  {#3}\ifx \@tempc \@empty \let \@tempc \@tempb \let \@tempb \@tempa \fi \ifx
  \@tempb \@empty \def\@tempb {arXiv}\fi \@ifundefined
  {mn@eprint@\@tempb}{\@tempb:\@tempc}{\expandafter \expandafter \csname
  mn@eprint@\@tempb\endcsname \expandafter{\@tempc}}}

\bibitem[\protect\citeauthoryear{{Adhikari} et~al.,}{{Adhikari}
  et~al.}{2021}]{Adhikari2020}
{Adhikari} S.,  et~al., 2021, \mn@doi [\apj] {10.3847/1538-4357/ac0bbc}, \href
  {https://ui.adsabs.harvard.edu/abs/2021ApJ...923...37A} {923, 37}

\bibitem[\protect\citeauthoryear{{Anbajagane}, {Evrard}  \&
  {Farahi}}{{Anbajagane} et~al.}{2022}]{Anbajagane2022}
{Anbajagane} D.,  {Evrard} A.~E.,   {Farahi} A.,  2022, \mn@doi [\mnras]
  {10.1093/mnras/stab3177}, \href
  {https://ui.adsabs.harvard.edu/abs/2022MNRAS.509.3441A} {509, 3441}

\bibitem[\protect\citeauthoryear{{Angulo}, {Lacey}, {Baugh}  \&
  {Frenk}}{{Angulo} et~al.}{2009}]{Angulo2009}
{Angulo} R.~E.,  {Lacey} C.~G.,  {Baugh} C.~M.,   {Frenk} C.~S.,  2009, \mn@doi
  [\mnras] {10.1111/j.1365-2966.2009.15333.x}, \href
  {https://ui.adsabs.harvard.edu/abs/2009MNRAS.399..983A} {399, 983}

\bibitem[\protect\citeauthoryear{{Ayromlou}, {Nelson}, {Yates}, {Kauffmann}  \&
  {White}}{{Ayromlou} et~al.}{2019}]{Ayromlou2019}
{Ayromlou} M.,  {Nelson} D.,  {Yates} R.~M.,  {Kauffmann} G.,   {White} S.
  D.~M.,  2019, \mn@doi [\mnras] {10.1093/mnras/stz1549}, \href
  {https://ui.adsabs.harvard.edu/abs/2019MNRAS.487.4313A} {487, 4313}

\bibitem[\protect\citeauthoryear{{Ayromlou}, {Nelson}, {Yates}, {Kauffmann},
  {Renneby}  \& {White}}{{Ayromlou} et~al.}{2021}]{Ayromlou2021}
{Ayromlou} M.,  {Nelson} D.,  {Yates} R.~M.,  {Kauffmann} G.,  {Renneby} M.,
  {White} S. D.~M.,  2021, \mn@doi [\mnras] {10.1093/mnras/staa4011}, \href
  {https://ui.adsabs.harvard.edu/abs/2021MNRAS.502.1051A} {502, 1051}

\bibitem[\protect\citeauthoryear{{Bah{\'e}} et~al.,}{{Bah{\'e}}
  et~al.}{2019}]{Bahe2019}
{Bah{\'e}} Y.~M.,  et~al., 2019, \mn@doi [\mnras] {10.1093/mnras/stz361}, \href
  {https://ui.adsabs.harvard.edu/abs/2019MNRAS.485.2287B} {485, 2287}

\bibitem[\protect\citeauthoryear{{Baldry} et~al.,}{{Baldry}
  et~al.}{2018}]{Baldry2018}
{Baldry} I.~K.,  et~al., 2018, \mn@doi [\mnras] {10.1093/mnras/stx3042}, \href
  {https://ui.adsabs.harvard.edu/abs/2018MNRAS.474.3875B} {474, 3875}

\bibitem[\protect\citeauthoryear{{Behroozi}, {Wechsler}, {Hearin}  \&
  {Conroy}}{{Behroozi} et~al.}{2019}]{Behroozi2019}
{Behroozi} P.,  {Wechsler} R.~H.,  {Hearin} A.~P.,   {Conroy} C.,  2019,
  \mn@doi [\mnras] {10.1093/mnras/stz1182}, \href
  {https://ui.adsabs.harvard.edu/abs/2019MNRAS.488.3143B} {488, 3143}

\bibitem[\protect\citeauthoryear{{Binney} \& {Tremaine}}{{Binney} \&
  {Tremaine}}{1987}]{Binney1987}
{Binney} J.,  {Tremaine} S.,  1987, {Galactic dynamics}.
{Princeton, N.J.: Princeton University Press}

\bibitem[\protect\citeauthoryear{{Bose}, {Eisenstein}, {Hernquist},
  {Pillepich}, {Nelson}, {Marinacci}, {Springel}  \& {Vogelsberger}}{{Bose}
  et~al.}{2019}]{Bose2019}
{Bose} S.,  {Eisenstein} D.~J.,  {Hernquist} L.,  {Pillepich} A.,  {Nelson} D.,
   {Marinacci} F.,  {Springel} V.,   {Vogelsberger} M.,  2019, \mn@doi [\mnras]
  {10.1093/mnras/stz2546}, \href
  {https://ui.adsabs.harvard.edu/abs/2019MNRAS.490.5693B} {490, 5693}

\bibitem[\protect\citeauthoryear{{Bose}, {Deason}, {Belokurov}  \&
  {Frenk}}{{Bose} et~al.}{2020}]{Bose2020}
{Bose} S.,  {Deason} A.~J.,  {Belokurov} V.,   {Frenk} C.~S.,  2020, \mn@doi
  [\mnras] {10.1093/mnras/staa1199}, \href
  {https://ui.adsabs.harvard.edu/abs/2020MNRAS.495..743B} {495, 743}

\bibitem[\protect\citeauthoryear{{Bower}, {Benson}, {Malbon}, {Helly}, {Frenk},
  {Baugh}, {Cole}  \& {Lacey}}{{Bower} et~al.}{2006}]{Bower2006}
{Bower} R.~G.,  {Benson} A.~J.,  {Malbon} R.,  {Helly} J.~C.,  {Frenk} C.~S.,
  {Baugh} C.~M.,  {Cole} S.,   {Lacey} C.~G.,  2006, \mn@doi [\mnras]
  {10.1111/j.1365-2966.2006.10519.x}, \href
  {https://ui.adsabs.harvard.edu/abs/2006MNRAS.370..645B} {370, 645}

\bibitem[\protect\citeauthoryear{{Bryan}, {Kay}, {Duffy}, {Schaye}, {Dalla
  Vecchia}  \& {Booth}}{{Bryan} et~al.}{2013}]{Bryan2013}
{Bryan} S.~E.,  {Kay} S.~T.,  {Duffy} A.~R.,  {Schaye} J.,  {Dalla Vecchia} C.,
    {Booth} C.~M.,  2013, \mn@doi [\mnras] {10.1093/mnras/sts587}, \href
  {https://ui.adsabs.harvard.edu/abs/2013MNRAS.429.3316B} {429, 3316}

\bibitem[\protect\citeauthoryear{{Budzynski}, {Koposov}, {McCarthy}, {McGee}
  \& {Belokurov}}{{Budzynski} et~al.}{2012}]{Budzynski2012}
{Budzynski} J.~M.,  {Koposov} S.~E.,  {McCarthy} I.~G.,  {McGee} S.~L.,
  {Belokurov} V.,  2012, \mn@doi [\mnras] {10.1111/j.1365-2966.2012.20663.x},
  \href {https://ui.adsabs.harvard.edu/abs/2012MNRAS.423..104B} {423, 104}

\bibitem[\protect\citeauthoryear{{Chua}, {Pillepich}, {Rodriguez-Gomez},
  {Vogelsberger}, {Bird}  \& {Hernquist}}{{Chua} et~al.}{2017}]{Chua2017}
{Chua} K. T.~E.,  {Pillepich} A.,  {Rodriguez-Gomez} V.,  {Vogelsberger} M.,
  {Bird} S.,   {Hernquist} L.,  2017, \mn@doi [\mnras] {10.1093/mnras/stx2238},
  \href {https://ui.adsabs.harvard.edu/abs/2017MNRAS.472.4343C} {472, 4343}

\bibitem[\protect\citeauthoryear{{Chua}, {Pillepich}, {Vogelsberger}  \&
  {Hernquist}}{{Chua} et~al.}{2019}]{Chua2019}
{Chua} K. T.~E.,  {Pillepich} A.,  {Vogelsberger} M.,   {Hernquist} L.,  2019,
  \mn@doi [\mnras] {10.1093/mnras/sty3531}, \href
  {https://ui.adsabs.harvard.edu/abs/2019MNRAS.484..476C} {484, 476}

\bibitem[\protect\citeauthoryear{{Chua}, {Vogelsberger}, {Pillepich}  \&
  {Hernquist}}{{Chua} et~al.}{2021}]{Chua2021}
{Chua} K. T.~E.,  {Vogelsberger} M.,  {Pillepich} A.,   {Hernquist} L.,  2021,
  arXiv e-prints, \href {https://ui.adsabs.harvard.edu/abs/2021arXiv210900012C}
  {p. arXiv:2109.00012}

\bibitem[\protect\citeauthoryear{{Crain} et~al.,}{{Crain}
  et~al.}{2015}]{Crain2015}
{Crain} R.~A.,  et~al., 2015, \mn@doi [\mnras] {10.1093/mnras/stv725}, \href
  {https://ui.adsabs.harvard.edu/abs/2015MNRAS.450.1937C} {450, 1937}

\bibitem[\protect\citeauthoryear{{Despali} \& {Vegetti}}{{Despali} \&
  {Vegetti}}{2017}]{Despali2017}
{Despali} G.,  {Vegetti} S.,  2017, \mn@doi [\mnras] {10.1093/mnras/stx966},
  \href {https://ui.adsabs.harvard.edu/abs/2017MNRAS.469.1997D} {469, 1997}

\bibitem[\protect\citeauthoryear{{Dolag}, {Borgani}, {Murante}  \&
  {Springel}}{{Dolag} et~al.}{2009}]{Dolag2009}
{Dolag} K.,  {Borgani} S.,  {Murante} G.,   {Springel} V.,  2009, \mn@doi
  [\mnras] {10.1111/j.1365-2966.2009.15034.x}, \href
  {https://ui.adsabs.harvard.edu/abs/2009MNRAS.399..497D} {399, 497}

\bibitem[\protect\citeauthoryear{{Donnari} et~al.,}{{Donnari}
  et~al.}{2021}]{Donnari2021}
{Donnari} M.,  et~al., 2021, \mn@doi [\mnras] {10.1093/mnras/staa3006}, \href
  {https://ui.adsabs.harvard.edu/abs/2021MNRAS.500.4004D} {500, 4004}

\bibitem[\protect\citeauthoryear{{Driver} et~al.,}{{Driver}
  et~al.}{2009}]{Driver2009}
{Driver} S.~P.,  et~al., 2009, \mn@doi [Astronomy and Geophysics]
  {10.1111/j.1468-4004.2009.50512.x}, \href
  {https://ui.adsabs.harvard.edu/abs/2009A&G....50e..12D} {50, 5.12}

\bibitem[\protect\citeauthoryear{{Driver} et~al.,}{{Driver}
  et~al.}{2011}]{Driver2011}
{Driver} S.~P.,  et~al., 2011, \mn@doi [\mnras]
  {10.1111/j.1365-2966.2010.18188.x}, \href
  {https://ui.adsabs.harvard.edu/abs/2011MNRAS.413..971D} {413, 971}

\bibitem[\protect\citeauthoryear{{Driver} et~al.,}{{Driver}
  et~al.}{2019}]{Driver2019}
{Driver} S.~P.,  et~al., 2019, \mn@doi [The Messenger]
  {10.18727/0722-6691/5126}, \href
  {https://ui.adsabs.harvard.edu/abs/2019Msngr.175...46D} {175, 46}

\bibitem[\protect\citeauthoryear{{Driver} et~al.,}{{Driver}
  et~al.}{2022a}]{Driver2022b}
{Driver} S.~P.,  et~al., 2022a, arXiv e-prints, \href
  {https://ui.adsabs.harvard.edu/abs/2022arXiv220308540D} {p. arXiv:2203.08540}

\bibitem[\protect\citeauthoryear{{Driver} et~al.,}{{Driver}
  et~al.}{2022b}]{Driver2022a}
{Driver} S.~P.,  et~al., 2022b, \mn@doi [\mnras] {10.1093/mnras/stac472}, \href
  {https://ui.adsabs.harvard.edu/abs/2022MNRAS.513..439D} {513, 439}

\bibitem[\protect\citeauthoryear{{Einasto}}{{Einasto}}{1965}]{Einasto1965}
{Einasto} J.,  1965, Trudy Astrofizicheskogo Instituta Alma-Ata, \href
  {https://ui.adsabs.harvard.edu/abs/1965TrAlm...5...87E} {5, 87}

\bibitem[\protect\citeauthoryear{{Elahi}, {Ca{\~n}as}, {Poulton}, {Tobar},
  {Willis}, {Lagos}, {Power}  \& {Robotham}}{{Elahi} et~al.}{2019}]{Elahi2019a}
{Elahi} P.~J.,  {Ca{\~n}as} R.,  {Poulton} R. J.~J.,  {Tobar} R.~J.,  {Willis}
  J.~S.,  {Lagos} C. d.~P.,  {Power} C.,   {Robotham} A. S.~G.,  2019, \mn@doi
  [\pasa] {10.1017/pasa.2019.12}, \href
  {https://ui.adsabs.harvard.edu/abs/2019PASA...36...21E} {36, e021}

\bibitem[\protect\citeauthoryear{{Emami} et~al.,}{{Emami}
  et~al.}{2021}]{Emami2021}
{Emami} R.,  et~al., 2021, \mn@doi [\apj] {10.3847/1538-4357/abf147}, \href
  {https://ui.adsabs.harvard.edu/abs/2021ApJ...913...36E} {913, 36}

\bibitem[\protect\citeauthoryear{{Engler} et~al.,}{{Engler}
  et~al.}{2021}]{Engler2021a}
{Engler} C.,  et~al., 2021, \mn@doi [\mnras] {10.1093/mnras/staa3505}, \href
  {https://ui.adsabs.harvard.edu/abs/2021MNRAS.500.3957E} {500, 3957}

\bibitem[\protect\citeauthoryear{{Fisher}, {Davis}, {Strauss}, {Yahil}  \&
  {Huchra}}{{Fisher} et~al.}{1994}]{Fisher1994}
{Fisher} K.~B.,  {Davis} M.,  {Strauss} M.~A.,  {Yahil} A.,   {Huchra} J.,
  1994, \mn@doi [\mnras] {10.1093/mnras/266.1.50}, \href
  {https://ui.adsabs.harvard.edu/abs/1994MNRAS.266...50F} {266, 50}

\bibitem[\protect\citeauthoryear{{Garrison-Kimmel} et~al.,}{{Garrison-Kimmel}
  et~al.}{2017}]{GarrisonKimmel2017}
{Garrison-Kimmel} S.,  et~al., 2017, \mn@doi [\mnras] {10.1093/mnras/stx1710},
  \href {https://ui.adsabs.harvard.edu/abs/2017MNRAS.471.1709G} {471, 1709}

\bibitem[\protect\citeauthoryear{{Grand} et~al.,}{{Grand}
  et~al.}{2021}]{Grand2021}
{Grand} R. J.~J.,  et~al., 2021, \mn@doi [\mnras] {10.1093/mnras/stab2492},
  \href {https://ui.adsabs.harvard.edu/abs/2021MNRAS.507.4953G} {507, 4953}

\bibitem[\protect\citeauthoryear{{Gu} et~al.,}{{Gu} et~al.}{2016}]{Gu2016}
{Gu} L.,  et~al., 2016, \mn@doi [\apj] {10.3847/0004-637X/826/1/72}, \href
  {https://ui.adsabs.harvard.edu/abs/2016ApJ...826...72G} {826, 72}

\bibitem[\protect\citeauthoryear{{Guo}, {Cole}, {Eke}  \& {Frenk}}{{Guo}
  et~al.}{2012}]{Guo2012}
{Guo} Q.,  {Cole} S.,  {Eke} V.,   {Frenk} C.,  2012, \mn@doi [\mnras]
  {10.1111/j.1365-2966.2012.21882.x}, \href
  {https://ui.adsabs.harvard.edu/abs/2012MNRAS.427..428G} {427, 428}

\bibitem[\protect\citeauthoryear{{Hadzhiyska}, {Bose}, {Eisenstein},
  {Hernquist}  \& {Spergel}}{{Hadzhiyska} et~al.}{2020}]{Hadzhiyska2020}
{Hadzhiyska} B.,  {Bose} S.,  {Eisenstein} D.,  {Hernquist} L.,   {Spergel}
  D.~N.,  2020, \mn@doi [\mnras] {10.1093/mnras/staa623}, \href
  {https://ui.adsabs.harvard.edu/abs/2020MNRAS.493.5506H} {493, 5506}

\bibitem[\protect\citeauthoryear{{Haggar}, {Pearce}, {Gray}, {Knebe}  \&
  {Yepes}}{{Haggar} et~al.}{2021}]{Haggar2021}
{Haggar} R.,  {Pearce} F.~R.,  {Gray} M.~E.,  {Knebe} A.,   {Yepes} G.,  2021,
  \mn@doi [\mnras] {10.1093/mnras/stab064}, \href
  {https://ui.adsabs.harvard.edu/abs/2021MNRAS.502.1191H} {502, 1191}

\bibitem[\protect\citeauthoryear{{Hansen}, {McKay}, {Wechsler}, {Annis},
  {Sheldon}  \& {Kimball}}{{Hansen} et~al.}{2005}]{Hansen2005}
{Hansen} S.~M.,  {McKay} T.~A.,  {Wechsler} R.~H.,  {Annis} J.,  {Sheldon}
  E.~S.,   {Kimball} A.,  2005, \mn@doi [\apj] {10.1086/444554}, \href
  {https://ui.adsabs.harvard.edu/abs/2005ApJ...633..122H} {633, 122}

\bibitem[\protect\citeauthoryear{{Henriques}, {White}, {Thomas}, {Angulo},
  {Guo}, {Lemson}, {Springel}  \& {Overzier}}{{Henriques}
  et~al.}{2015}]{Henriques2015}
{Henriques} B. M.~B.,  {White} S. D.~M.,  {Thomas} P.~A.,  {Angulo} R.,  {Guo}
  Q.,  {Lemson} G.,  {Springel} V.,   {Overzier} R.,  2015, \mn@doi [\mnras]
  {10.1093/mnras/stv705}, \href
  {https://ui.adsabs.harvard.edu/abs/2015MNRAS.451.2663H} {451, 2663}

\bibitem[\protect\citeauthoryear{{Joshi}, {Parker}, {Wadsley}  \&
  {Keller}}{{Joshi} et~al.}{2019}]{Joshi2019}
{Joshi} G.~D.,  {Parker} L.~C.,  {Wadsley} J.,   {Keller} B.~W.,  2019, \mn@doi
  [\mnras] {10.1093/mnras/sty3119}, \href
  {https://ui.adsabs.harvard.edu/abs/2019MNRAS.483..235J} {483, 235}

\bibitem[\protect\citeauthoryear{{Kafle} et~al.,}{{Kafle}
  et~al.}{2016}]{Kafle2016}
{Kafle} P.~R.,  et~al., 2016, \mn@doi [\mnras] {10.1093/mnras/stw2290}, \href
  {https://ui.adsabs.harvard.edu/abs/2016MNRAS.463.4194K} {463, 4194}

\bibitem[\protect\citeauthoryear{{Kelley}, {Bullock}, {Garrison-Kimmel},
  {Boylan-Kolchin}, {Pawlowski}  \& {Graus}}{{Kelley}
  et~al.}{2019}]{Kelley2019}
{Kelley} T.,  {Bullock} J.~S.,  {Garrison-Kimmel} S.,  {Boylan-Kolchin} M.,
  {Pawlowski} M.~S.,   {Graus} A.~S.,  2019, \mn@doi [\mnras]
  {10.1093/mnras/stz1553}, \href
  {https://ui.adsabs.harvard.edu/abs/2019MNRAS.487.4409K} {487, 4409}

\bibitem[\protect\citeauthoryear{{Krone-Martins}, {Ishida}  \& {de
  Souza}}{{Krone-Martins} et~al.}{2014}]{KroneMartins2014}
{Krone-Martins} A.,  {Ishida} E.~E.~O.,   {de Souza} R.~S.,  2014, \mn@doi
  [\mnras] {10.1093/mnrasl/slu067}, \href
  {https://ui.adsabs.harvard.edu/abs/2014MNRAS.443L..34K} {443, L34}

\bibitem[\protect\citeauthoryear{{Lacey} et~al.,}{{Lacey}
  et~al.}{2016}]{Lacey2016}
{Lacey} C.~G.,  et~al., 2016, \mn@doi [\mnras] {10.1093/mnras/stw1888}, \href
  {https://ui.adsabs.harvard.edu/abs/2016MNRAS.462.3854L} {462, 3854}

\bibitem[\protect\citeauthoryear{{Lagos}, {Tobar}, {Robotham}, {Obreschkow},
  {Mitchell}, {Power}  \& {Elahi}}{{Lagos} et~al.}{2018}]{Lagos2018}
{Lagos} C. d.~P.,  {Tobar} R.~J.,  {Robotham} A. S.~G.,  {Obreschkow} D.,
  {Mitchell} P.~D.,  {Power} C.,   {Elahi} P.~J.,  2018, \mn@doi [\mnras]
  {10.1093/mnras/sty2440}, \href
  {https://ui.adsabs.harvard.edu/abs/2018MNRAS.481.3573L} {481, 3573}

\bibitem[\protect\citeauthoryear{{Lin}, {Jing}, {Mao}, {Gao}  \&
  {McCarthy}}{{Lin} et~al.}{2006}]{Lin2006}
{Lin} W.~P.,  {Jing} Y.~P.,  {Mao} S.,  {Gao} L.,   {McCarthy} I.~G.,  2006,
  \mn@doi [\apj] {10.1086/508052}, \href
  {https://ui.adsabs.harvard.edu/abs/2006ApJ...651..636L} {651, 636}

\bibitem[\protect\citeauthoryear{{Liske} et~al.,}{{Liske}
  et~al.}{2015}]{Liske2015}
{Liske} J.,  et~al., 2015, \mn@doi [\mnras] {10.1093/mnras/stv1436}, \href
  {https://ui.adsabs.harvard.edu/abs/2015MNRAS.452.2087L} {452, 2087}

\bibitem[\protect\citeauthoryear{{{\L}okas}}{{{\L}okas}}{2020}]{Lokas2020}
{{\L}okas} E.~L.,  2020, \mn@doi [\aap] {10.1051/0004-6361/202037643}, \href
  {https://ui.adsabs.harvard.edu/abs/2020A&A...638A.133L} {638, A133}

\bibitem[\protect\citeauthoryear{{Lovell} et~al.,}{{Lovell}
  et~al.}{2018}]{Lovell2018}
{Lovell} M.~R.,  et~al., 2018, \mn@doi [\mnras] {10.1093/mnras/sty2339}, \href
  {https://ui.adsabs.harvard.edu/abs/2018MNRAS.481.1950L} {481, 1950}

\bibitem[\protect\citeauthoryear{{Manwadkar} \& {Kravtsov}}{{Manwadkar} \&
  {Kravtsov}}{2021}]{Manwadkar2021}
{Manwadkar} V.,  {Kravtsov} A.,  2021, arXiv e-prints, \href
  {https://ui.adsabs.harvard.edu/abs/2021arXiv211204511M} {p. arXiv:2112.04511}

\bibitem[\protect\citeauthoryear{{Marinacci} et~al.,}{{Marinacci}
  et~al.}{2018}]{Marinacci2018}
{Marinacci} F.,  et~al., 2018, \mn@doi [\mnras] {10.1093/mnras/sty2206}, \href
  {https://ui.adsabs.harvard.edu/abs/2018MNRAS.480.5113M} {480, 5113}

\bibitem[\protect\citeauthoryear{{Marini} et~al.,}{{Marini}
  et~al.}{2021}]{Marini2021}
{Marini} I.,  et~al., 2021, \mn@doi [\mnras] {10.1093/mnras/staa3486}, \href
  {https://ui.adsabs.harvard.edu/abs/2021MNRAS.500.3462M} {500, 3462}

\bibitem[\protect\citeauthoryear{{Nagai} \& {Kravtsov}}{{Nagai} \&
  {Kravtsov}}{2005}]{Nagai2005}
{Nagai} D.,  {Kravtsov} A.~V.,  2005, \mn@doi [\apj] {10.1086/426016}, \href
  {https://ui.adsabs.harvard.edu/abs/2005ApJ...618..557N} {618, 557}

\bibitem[\protect\citeauthoryear{{Naiman} et~al.,}{{Naiman}
  et~al.}{2018}]{Naiman2018}
{Naiman} J.~P.,  et~al., 2018, \mn@doi [\mnras] {10.1093/mnras/sty618}, \href
  {https://ui.adsabs.harvard.edu/abs/2018MNRAS.477.1206N} {477, 1206}

\bibitem[\protect\citeauthoryear{{Nelson} et~al.,}{{Nelson}
  et~al.}{2015}]{Nelson2015}
{Nelson} D.,  et~al., 2015, \mn@doi [Astronomy and Computing]
  {10.1016/j.ascom.2015.09.003}, \href
  {https://ui.adsabs.harvard.edu/abs/2015A&C....13...12N} {13, 12}

\bibitem[\protect\citeauthoryear{{Nelson} et~al.,}{{Nelson}
  et~al.}{2018}]{Nelson2018}
{Nelson} D.,  et~al., 2018, \mn@doi [\mnras] {10.1093/mnras/stx3040}, \href
  {https://ui.adsabs.harvard.edu/abs/2018MNRAS.475..624N} {475, 624}

\bibitem[\protect\citeauthoryear{{Nelson} et~al.,}{{Nelson}
  et~al.}{2019a}]{Nelson2019a}
{Nelson} D.,  et~al., 2019a, \mn@doi [Computational Astrophysics and Cosmology]
  {10.1186/s40668-019-0028-x}, \href
  {https://ui.adsabs.harvard.edu/abs/2019ComAC...6....2N} {6, 2}

\bibitem[\protect\citeauthoryear{{Nelson} et~al.,}{{Nelson}
  et~al.}{2019b}]{Nelson2019b}
{Nelson} D.,  et~al., 2019b, \mn@doi [\mnras] {10.1093/mnras/stz2306}, \href
  {https://ui.adsabs.harvard.edu/abs/2019MNRAS.490.3234N} {490, 3234}

\bibitem[\protect\citeauthoryear{{Onions} et~al.,}{{Onions}
  et~al.}{2012}]{Onions2012}
{Onions} J.,  et~al., 2012, \mn@doi [\mnras]
  {10.1111/j.1365-2966.2012.20947.x}, \href
  {https://ui.adsabs.harvard.edu/abs/2012MNRAS.423.1200O} {423, 1200}

\bibitem[\protect\citeauthoryear{{Pillepich} et~al.,}{{Pillepich}
  et~al.}{2018a}]{Pillepich2018a}
{Pillepich} A.,  et~al., 2018a, \mn@doi [\mnras] {10.1093/mnras/stx2656}, \href
  {https://ui.adsabs.harvard.edu/abs/2018MNRAS.473.4077P} {473, 4077}

\bibitem[\protect\citeauthoryear{{Pillepich} et~al.,}{{Pillepich}
  et~al.}{2018b}]{Pillepich2018}
{Pillepich} A.,  et~al., 2018b, \mn@doi [\mnras] {10.1093/mnras/stx3112}, \href
  {https://ui.adsabs.harvard.edu/abs/2018MNRAS.475..648P} {475, 648}

\bibitem[\protect\citeauthoryear{{Pillepich} et~al.,}{{Pillepich}
  et~al.}{2019}]{Pillepich2019}
{Pillepich} A.,  et~al., 2019, \mn@doi [\mnras] {10.1093/mnras/stz2338}, \href
  {https://ui.adsabs.harvard.edu/abs/2019MNRAS.490.3196P} {490, 3196}

\bibitem[\protect\citeauthoryear{{Pujol} et~al.,}{{Pujol}
  et~al.}{2017}]{Pujol2017}
{Pujol} A.,  et~al., 2017, \mn@doi [\mnras] {10.1093/mnras/stx913}, \href
  {https://ui.adsabs.harvard.edu/abs/2017MNRAS.469..749P} {469, 749}

\bibitem[\protect\citeauthoryear{{Rana}, {More}, {Miyatake}, {Nishimichi},
  {Takada}, {Robotham}, {Hopkins}  \& {Holwerda}}{{Rana}
  et~al.}{2022}]{Rana2022}
{Rana} D.,  {More} S.,  {Miyatake} H.,  {Nishimichi} T.,  {Takada} M.,
  {Robotham} A. S.~G.,  {Hopkins} A.~M.,   {Holwerda} B.~W.,  2022, \mn@doi
  [\mnras] {10.1093/mnras/stac007}, \href
  {https://ui.adsabs.harvard.edu/abs/2022MNRAS.510.5408R} {510, 5408}

\bibitem[\protect\citeauthoryear{{Rasia}, {Tormen}  \& {Moscardini}}{{Rasia}
  et~al.}{2004}]{Rasia2004}
{Rasia} E.,  {Tormen} G.,   {Moscardini} L.,  2004, \mn@doi [\mnras]
  {10.1111/j.1365-2966.2004.07775.x}, \href
  {https://ui.adsabs.harvard.edu/abs/2004MNRAS.351..237R} {351, 237}

\bibitem[\protect\citeauthoryear{{Renneby}, {Henriques}, {Hilbert}, {Nelson},
  {Vogelsberger}, {Angulo}, {Springel}  \& {Hernquist}}{{Renneby}
  et~al.}{2020}]{Renneby2020}
{Renneby} M.,  {Henriques} B. M.~B.,  {Hilbert} S.,  {Nelson} D.,
  {Vogelsberger} M.,  {Angulo} R.~E.,  {Springel} V.,   {Hernquist} L.,  2020,
  \mn@doi [\mnras] {10.1093/mnras/staa2675}, \href
  {https://ui.adsabs.harvard.edu/abs/2020MNRAS.498.5804R} {498, 5804}

\bibitem[\protect\citeauthoryear{{Rhee}, {Smith}, {Choi}, {Yi}, {Jaff{\'e}},
  {Candlish}  \& {S{\'a}nchez-J{\'a}nssen}}{{Rhee} et~al.}{2017}]{Rhee2017}
{Rhee} J.,  {Smith} R.,  {Choi} H.,  {Yi} S.~K.,  {Jaff{\'e}} Y.,  {Candlish}
  G.,   {S{\'a}nchez-J{\'a}nssen} R.,  2017, \mn@doi [\apj]
  {10.3847/1538-4357/aa6d6c}, \href
  {https://ui.adsabs.harvard.edu/abs/2017ApJ...843..128R} {843, 128}

\bibitem[\protect\citeauthoryear{{Riggs}, {Barbhuiyan}, {Loveday}, {Brough},
  {Holwerda}, {Hopkins}  \& {Phillipps}}{{Riggs} et~al.}{2021}]{Riggs2021}
{Riggs} S.~D.,  {Barbhuiyan} R.~W.~Y.~M.,  {Loveday} J.,  {Brough} S.,
  {Holwerda} B.~W.,  {Hopkins} A.~M.,   {Phillipps} S.,  2021, \mn@doi [\mnras]
  {10.1093/mnras/stab1697}, \href
  {https://ui.adsabs.harvard.edu/abs/2021MNRAS.506...21R} {506, 21}

\bibitem[\protect\citeauthoryear{{Robotham} et~al.,}{{Robotham}
  et~al.}{2011}]{Robotham2011}
{Robotham} A.~S.~G.,  et~al., 2011, \mn@doi [\mnras]
  {10.1111/j.1365-2966.2011.19217.x}, \href
  {https://ui.adsabs.harvard.edu/abs/2011MNRAS.416.2640R} {416, 2640}

\bibitem[\protect\citeauthoryear{{Rodriguez-Gomez} et~al.,}{{Rodriguez-Gomez}
  et~al.}{2015}]{RodriquezGomez2015}
{Rodriguez-Gomez} V.,  et~al., 2015, \mn@doi [\mnras] {10.1093/mnras/stv264},
  \href {https://ui.adsabs.harvard.edu/abs/2015MNRAS.449...49R} {449, 49}

\bibitem[\protect\citeauthoryear{{Roper}, {Thomas}  \& {Srisawat}}{{Roper}
  et~al.}{2020}]{Roper2020}
{Roper} W.~J.,  {Thomas} P.~A.,   {Srisawat} C.,  2020, \mn@doi [\mnras]
  {10.1093/mnras/staa982}, \href
  {https://ui.adsabs.harvard.edu/abs/2020MNRAS.494.4509R} {494, 4509}

\bibitem[\protect\citeauthoryear{{Sawala}, {Frenk}, {Crain}, {Jenkins},
  {Schaye}, {Theuns}  \& {Zavala}}{{Sawala} et~al.}{2013}]{Sawala2013}
{Sawala} T.,  {Frenk} C.~S.,  {Crain} R.~A.,  {Jenkins} A.,  {Schaye} J.,
  {Theuns} T.,   {Zavala} J.,  2013, \mn@doi [\mnras] {10.1093/mnras/stt259},
  \href {https://ui.adsabs.harvard.edu/abs/2013MNRAS.431.1366S} {431, 1366}

\bibitem[\protect\citeauthoryear{{Schaller} et~al.,}{{Schaller}
  et~al.}{2015}]{Schaller2015}
{Schaller} M.,  et~al., 2015, \mn@doi [\mnras] {10.1093/mnras/stv1067}, \href
  {https://ui.adsabs.harvard.edu/abs/2015MNRAS.451.1247S} {451, 1247}

\bibitem[\protect\citeauthoryear{{Schaye} et~al.,}{{Schaye}
  et~al.}{2015}]{Schaye2015}
{Schaye} J.,  et~al., 2015, \mn@doi [\mnras] {10.1093/mnras/stu2058}, \href
  {https://ui.adsabs.harvard.edu/abs/2015MNRAS.446..521S} {446, 521}

\bibitem[\protect\citeauthoryear{{Sheth} \& {Tormen}}{{Sheth} \&
  {Tormen}}{2004}]{Sheth2004}
{Sheth} R.~K.,  {Tormen} G.,  2004, \mn@doi [\mnras]
  {10.1111/j.1365-2966.2004.07733.x}, \href
  {https://ui.adsabs.harvard.edu/abs/2004MNRAS.350.1385S} {350, 1385}

\bibitem[\protect\citeauthoryear{{Shi} et~al.,}{{Shi} et~al.}{2020}]{Shi2020}
{Shi} J.,  et~al., 2020, \mn@doi [\apj] {10.3847/1538-4357/ab8464}, \href
  {https://ui.adsabs.harvard.edu/abs/2020ApJ...893..139S} {893, 139}

\bibitem[\protect\citeauthoryear{{Smith}, {Choi}, {Lee}, {Rhee},
  {Sanchez-Janssen}  \& {Yi}}{{Smith} et~al.}{2016}]{Smith2016}
{Smith} R.,  {Choi} H.,  {Lee} J.,  {Rhee} J.,  {Sanchez-Janssen} R.,   {Yi}
  S.~K.,  2016, \mn@doi [\apj] {10.3847/1538-4357/833/1/109}, \href
  {https://ui.adsabs.harvard.edu/abs/2016ApJ...833..109S} {833, 109}

\bibitem[\protect\citeauthoryear{{Springel}}{{Springel}}{2010}]{Springel2010}
{Springel} V.,  2010, \mn@doi [\mnras] {10.1111/j.1365-2966.2009.15715.x},
  \href {https://ui.adsabs.harvard.edu/abs/2010MNRAS.401..791S} {401, 791}

\bibitem[\protect\citeauthoryear{{Springel}, {White}, {Tormen}  \&
  {Kauffmann}}{{Springel} et~al.}{2001}]{Springel2001}
{Springel} V.,  {White} S. D.~M.,  {Tormen} G.,   {Kauffmann} G.,  2001,
  \mn@doi [\mnras] {10.1046/j.1365-8711.2001.04912.x}, \href
  {https://ui.adsabs.harvard.edu/abs/2001MNRAS.328..726S} {328, 726}

\bibitem[\protect\citeauthoryear{{Springel} et~al.,}{{Springel}
  et~al.}{2005}]{Springel2005}
{Springel} V.,  et~al., 2005, \mn@doi [\nat] {10.1038/nature03597}, \href
  {https://ui.adsabs.harvard.edu/abs/2005Natur.435..629S} {435, 629}

\bibitem[\protect\citeauthoryear{{Springel} et~al.,}{{Springel}
  et~al.}{2018}]{Springel2018}
{Springel} V.,  et~al., 2018, \mn@doi [\mnras] {10.1093/mnras/stx3304}, \href
  {https://ui.adsabs.harvard.edu/abs/2018MNRAS.475..676S} {475, 676}

\bibitem[\protect\citeauthoryear{{Springel}, {Pakmor}, {Zier}  \&
  {Reinecke}}{{Springel} et~al.}{2021}]{Springel2021}
{Springel} V.,  {Pakmor} R.,  {Zier} O.,   {Reinecke} M.,  2021, \mn@doi
  [\mnras] {10.1093/mnras/stab1855}, \href
  {https://ui.adsabs.harvard.edu/abs/2021MNRAS.506.2871S} {506, 2871}

\bibitem[\protect\citeauthoryear{{Tal}, {Wake}  \& {van Dokkum}}{{Tal}
  et~al.}{2012}]{Tal2012}
{Tal} T.,  {Wake} D.~A.,   {van Dokkum} P.~G.,  2012, \mn@doi [\apjl]
  {10.1088/2041-8205/751/1/L5}, \href
  {https://ui.adsabs.harvard.edu/abs/2012ApJ...751L...5T} {751, L5}

\bibitem[\protect\citeauthoryear{\VAN{Bosch}{Van}{van}{ den Bosch}, {Ogiya},
  {Hahn}  \& {Burkert}}{\VAN{Bosch}{Van}{van}{ den Bosch}
  et~al.}{2018}]{VanDenBosch2018a}
\VAN{Bosch}{Van}{van}{ den Bosch} F.~C.,  {Ogiya} G.,  {Hahn} O.,   {Burkert}
  A.,  2018, \mn@doi [\mnras] {10.1093/mnras/stx2956}, \href
  {https://ui.adsabs.harvard.edu/abs/2018MNRAS.474.3043V} {474, 3043}

\bibitem[\protect\citeauthoryear{{V{\'a}zquez-Mata} et~al.,}{{V{\'a}zquez-Mata}
  et~al.}{2020}]{VazquezMata2020}
{V{\'a}zquez-Mata} J.~A.,  et~al., 2020, \mn@doi [\mnras]
  {10.1093/mnras/staa2889}, \href
  {https://ui.adsabs.harvard.edu/abs/2020MNRAS.499..631V} {499, 631}

\bibitem[\protect\citeauthoryear{{Viola} et~al.,}{{Viola}
  et~al.}{2015}]{Viola2015}
{Viola} M.,  et~al., 2015, \mn@doi [\mnras] {10.1093/mnras/stv1447}, \href
  {https://ui.adsabs.harvard.edu/abs/2015MNRAS.452.3529V} {452, 3529}

\bibitem[\protect\citeauthoryear{{Vogelsberger} et~al.,}{{Vogelsberger}
  et~al.}{2014}]{Vogelsberger2014a}
{Vogelsberger} M.,  et~al., 2014, \mn@doi [\nat] {10.1038/nature13316}, \href
  {https://ui.adsabs.harvard.edu/abs/2014Natur.509..177V} {509, 177}

\bibitem[\protect\citeauthoryear{{Wang}, {Sales}, {Henriques}  \&
  {White}}{{Wang} et~al.}{2014}]{Wang2014}
{Wang} W.,  {Sales} L.~V.,  {Henriques} B. M.~B.,   {White} S. D.~M.,  2014,
  \mn@doi [\mnras] {10.1093/mnras/stu988}, \href
  {https://ui.adsabs.harvard.edu/abs/2014MNRAS.442.1363W} {442, 1363}

\bibitem[\protect\citeauthoryear{{Weinberg}, {Colombi}, {Dav{\'e}}  \&
  {Katz}}{{Weinberg} et~al.}{2008}]{Weinberg2008}
{Weinberg} D.~H.,  {Colombi} S.,  {Dav{\'e}} R.,   {Katz} N.,  2008, \mn@doi
  [\apj] {10.1086/524646}, \href
  {https://ui.adsabs.harvard.edu/abs/2008ApJ...678....6W} {678, 6}

\bibitem[\protect\citeauthoryear{{Weinberger} et~al.,}{{Weinberger}
  et~al.}{2017}]{Weinberger2017}
{Weinberger} R.,  et~al., 2017, \mn@doi [\mnras] {10.1093/mnras/stw2944}, \href
  {https://ui.adsabs.harvard.edu/abs/2017MNRAS.465.3291W} {465, 3291}

\bibitem[\protect\citeauthoryear{{Xu}, {Zehavi}  \& {Contreras}}{{Xu}
  et~al.}{2021}]{Xu2021}
{Xu} X.,  {Zehavi} I.,   {Contreras} S.,  2021, \mn@doi [\mnras]
  {10.1093/mnras/stab100}, \href
  {https://ui.adsabs.harvard.edu/abs/2021MNRAS.502.3242X} {502, 3242}

\bibitem[\protect\citeauthoryear{{Zehavi}, {Kerby}, {Contreras}, {Jim{\'e}nez},
  {Padilla}  \& {Baugh}}{{Zehavi} et~al.}{2019}]{Zehavi2019}
{Zehavi} I.,  {Kerby} S.~E.,  {Contreras} S.,  {Jim{\'e}nez} E.,  {Padilla} N.,
    {Baugh} C.~M.,  2019, \mn@doi [\apj] {10.3847/1538-4357/ab4d4d}, \href
  {https://ui.adsabs.harvard.edu/abs/2019ApJ...887...17Z} {887, 17}

\makeatother
\end{thebibliography}
